\documentclass[11pt, oneside]{article}
\usepackage{graphicx}
\usepackage{amsmath}
\usepackage{amsfonts}
\usepackage{amssymb}
\usepackage{amsbsy}
\usepackage{wrapfig}
\usepackage{times}
\usepackage{multirow}
\usepackage{footmisc}
\usepackage{rotating}
\usepackage{subfigure}
\usepackage{color}
\usepackage{setspace}
\usepackage{hyperref}
\usepackage{natbib}
\bibpunct{(}{)}{;}{a}{,}{,}

\newlength{\bigfigheight}
\newlength{\smallfigheight}

\title{Direct statistical simulation of low-order dynamo systems}
\author{Kuan Li$^1$, J.B. Marston$^2$, and Steven M. Tobias$^1$}
\date{}

\begin{document}

\maketitle

{\centering
1. Department of Applied Mathematics, University of Leeds, Leeds, LS2 9JT, UK \\
2. Brown Theoretical Physics Center and Department of Physics, \\
Box 1843, Brown University, Providence, Rhode Island 02912-1843, USA \\
}

\section*{Abstract}

In this paper we investigate the effectiveness of direct statistical simulation (DSS) for two low-order models of dynamo action. The first model, which is a simple model of solar and stellar dynamo action, is third-order and has cubic nonlinearities whilst the second has only quadratic nonlinearities and describes the interaction of convection and an aperiodically reversing magnetic field. We show how DSS can be utilised to solve for the statistics of these systems of equations both in the presence and the absence of stochastic terms, by truncating the cumulant hierarchy at either second or third order. We compare two different techniques for solving for the statistics, timestepping --- which is able to locate only stable solutions of the equations for the statistics and direct detection of the fixed points.  We develop a complete methodology and symbolic package in Python for deriving the statistical equations governing the Low-order dynamic systems in cumulant expansions. We demonstrate that although direct detection of the fixed points is efficient and accurate for DSS truncated at second order, the addition of higher order terms leads to the inclusion of many unstable fixed points that may be found by direct detection of the fixed point by iterative methods. In those cases timestepping is a more robust protocol for finding meaninful solutions to DSS.

\section{Introduction}

Magnetic fields in planets stars and galaxies are believed to be generated by the interaction of turbulent motions of electrically conducting fluids, rotation and magnetic fields. This complicated nonlinear interaction can, in theory, be modelled by the solution of nonlinear partial differential equations. However the need for a description on a vast range of spatial and temporal scales means that the pertinent parameter regime lies far beyond the capabilities of even modern supercomputers \citep[see e.g.][]{tobias_2021}. For this reason alternative approaches and descriptions are being investigated.

One such approach involves deriving and solving equations for the low-order statistics of the underlying system. This approach, termed Direct Statistical Simulation (DSS), has the advantage that the the low-order statistics are smoother in space  than the detailed dynamics; hence fewer spatial modes may be required for an accurate description of the statistics than the dynamics. Moreover the statistics are  more slowly-varying than the complicated dynamics and so may be described by the evolution on a slow-manifold (or perhaps even by a fixed point of the dynamical system describing DSS). However the derivation of the equations for DSS from the relevant PDEs is complicated, requiring either repeated differentiation of the Hopf functional equation \citep{hopf_1952} or repeated integrations to find the evolution of high-order correlations. Moreover, DSS suffers from the ``curse of dimensionality'', involving the solution of a hierarchy of equations for cumulants that have a higher dimension than the underlying fields described by the original PDEs. For these reasons, truncation of the relevant cumulant hierarchy needs to be effected as soon as convenient and  efficient methods for the solution of the system need to be developed. 

DSS has previously been utilised for a range of PDE models involving the interaction of mean flows and magnetic fields with turbulence. In most cases the systems described the interaction of turbulence with  a zonal mean flow or magnetic field; examples include models of the the driving of (magnetised) zonal barotropic jets by stochastic driving \citep{tdm2011}, driving of zonal flows in plasmas \citep{Weinstock_1969} and the turbulence and dynamo originating from the magnetorotational instability \citep{sb2015}. Because of the technical challenges of deriving and solving the DSS system, often the most straightforward implementation of DSS, termed CE2, has been performed and a complete investigation of the system for a range of parameters was not possible. CE2 has been shown to give an accurate description of the low-order statistics for systems with significant mean flows and fields that are close to statistical equilibrium. However, in certain cases away from statistical equilibrium CE2 is less accurate and fails to give a valid description \citep{tm2013}; higher order truncations are needed.

In this paper we investigate the effectiveness of DSS for two model problems in dynamo theory. We consider the simple case where the dynamics is described by the evolution of systems of ordinary differential equations (ODEs). Though these systems are obviously simplifications of the dynamics of the full geophysical or astrophysical dynamos, they serve as useful testbeds for evaluating methods of DSS. The simplicity of the systems allows the development of strategies for the implementation of DSS and of an understanding of the nature of the approximations used.  The two problems we consider display different model dynamics. In the first \citep{W_Smith_2005}, relevant to the solar dynamo, the initial instability is to oscillatory dynamo action, and oscillatory magnetic fields may then be modulated aperiodically. The second system models the interaction of convection with dynamo action in a geodynamo setting \citep{cm1993}. Here the initial dynamo bifurcation is stationary and chaos sets in via a global bifurcation that leads to reversals of the magnetic field reminiscent of that exhibited by the Earth.

This paper is organised as follows. In the next section the general formulation of Direct Statistical Simulation is described. This includes the method of derivation of the relevant equations in addition to the various methods of solution. In section 3 we introduce the low-order solar dynamo system, giving a description of the dynamics before describing how well our DSS strategies are able to provide an accurate description of the low-order statistics. In section 4 we perform a similar analysis of the convective dynamo system. We conclude in the Discussion section by suggesting the implications of  our results for a programme of DSS for PDE models.

\section{The cumulant representation of low-order magnetohydrodynamical systems}

In this section we describe how the low-order statistics of systems of ordinary differential equations can be accessed via DSS. The models we consider in the paper are simplified models of dynamo action and will be described in subsequent sections. In general such low-order models are derived either via a Galerkin truncation of the partial differential equation (PDE) system, e.g., see \citep{holmes_etal_2012}, \citep{Passos_2008} or via a normal form analysis \citep[see e.g.][]{Tobias_1995}


Both of the low-order systems we consider may be represented by a set of ordinary differential equations with up to cubic nonlinearities;  the $i$-th component of a cubic nonlinear system is written as
\begin{equation}
	d_t x_i =  \sum_j {\cal L}_{ij} \ x_j +   \sum_{j,k} {\cal Q}_{ijk} \ x_j x_k + \sum_{j,k,l} {\cal C}_{ijkl} \ x_j x_k x_l + f_i,
\label{govEq}
\end{equation}
where the coefficients of the cubic and quadratic nonlinear interactions are given by ${\cal C}_{ijkl}$ and ${\cal Q}_{ijk}$ respectively and ${\cal L}_{ij}$ is for the linear term.  Here we also include the possibility of a stochastic forcing, $f_i$,  that is assumed to be the independent Gaussian, $f_i \sim{\cal N}(\mu_i,  \sigma_i^2)$, is introduced to synthesize the unmodelled physical processes, where $\mu_i$ and $\sigma_i^2$ are the statistical mean and variance of $f_i$, respectively \citep[see e.g.][]{allawala_2016}.

\subsection{The cumulant expansion of low-order dynamical systems}

For a dissipative low-order dynamical system of the form given in equation~\ref{govEq} in the absence of noise, the solution often takes the form of an attractor given by a fixed point, periodic solution or chaotic attractor. For such solutions, and in the presence of noise, this attractor can be characterised by the calculation of a probability density function (PDF). The low-order statistics of the PDF may then be represented by the cumulants of the distribution  
\citep{Kendall_1987}. Consider separating each state variable via a
Reynolds decomposition, that is we represent the unknown field, $x_i$, as the sum of the coherent component, $C_{x_i} = \langle x_i \rangle $, and a non-coherent counterpart, $\delta x_i$, so that 
	\begin{equation}
			x_i = C_{x_i} + \delta {x_i},
		\label{rede}
	\end{equation}
where the statistical average of the fluctuation vanishes, i,e., $\langle \delta x_i \rangle =0$. We also assume that the statistical average further satisfies of the Reynolds averaging rules, i.e.,
\begin{equation}
\langle \delta x_i \rangle = 0, \hspace{0.05\hsize} \langle \delta x_i C_{x_j} \rangle = 0, \hspace{0.02\hsize} \text{and} \hspace{0.02\hsize} \langle x_i C_{x_j} \rangle = C_{x_i} C_{x_j}.
\end{equation}
In this paper, the ensemble average is employed to derive the cumulant equations of the low-order dynamical systems and is noted as $\langle \bullet \rangle$. It is then useful to define the higher cumulants  measuring the shape of the probability density function (PDF) using this averaging procedure \citet{Kendall_1987}. The first three cumulants are defined as  
\begin{equation}
C_{x_i} = \langle x_i  \rangle, \hspace{0.05\hsize} C_{x_i x_j} = \langle \delta x_i \delta x_j \rangle \hspace{0.02\hsize} \text{and} \hspace{0.02\hsize} C_{x_i x_j x_k} = \langle \delta x_i \delta x_j \delta x_k \rangle.
\end{equation}
In this study, we explicitly use cumulant expansions up to fourth order, where the fourth cumulant is defined as
\begin{equation}
C_{x_ix_jx_kx_l} = \langle \delta x_i \delta x_j \delta x_k \delta x_l \rangle - C_{x_ix_j} C_{x_kx_l}- C_{x_ix_k} C_{x_jx_l} - C_{x_jx_k} C_{x_ix_l}. 
\end{equation}

In this paper we derive the cumulant expansions of the governing dynamical equation (\ref{govEq}) directly from the definition of cumulants. Alternatively the cumulant equations can be obtained via the Hopf functional approach \citep{frisch_1995}. 

Taking the ensemble average of Eq. (\ref{govEq}), we obtain the first order cumulant equation for the coherent component, $C_{x_i}$,
\begin{eqnarray}
	d_t C_{x_i} = d_t \langle x_i \rangle &=& \sum_{j,k,l} {\cal C}_{ijkl} \ \langle x_j x_k x_l \rangle+ \sum_{j,k} {\cal Q}_{ijk} \ \langle x_j x_k \rangle + \sum_j {\cal L}_{ij} \ \langle x_j \rangle+ \langle f_i \rangle, \nonumber \\
	&=& \sum_{j,k,l} {\cal C}_{ijkl} \left[ C_{x_j} C_{x_k} C_{x_l} + C_{x_j} C_{x_kx_l} + C_{x_jx_k} C_{x_l} + C_{{x_j}{x_l}}C_{x_k} + C_{x_jx_kx_l} \right] \nonumber\\
	&+& \sum_{j,k} {\cal Q}_{ijk} \ \left[C_{x_j} C_{x_k}+C_{x_j x_k} \right] + \sum_j {\cal L}_{ij} \ C_{x_j}+ \mu_i.
\label{ce1}
\end{eqnarray}
The evolution of the fluctuation, $\delta x_i$, that is determined by subtracting Eq.  (\ref{ce1}) from (\ref{govEq}) is used to derive the high order cumulant equations. By multiplying $\delta x_m$ by the governing equation of $d_t \delta x_i$, $\delta x_i$ with $d_t \delta x_m$ and taking the ensemble average, we find that the second order equation satisfies
\begin{eqnarray}
d_t C_{x_ix_m} &=& \left\langle \delta x_m \  \frac{d}{dt} \delta x_i  + \delta x_i \  \frac{d}{dt} \delta x_m \right\rangle =\left\lbrace \left\langle \delta x_m \  \frac{d}{dt}  \ (x_i-C_{x_i})  \right\rangle \right\rbrace \nonumber \\
&=&  \left\lbrace \sum_{j,k,l} {\cal C}_{ijkl} \left[  C_{x_jx_kx_lx_m} + C_{x_j}C_{x_kx_lx_m} + C_{x_jx_kx_m}C_{x_l}+ C_{x_jx_lx_m}C_{x_k} + C_{x_j}C_{x_k}C_{x_lx_m} \right. \right. \nonumber \\
 && \left. + C_{x_j}C_{x_kx_m}C_{x_l}  + C_{x_jx_m}C_{x_k}C_{x_l} - C_{x_jx_k}C_{x_lx_m}  - C_{x_jx_l}C_{x_kx_m}   - C_{x_jx_m}C_{x_kx_l}
\right] \nonumber \\
&& +\sum_{j,k} {\cal Q}_{ijk} \left[ C_{x_j}C_{x_kx_m}+ C_{x_jx_m}C_{x_k} + C_{x_jx_kx_m} \right] + \left. \sum_j {\cal L}_{ij} \ C_{x_jx_m} \right\rbrace + 2 \left\langle \delta f_i \delta f_m \right\rangle,
 \label{ce2}
\end{eqnarray}
where the symbol $\lbrace \bullet \rbrace$ notes the symmetrization procedure by swamping the field $\delta x_i$ and $\delta x_m$ of Eq. (\ref{ce2}). Note that the stochastic force, $f_i$ is assumed to be  Gaussian and independent and so the stochastic force is only self-correlated in the second order equation, e.g., $\langle \delta f_i^2\rangle = \sigma_i^2$, $\langle \delta f_j \delta f_i\rangle = \langle \delta f_j \rangle \langle \delta f_i\rangle = 0$ and $\langle \delta x_i \delta f_i\rangle = \langle \delta x_i \rangle \langle \delta f_i\rangle = 0$.
The procedure for deriving the third order equation follows the same fashion as for the second order Eq. (\ref{ce2}), i.e.,
\begin{eqnarray}
d_t C_{x_ix_mx_n} &=& \left\langle \delta x_m \delta x_n \  \frac{d}{dt} \delta x_i  + \delta x_n \delta x_i \  \frac{d}{dt} \delta x_m + \delta x_i \delta x_m \  \frac{d}{dt} \delta x_n \right\rangle =\left\lbrace \left\langle \delta x_m \delta x_n \  \frac{d}{dt}  \ (x_i-C_{x_i}) \right\rangle \right\rbrace \nonumber \\
&=& \left\lbrace \sum_{j,k,l} {\cal C}_{ijkl} \  S_c+ \sum_{j,k} {\cal Q}_{ijk}  \ S_q + \sum_j {\cal L}_{ij} \ C_{x_jx_mx_n} \right\rbrace,
 \label{ce3}
\end{eqnarray}
where the symmetrization procedure noted by $\lbrace \bullet \rbrace$ involves all permutations of $\delta x_i$, $\delta x_m$ and $\delta x_n$. The terms, $S_c$ that couples with the cumulants of the first five orders and $S_q$ that couples with the first four represent the third order cumulant expansions of the cubic and quadratic terms. The derivation of these terms is tedious and detailed in Appendix \ref{demo}. The complexity of the cumulant expansions of the nonlinear terms increases rapidly as increasing the degree of nonlinearity and the truncation order. For this reason we have developed software to automate the derivation of the cumulant hierarchy for any ODE system (with up to cubic nonlinearity). This {\it Python} software for deriving the cumulant equations is available in the Supplementary material of this paper. The symbolic representation of the cumulant equations are further converted into the numerical functions for computation via {\it lambdify} in {\it sympy}. Using this approach the cumulant equations with sparse representations are solved with fewer computations than using the conventional method via the dense matrix-vector multiplications \citep[e.g., see][]{allawala_2016}.

\subsection{The statistical closures of the cumulant equations}
In a cumulant hierarchy, the expansion of the dynamical equation (\ref{govEq}) leads to an infinite set of coupled equations. Hence, a proper statistical closure must be chosen to truncate the cumulant expansion at the lowest possible order. For Gaussian distributions 
the cumulant hierarchy naturally truncates at second order; all statistics of order greater than two are zero.  For this case the cumulant equations describing the evolution of the first and second cumulant in Eq. (\ref{ce1}) and (\ref{ce2}) are called the CE2 system, where all higher order terms greater than two are neglected \citep[see e.g.][]{mqt2019}.

However it is certainly possible that the statistics of a dynamical systems is poorly represented by a Gaussian PDF. Many distributions exhibit strong asymmetry (skewness) or long tails (flatness) as we will see in \$\ref{solar} and \$\ref{disc}. For these problems, one may have to take the third order cumulants into consideration, setting the fourth order cumulant to zero $C_{x_ix_jx_kx_l}=0$, \citep{orszag_1970} i.e.,
\begin{equation}
0 = C_{x_ix_jx_kx_l} = \langle \delta x_i \delta x_j \delta x_k \delta x_l \rangle - \left( C_{x_ix_j} C_{x_kx_l} + C_{x_ix_k} C_{x_jx_l} + C_{x_jx_k} C_{x_ix_l}\right).
\end{equation}
Here the effects of the fourth order cumulants that are proportional to the rate of change (gradient) of $x_i$ \citep{monin_1975} are further modelled by a diffusion process, $-C_{x_ix_mx_n}/\tau_d$. The parameter, $\tau_d>0$, is known as the eddy damping parameter \citep{mqt2019}. For a cubic nonlinearity term, the third order expansion also involves the 5-point correlations that is set to zero for simplicity.   The third order cumulant equation (\ref{ce3}) may now be rewritten as 
\begin{eqnarray}
d_t C_{x_ix_mx_n} &=& \left\lbrace \sum_{j,k,l} {\cal C}_{ijkl} \  S_c + \sum_{j,k} {\cal Q}_{ijk}  \ S_q + \sum_j {\cal L}_{ij} \ C_{x_jx_mx_n} \right\rbrace - \frac{C_{x_ix_mx_n}}{\tau_d}.
 \label{ce3II}
\end{eqnarray}
The cumulant equation that consists of (\ref{ce1}), (\ref{ce2}) and (\ref{ce3II}) are known as CE3 approximation.

The CE3 equations are complicated and involve many interactions. They may be simplified slightly by assuming that the third cumulant evolves rapidly in comparison with the first and second cumulant. This 
means that Eq. (\ref{ce3II}) is further simplified to a diagnostic system by setting all time derivatives for the third cumulants to be zero, i.e. $d_t C_{x_ix_mx_n}=0$. A further simplification that leads to faster computation involved the neglect of  all terms involving the first order cumulants, $C_{x_i}$ in the equations for the third cumulant. The third order cumulants then are the solution of the diagnostic equation,
\begin{eqnarray}
0 = \left\lbrace \sum_{j,k,l} {\cal C}_{ijkl} \  S'_c + \sum_{j,k} {\cal Q}_{ijk}  \ S'_q + \sum_j {\cal L}_{ij} \ C_{x_jx_mx_n} \right\rbrace - \frac{C_{x_ix_mx_n}}{\tau_d}.
 \label{ce25}
\end{eqnarray}
This representation is then directly substituted into the cumulant equation of the second order in Eq. (\ref{ce2}), where $S'_c$ and $S'_q$ stand for the cumulant expansions, $S_c$ and $S_q$, without the first order cumulants.  This truncation that couples Eqs. (\ref{ce1}), (\ref{ce2}) and (\ref{ce25}) is named CE2.5 approximation \citep{mqt2019} and \citep{allawala_2020}.

We shall investigate how well the low-order statistics of the full distribution are captured by the solutions of the cumulant hierarchy in the approximations described above for two dynamo systems.

\subsection{The fixed points of the cumulant equations}

The cumulant hierarchy derived symbolically comprises ODEs that can be integrated forward in time using standard timestepping methods. This approach will determine stable solutions to the cumulant equations. It may also be of interest to determine the fixed points of the cumulant system. 
These fixed points are the invariant solutions to the governing equations and in general are not unique. There may indeed be many invariant solutions --- particularly for the higher order truncations of the hierarchy.

Many of the time-invariant solutions of the cumulant system are either unstable as we integrate CE2/2.5/3 equations in time or statistically non-realizable. Realizability is ensured if  the second  cumulant satisfies the Cauchy–Schwarz inequality, $C_{x_i}^2C_{x_j}^2 \ge C_{x_ix_j}^2$. Hence, a fundamental problem is to determine the number of realizable fixed points that exist in the cumulant equations, whether or not they are stable, and how to obtain them efficiently.

In order to determine the fixed points, we utilise the misfit functional, ${\cal J}$, to measure the temporal variation of the cumulant equations, i.e., for CE2/2.5 and CE3 approximations, the misfit, ${\cal J}$, is defined as
\begin{eqnarray}
	{\cal J} &=& \frac{1}{2} \left[ \sum_i \left[\frac{d}{d t} C_{x_i}\right]^2 + \sum_{i,j}\left[\frac{d}{d t} C_{x_ix_j}\right]^2  \right], \nonumber\\
	{\cal J} &=& \frac{1}{2} \left[ \sum_i \left[\frac{d}{d t} C_{x_i}\right]^2 + \sum_{i,j}\left[\frac{d}{d t} C_{x_ix_j}\right]^2  +  \sum_{i,j,k}\left[\frac{d}{d t} C_{x_ix_jx_k}\right]^2\right],
\label{defmis}
\end{eqnarray}
for CE2 and CE3 respectively.

For a fixed point the temporal variation of the cumulant equations vanishes and the misfit converges to zero, i.e. ${\cal J}=0$. As discussed, the time stepping method is an effective scheme to find the {\textit stable} fixed points of the dynamical system.
In addition to  time stepping methods for minimizing ${\cal J}$, it is of great interest to determine the advantages and limitations of other methods for determining the time-invariant solutions of the cumulant equations via minimization of the misfits. Such methods may include  quasi-newton methods, conjugate-gradient ({\it CG}) methods, the {\it trust-region} method \citep{Nocedal_06} or sequential quadratic programming ({\it SLSQP}) \citep{Karft_1988}. We have experimented with all of these methods as discussed below for the dynamo models.

\section{The low-order solar dynamo}
\label{solar}

The first model of dynamo action that we consider here is one relevant to solar and stellar dynamo action \citep{Tobias_1995,W_Smith_2005}. The third-order model is derived using a normal form analysis and undergoes the same bifurcation sequence as more complicated, so-called $\alpha$-$\omega$ PDE models \citep[see e.g.][]{tobias_2002}. The system takes the form of a third-order model with up to cubic nonlinearity, where the evolution equations are given by  
\begin{eqnarray}
	 d_t x &=& \lambda x - \omega y + a z x + d(x^3-3xy^2) + f_x \nonumber \\
	 d_t y &=& \lambda y + \omega x + a z y + d(3x^2y-y^3) + f_y \nonumber \\
	 d_t z &=& \mu - z^2 -(x^2+y^2)+cz^3 + f_z,
	 \label{LowOrderD}
\end{eqnarray}
where the dimensionless functions $x$ and $y$ represent the toroidal and poloidal components of the magnetic field and $z$ represents the velocity field. The control parameters, $\omega$ and $\lambda$ stand for the basic linear cycle frequency and growth rate of the magnetic field $x$ and $y$, $a$ and $c$ do not have physical interpretation but are used to remove the degeneracy of the secondary Hopf bifurcation \citet{W_Smith_2005} that leads to modulation of the basic cycle. The stochastic forces, $f_x, f_y$ and $f_z$ are introduced to represent unmodelled physical processes.

We begin by deriving the cumulant equations for this system using the symbolic software. Substituting the governing dynamical equation, Eq. (\ref{LowOrderD}), into the cumulant expansions in Eq. (\ref{ce1}), (\ref{ce2}) and (\ref{ce3}), we obtain the low-order statistical approximations of the solar dynamo system, where the first order equations read
\begin{eqnarray}
d_t C_x &=& d (C_x^3 + 3C_xC_{xx} - 3C_xC_y^2 - 3C_xC_{yy}+ C_{xxx} - 6C_{xy}C_y - 3C_{xyy}) \nonumber \\
&+& a (C_x C_z + C_{xz}) + \lambda C_x - \omega C_y + \mu_x ,\nonumber \\
d_t C_y &=& d (3 C_x^2 C_y + 6 C_x C_{xy}  + 3C_{xx}C_y + 3 C_{xxy} - C_y^3 - 3C_yC_{yy} - C_{yyy}) \nonumber \\
&+& a (C_yC_z +C_{yz})+ \lambda C_y  + \omega C_x + \mu_y, \nonumber \\
d_t C_z &=& c (C_z^3 + C_{zzz} + 3 C_zC_{zz})  -C_x^2 - C_{xx} - C_y^2 - C_{yy} - C_z^2 - C_{zz} + \mu_z.
\label{solarc1}
\end{eqnarray}
The term, $\mu_{x,y,z}$, is the statistical mean of the stochastic force, $f_{x,y,z}$. The second and third order equations have a complicated form and are detailed in Eqs. (\ref{solarc2}) and (\ref{solarc3}) in Appendix \ref{appsolar}. The cumulant equations are further truncated according to CE2/2.5/3 truncation rules, respectively.

Following \citet{W_Smith_2005}, we define a control parameter $\Omega$ and set $\mu =\sqrt{\Omega}$ and $\lambda = \frac{1}{4}\left\lbrace \left[ \ln(\Omega)+1/3 \right] e^{-\Omega/100}\right\rbrace$. We also set the other parameters, $a = 3$, $c=-d=-0.4$ and $\omega=10.25$, as in  \citet{W_Smith_2005}. 

 We choose two dynamical regimes of this system for comparison of the direct solution of the equations and DSS. In the first  where  $\Omega=1.8$, the dynamical system in the absence of noise yields a quasi-periodic solution where the basic dynamo cycle is modulated on a longer timescale via interaction of the magnetic field with the velocity ($z$). For the second choice
$\Omega=20$, the system settles into a chaotic state (in the absence of noise). Here the chaos arises through the dynamical break down of a two-torus as discussed in \citet{Tobias_1995}. 
We shall discuss how well DSS (via solution of equations (\ref{solarc1}), (\ref{solarc2}) and (\ref{solarc3})) performs for these two cases below.

\subsection{The low-order solar dynamo model in the quasiperiodic state}

\begin{figure}[htp]
	\centering
	\subfigure[]
	{
		\includegraphics[width=0.15\hsize]{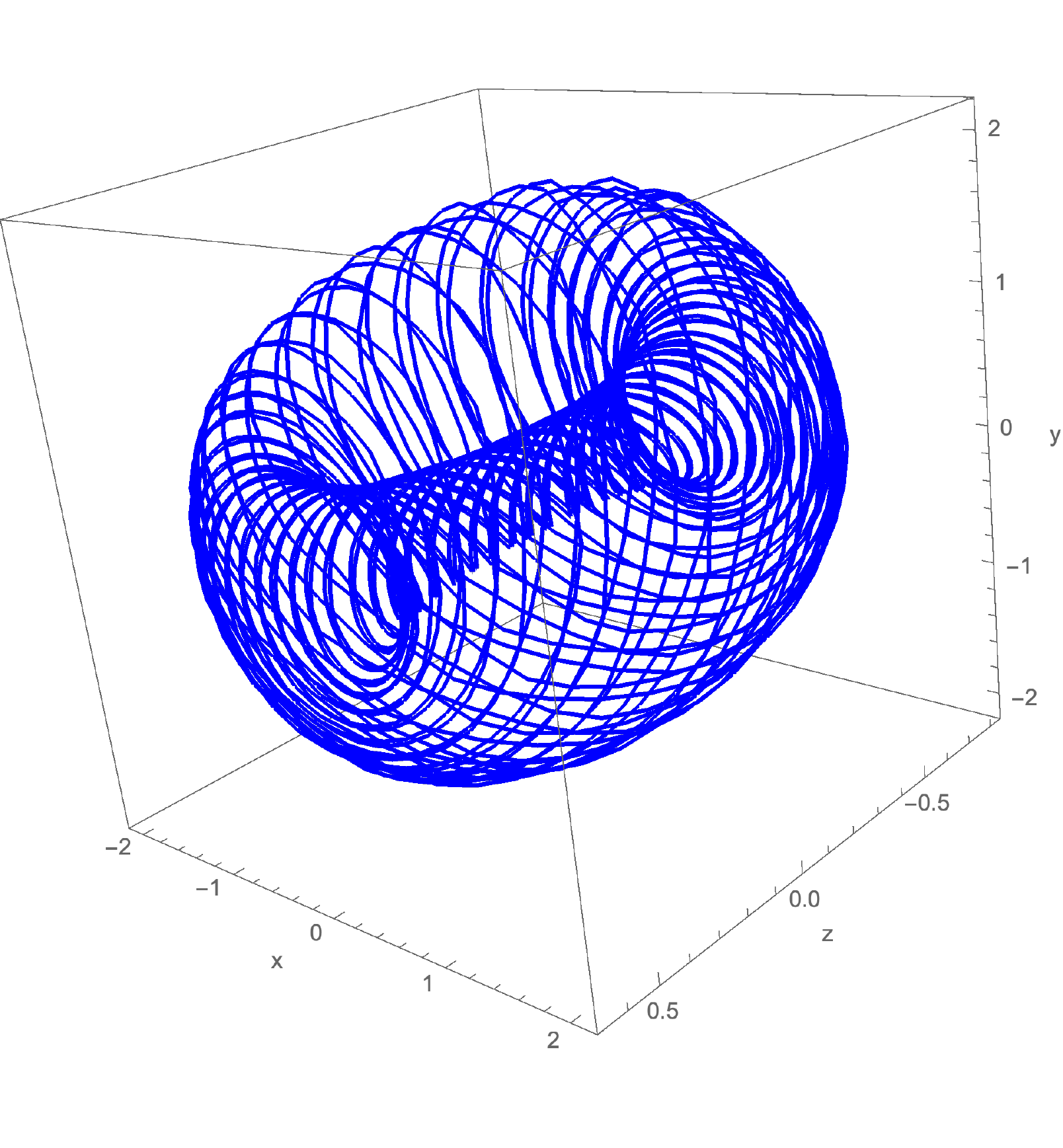}	
	}
	\subfigure[]
	{
		\includegraphics[width=0.15\hsize]{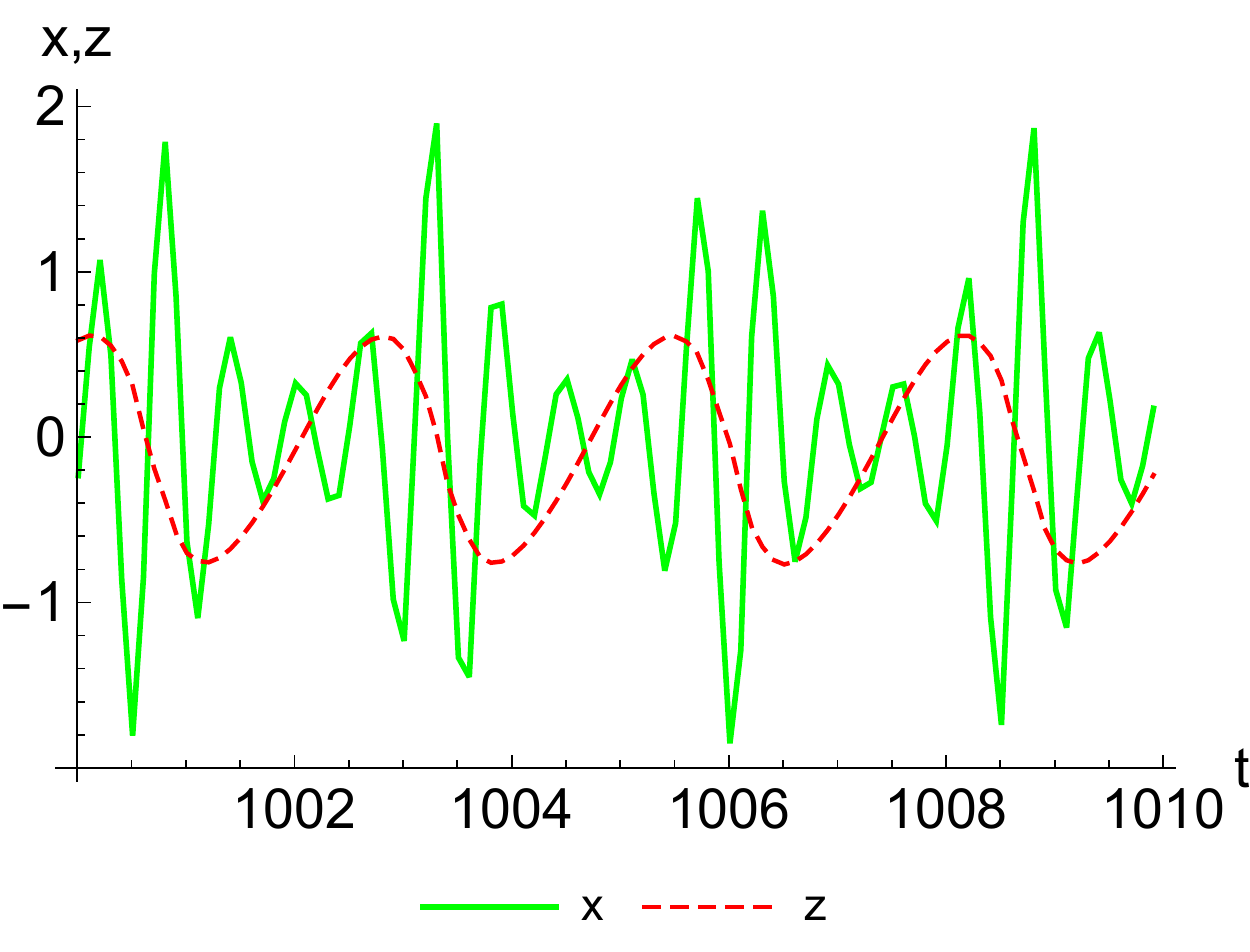}	
	}
	\subfigure[]
	{
		\includegraphics[width=0.15\hsize]{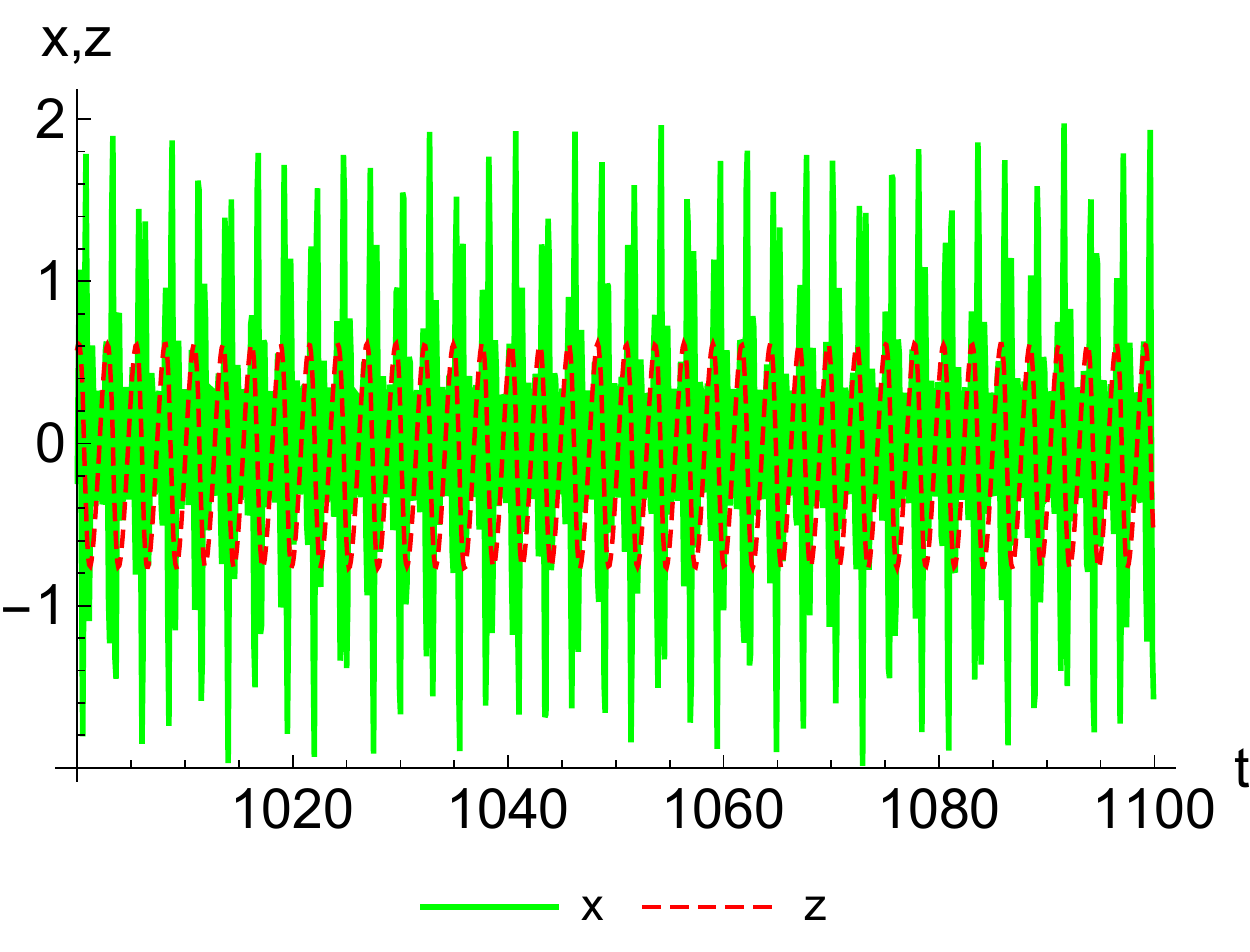}	
	}
		\subfigure[]
	{
		\includegraphics[width=0.15\hsize]{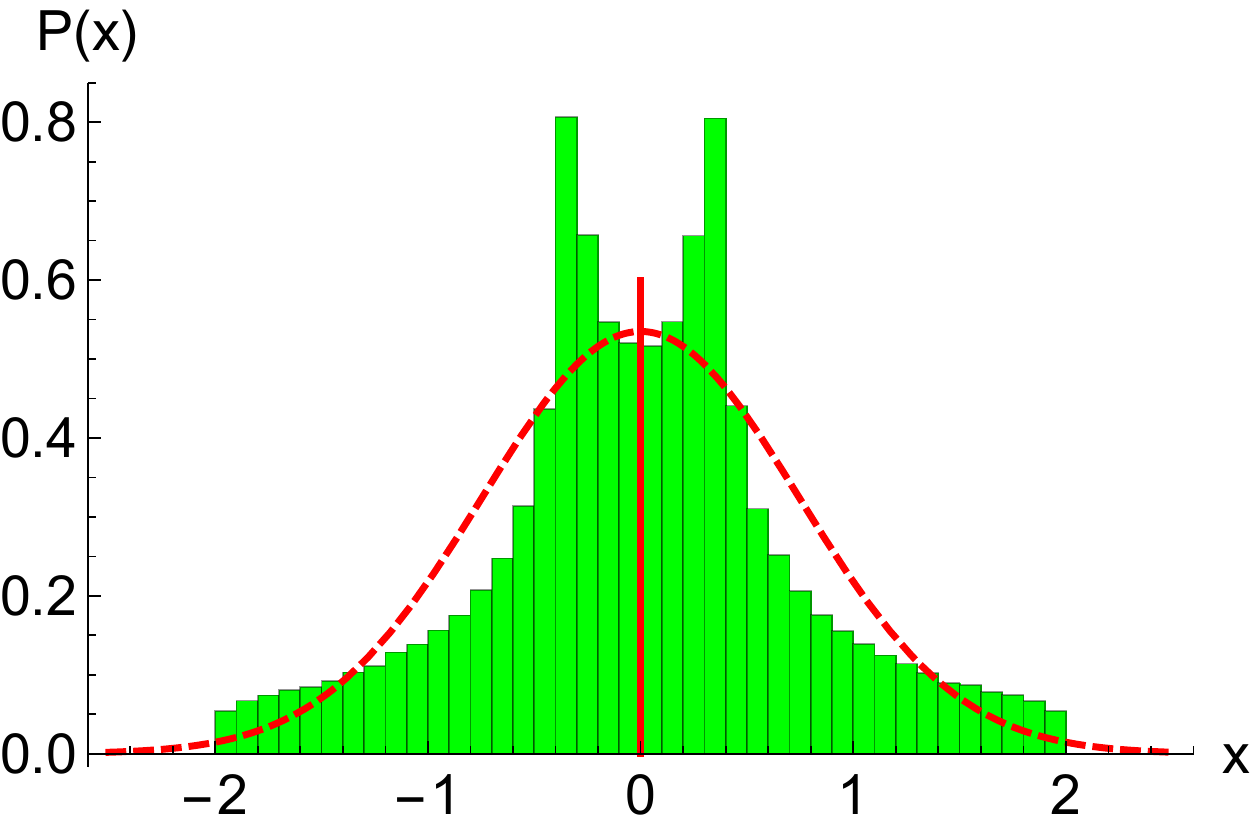}	
	}
	\subfigure[]
	{
		\includegraphics[width=0.15\hsize]{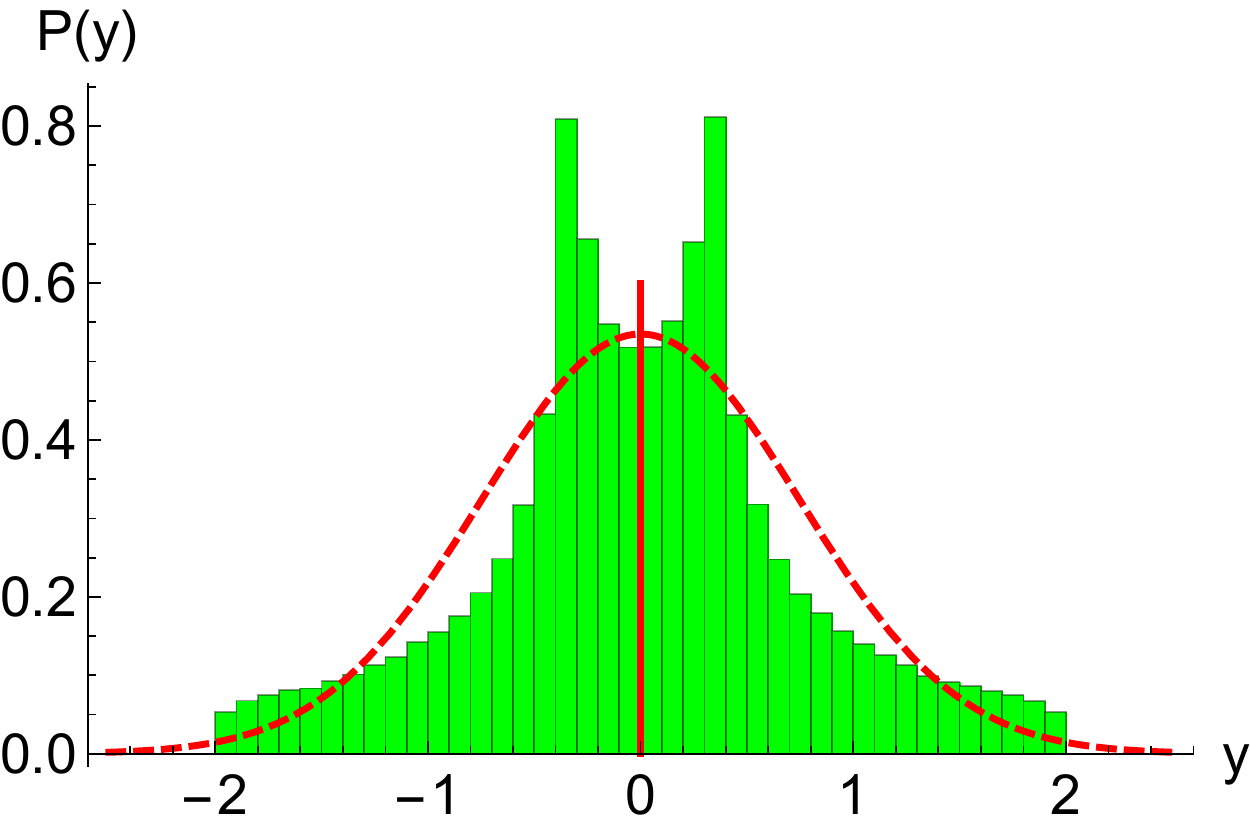}	
	}
	\subfigure[]
	{
		\includegraphics[width=0.15\hsize]{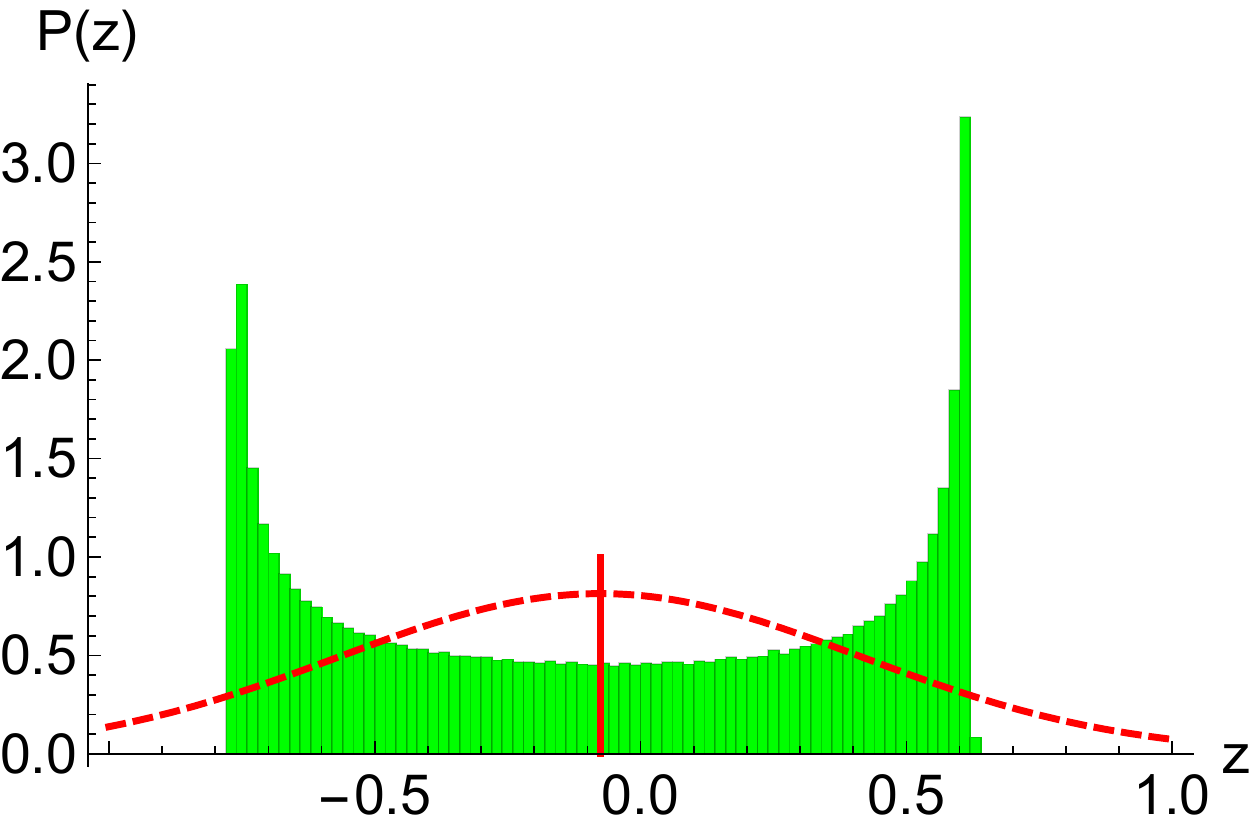}	
	}
\caption{Illustration of the trajectory, time series for a small and large time window and the probability distributions of the solar dynamo system in the quasiperiodic state for $\Omega=1.8$ and $f_{x,y,z}=0$, where in (d)--(f), the green histograms are the probability distributions, $P(x)$, $P(y)$ and $P(z)$, and the dashed red curves are  Gaussian distributions with the same mean and variance as for $P(x)$, $P(y)$ and $P(z)$, respectively.}
\label{solarpt1}
\end{figure}

For $\Omega=1.8$  we integrate the dynamical system, Eq. (\ref{LowOrderD}) (with the stochastic terms zero), forward in time from random initial conditions and find that the solution of the system always settles in the same quasiperiodic orbit shown in Fig. (\ref{solarpt1}).  
The solution takes the form of a two-torus where the amplitude of the periodic oscillation of the magnetic field is modulated but the velocity field remains periodic, see Fig. (\ref{solarpt1} b\& c). The probability distributions, $P(x)$, $P(y)$ and $P(z)$, that are shown in Fig. (\ref{solarpt1}d--f), have two peaks, which results from the periodic oscillation of the dynamo system, where the green histograms stand for probability distributions, $P(x)$, $P(y)$ and $P(z)$, and the dashed red curves are for the Gaussian distribution with the same mean and variance for comparison purposes. The PDF of the velocity field is asymmetric, which indicates the importance of the third order cumulant for $P(z)$.

\begin{figure}[htp]
	\centering
	\subfigure[]
	{
		\includegraphics[width=0.15\hsize]{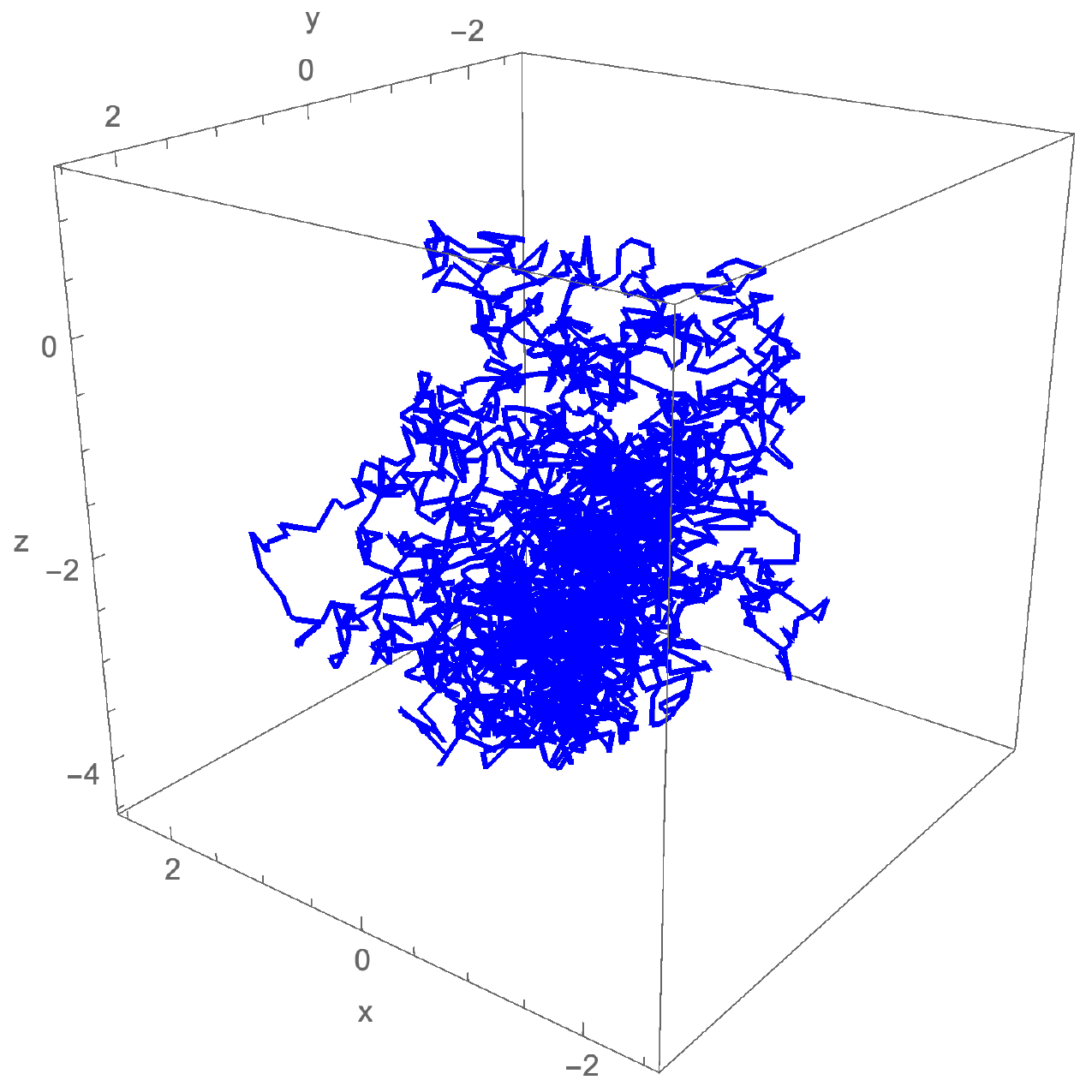}	
	}
	\subfigure[]
	{
		\includegraphics[width=0.15\hsize]{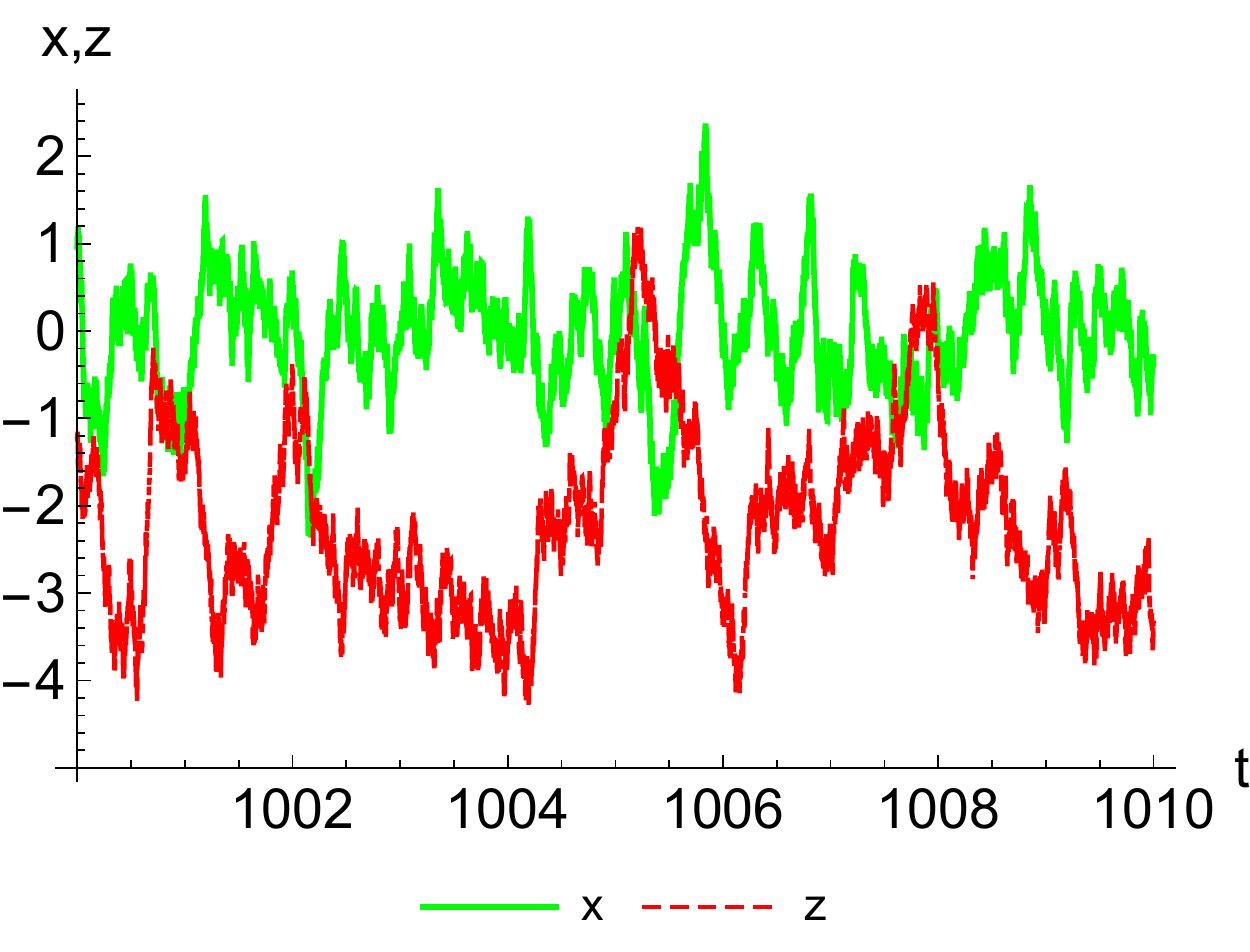}	
	}
	\subfigure[]
	{
		\includegraphics[width=0.15\hsize]{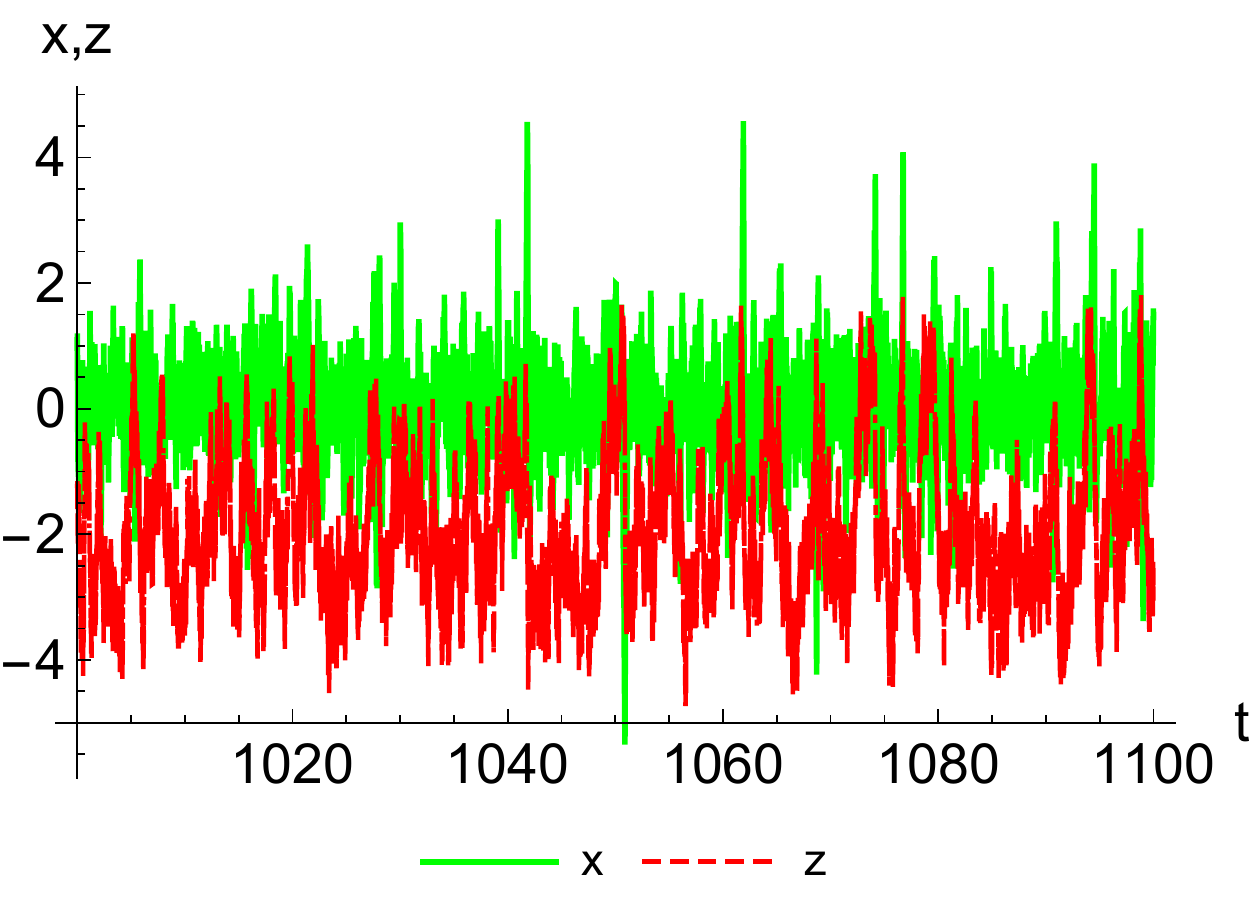}	
	}
	\subfigure[]
	{
		\includegraphics[width=0.15\hsize]{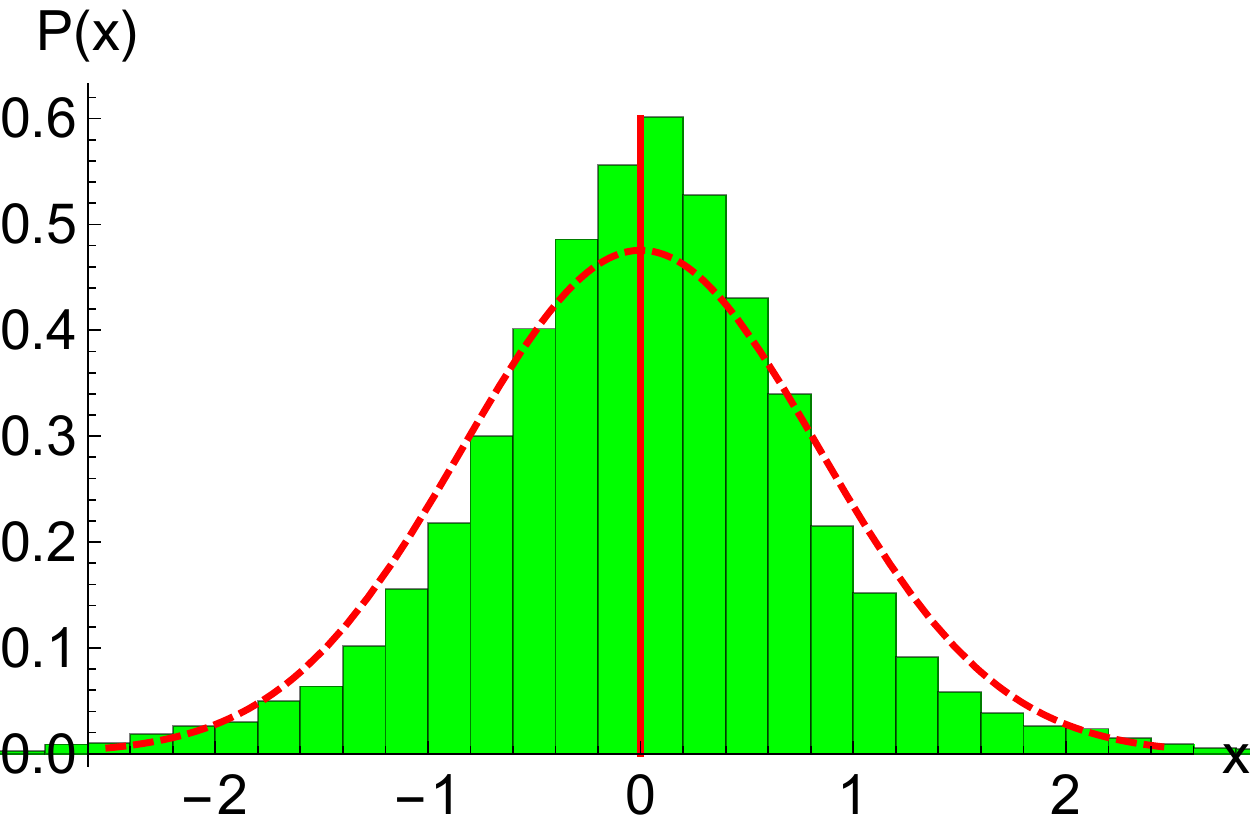}	
	}
	\subfigure[]
	{
		\includegraphics[width=0.15\hsize]{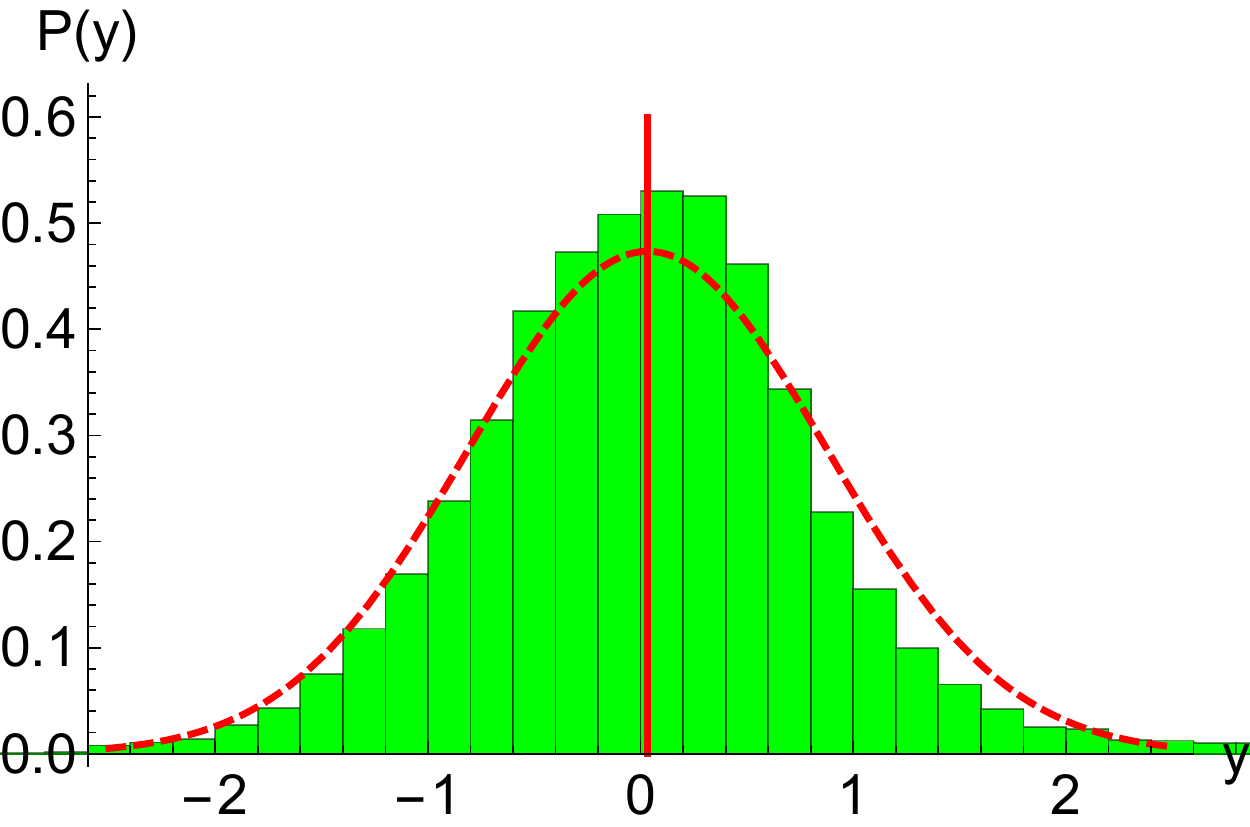}	
	}
	\subfigure[]
	{
		\includegraphics[width=0.15\hsize]{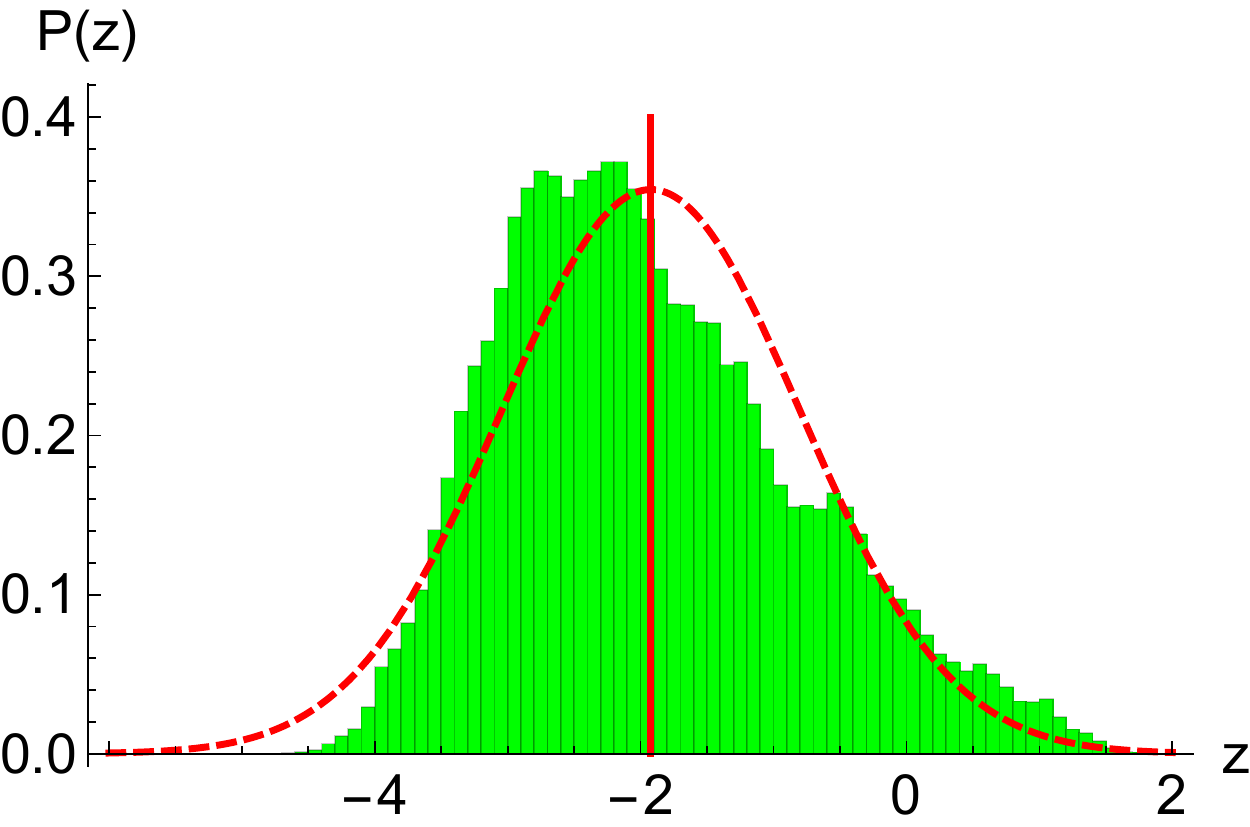}	
	}
\caption{Illustration of the trajectory, time series for small and large time window and the probability distributions of the solar dynamo system in the quasiperiodic state for $\Omega=1.8$ with stochastic forcing, $f_{x,y,z}\sim{\cal N}(0,\ 5)$, where the green histograms are the probability distributions, $P(x)$, $P(y)$ and $P(z)$, and the dashed red curves are for the Gaussian distribution with the same mean and variance as for $P(x)$, $P(y)$ and $P(z)$, respectively.}
\label{solarpt2}
\end{figure}

The dynamics of the deterministic dynamo system can be modifies by switching on the stochastic force, $f_{x,y,z}$. Illustrated in Fig. (\ref{solarpt2}) 
is the trajectory, time series and the PDFs of the dynamo system for $\Omega=1.8$ and $f_{x,y,z}\sim{\cal N}(0,\ 5)$, where in (\ref{solarpt2}d--f) the green histograms are the probability distributions, $P(x)$, $P(y)$ and $P(z)$, and the dashed red curves are for the Gaussian distribution with the same mean and variance as for $P(x)$, $P(y)$ and $P(z)$, respectively. This exceptionally strong stochastic term has significantly changed the dynamics of the dynamo system, i.e., the solution of the magnetic and velocity field oscillate randomly about a single attractor and the PDFs of $x$, $y$ and $z$ are converged to Gaussian. Here the stochastic term completely dominates the nonlinear dynamics.

\subsection{Direct statistical simulation of the solar dynamo in the quasiperiodic states}

Initially, we integrate the cumulant equation, starting from the random initial conditions, forwards in time to obtain the statistical equilibrium of the solar dynamo system for $\Omega=1.8$ without and with the stochastic force, $f_{x,y,z}$. The results are summarised, and compared with the statistics accumulated from DNS, in Table (\ref{kindynamo_ces}). 
\begin{table}[htp]
\centering
\begin{tabular}{l|c|c|c|c|c|c|c|c|c|c|c}
    & $f_{x,y,z}$  &$\tau_d^{-1}$ & $C_x$  & $C_y$  & $C_z$    & $C_{xx}$ & $C_{xy}$ & $C_{xz}$ & $C_{yy}$   & $C_{yz}$ & $C_{zz}$ \\\hline
DNS & $0$           &                & $0$ & $0$  & $-0.076$ & $0.59$ & $0.001$ & $0$& $0.56$ & $0$ & $0.24$ \\\hline
CE2.5 &               &   $20$      & $0$ & $0$  & $-0.071$ & $0.59$ & $0$ & $0$& $0.59$ & $0$ & $0.18$ \\\hline
CE3 &                  &   $10$      & $0$ & $0$  & $-0.078$ & $0.58$ & $0$ & $0$& $0.58$ & $0$ & $0.20$ \\\hline
CE3 &                  &   $20$      & $0$ & $0$  & $-0.070$ & $0.59$ & $0$ & $0$& $0.59$ & $0$ & $0.18$ \\\hline
CE3 &                  &   $50$      & $0$ & $0$  & $-0.070$ & $0.60$ & $0$ & $0$& $0.60$ & $0$ &  $0.16$ \\\hline\hline
DNS & ${\cal N}(0,\ 2)$ & & $0$ & $0$  & $-1.40$ & $0.55$ & $0$ & $0$& $0.53$ & $0$ & $1.05$ \\\hline
CE2 &                   &          & $0$ & $0$  & $-1.14$ & $0.62$ & $0$  & $0$& $0.62$ & $0$ & $1.62$ \\\hline
CE2.5 &                &  $20$ & $-0.07$ & $-0.01$  & $-0.92$ & $0.46$ & $0.02$  & $-0.11$& $0.63$ & $0.22$ & $1.14$ \\\hline
CE3 &                   & $20$ & $0$ & $0$  & $-1.40$ & $0.64$ & $0$  & $0$& $0.64$ & $0$ & $1.49$ \\\hline
CE3 &                   & $50$ & $0$ & $0$  & $-1.26$ & $0.64$ & $0$  & $0$& $0.64$ & $0$ & $1.54$ \\\hline\hline
DNS & ${\cal N}(0,\ 5)$ & & $0$ & $0$  & $-1.79$ & $0.87$ & $-0.003$ & $0$& $0.83$ & $0$ & $1.37$ \\\hline
CE2 &                            & & $0$ & $0$  & $-1.54$ & $1.13$ & $0$ & $0$& $1.13$ & $0$ & $2.13$ \\\hline
CE2.5 &                & $10$& $-0.15$ & $0.05$  & $-1.53$ & $0.61$ & $0.74$ & $-0.08$& $0.97$ & $0.60$ & $0.97$ \\\hline
CE2.5 &                & $50$& $-0.15$ & $0.04$  & $-1.26$ & $0.53$ & $0.07$ & $-0.19$& $0.93$ & $0.53$ & $1.58$ \\\hline
CE3 &                            & $50$ & $0$ & $0$  & $-1.69$ & $1.16$ & $0$ & $0$& $1.16$ & $0$ & $2.05$ \\\hline\hline
DNS & ${\cal N}(0,\ 20)$ & & $0$ & $0$  & $-2.28$ & $2.21$ & $0.26$ & $0$& $2.27$ & $-0.01$ & $2.48$ \\\hline
CE2 &                               & & $0$ & $0$  & $-2.16$ & $3.20$ & $0$ & $0$& $3.20$ & $0$ & $3.59$ \\\hline
CE2.5 &                            & $20$& $-0.25$ & $0.26$  & $-1.94$ & $1.10$ & $0.30$ & $0.29$& $2.42$ & $1.70$ & $1.76$ \\\hline
CE3 &                               & $50$ & $0$ & $0$  & $-2.35$ & $3.30$ & $0$ & $0$& $3.30$ & $0$ & $3.70$ 
\end{tabular}
\caption{The low order statistics of the solar dynamo system in the quasiperiodic state for $\Omega=1.8$ with different stochastic forcing, $f_{x,y,z}$.}
\label{kindynamo_ces}
\end{table}
In the absence of noise, we observe that the dynamics of the dynamo system in this parameter regime can be accurately described by the CE3 approximation for a range of $\tau_d$.The solution of cumulant equations always converges the unique statistical equilibrium. In the absence of noise however CE2 does not converge.

\begin{figure}[htp]
	\centering
	\subfigure[$C_{x_i}$]
	{
		\includegraphics[width=0.25\hsize]{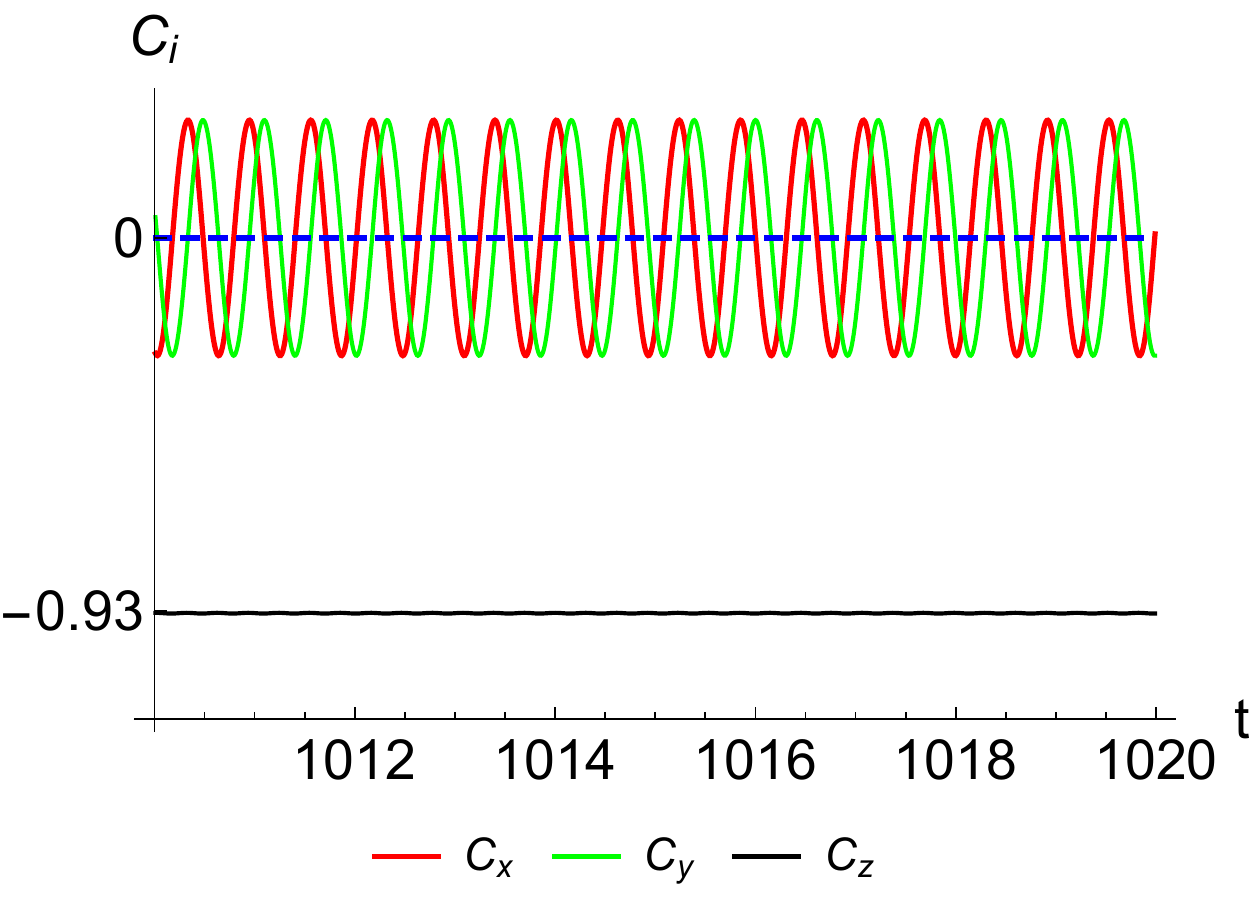}	
	}
	\subfigure[$C_{x_ix_j}$]
	{
		\includegraphics[width=0.25\hsize]{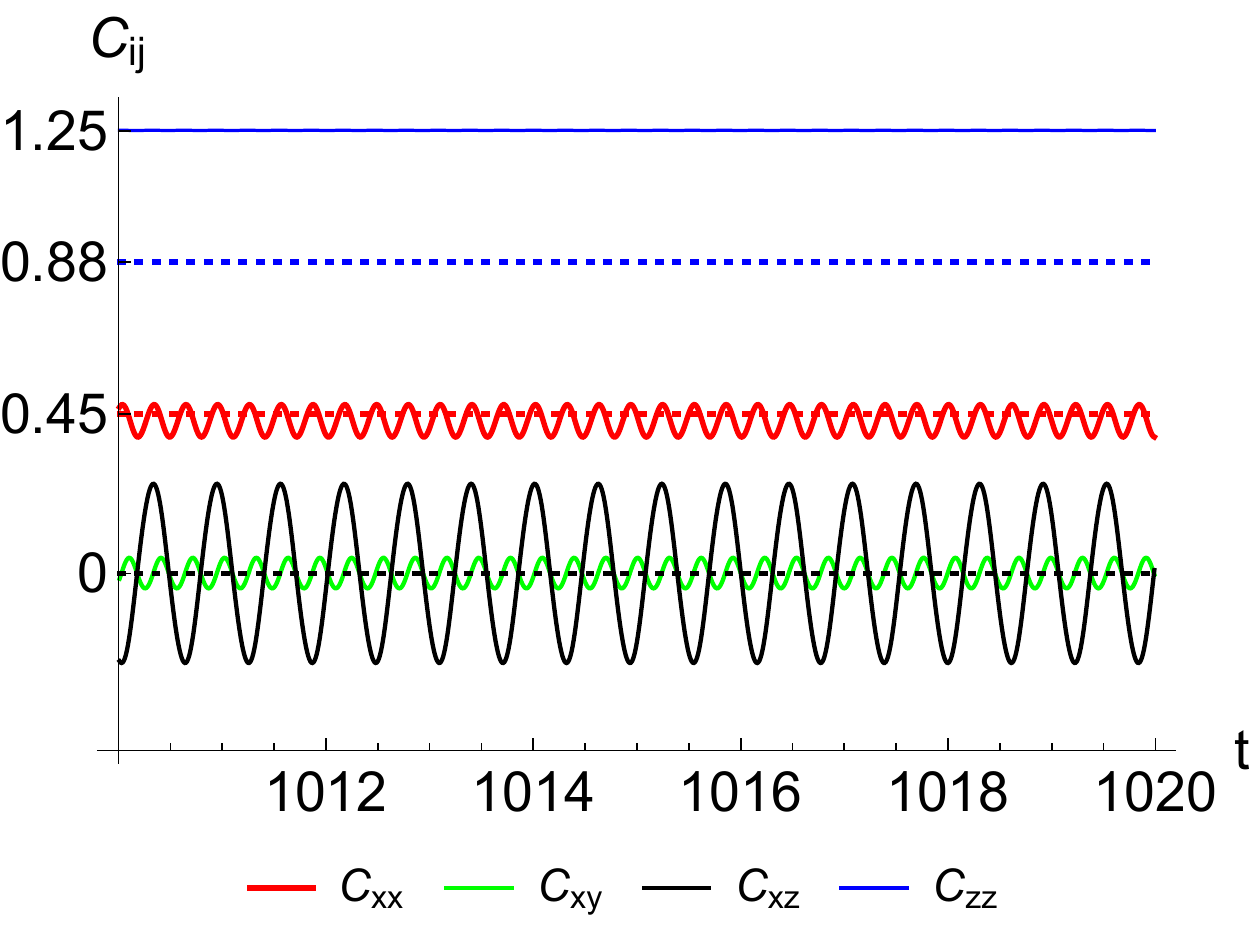}	
	}
\caption{Solution of CE2 approximation of the solar dynamo in the quasiperiodic state for $\Omega=1.8$ and $f_{x,y,z}\sim {\cal N}(0, \ 1)$,  where the solution of CE2 is shown in solid curves and the statistics via DNS are plotted in dashed solid lines.}
\label{solarpt3}
\end{figure}

When stochastic driving is also included, DSS becomes even more effective. As the  amplitude of the stochastic force is increased, e.g., see Fig (\ref{solarpt1}) and (\ref{solarpt2}), the PDFs of the magnetic and velocity field become more Gaussian. For $\sigma_{x,y,z}^2 \simeq 1$ the low-order statistics can be captured by both CE3 and CE2. The CE2 system has now become numerically stable ---  but the solution remains oscillatory. Fig., (\ref{solarpt3})
shows this behaviour for the solution of CE2 system for $f_{x,y,z}\sim {\cal N}(0, \ 1)$, where the solid curves are the solution of CE2 and the dashed straight lines are for the statistics obtained via the ensemble average of DNS solution. With increasing $\sigma_{x,y,z}^2$ the solution of CE2 becomes as accurate as the CE3 approximation. The eddy damping parameter, $\tau_d>0$,  must be introduced to stabilise the numerical integrations of CE3 system, where the accurate solutions are obtained for $\tau_d$ within the range, ${\cal O}(10^{-2})$ to ${\cal O}(10^{-1})$. We note that some cumulants are more accurately represented than others by the CE2 and CE3 approximations; the second order cumulant, $C_{zz}$, is the least accurately modelled by the CE2/3 approximations with maximum error about $50\%$. For stronger stochastic force however, which brings the PDF of $z$ close to Gaussian, this error is significantly reduced. 

The CE2.5 approximation is found numerically stable for all test cases for the eddy damping parameter, $\tau_d$ in the range from ${\cal O}(10^{-2})$ to ${\cal O}(10^{-1})$. This approximation assumes that the terms involving the first order cumulants, $C_{x_i}$, are statistically insignificant in the governing equation of the third order for $C_{x_ix_mx_n}$ and are neglected in the numerical computation. The solution of the CE2.5 approximation is found to be as accurate as  CE3 for the case of no noise ($f_{x,y,z}=0$) but becomes progressively less accurate as   $\sigma_{x,y,z}^2$ is increased, e.g, see Table (\ref{kindynamo_ces}) for the test case of $\sigma_{x,y,z}^2=20$. This is understandable as adding noise in this system increases the value of the first cumulant $C_z$. As CE2.5 neglects any interactions with the first cumulant to determine the third cumulant, it is to be expected that the approximation becomes less accurate as the influence of this cumulant grows.

\subsection{The fixed points of the cumulant hierarchy for solar dynamo in the quasiperiodic states}

Although timestepping allows the access of the stable solutions of the cumulant equations, other fixed points are possible solutions as discussed earlier. Here we assess the effectiveness of various methods for accessing these fixed points.

\begin{figure}[htp]
	\centering
	\subfigure[$C_{z} \sim {\cal J}$]
	{
		\includegraphics[width=0.25\hsize]{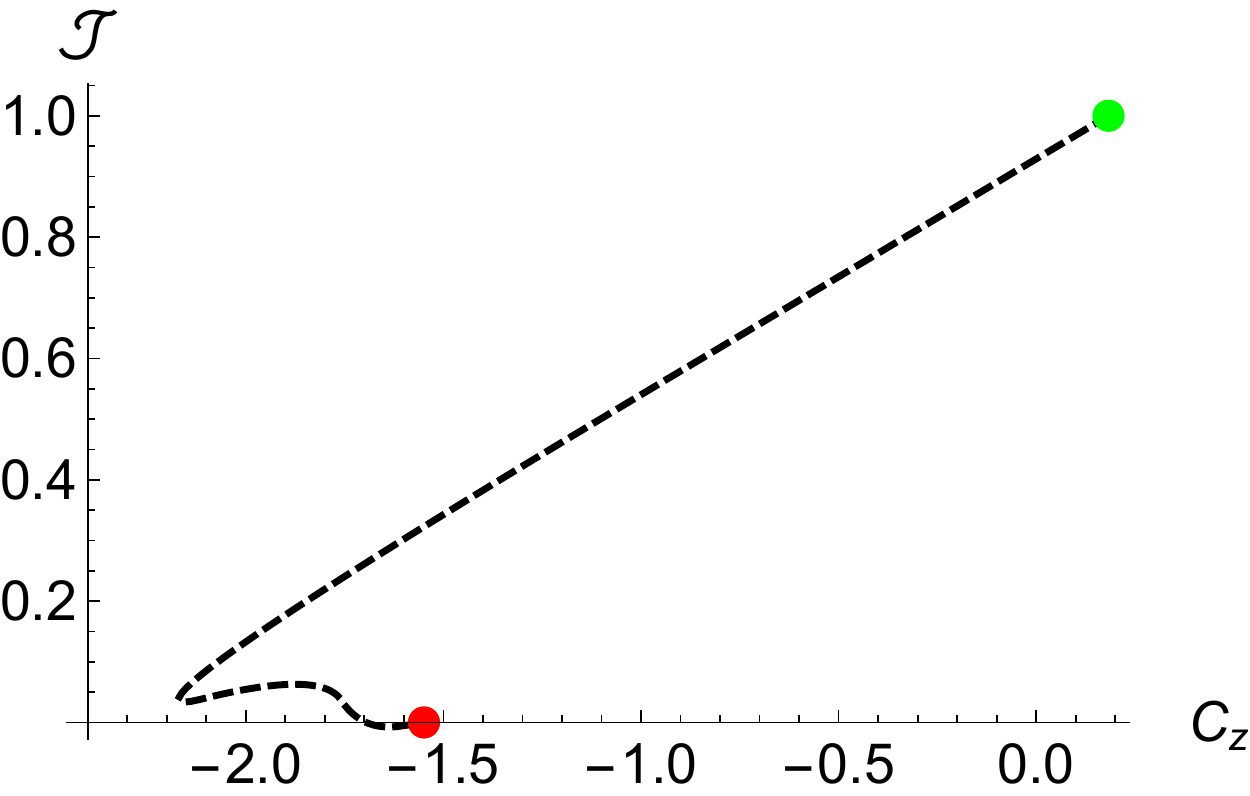}	
	}
	\subfigure[$C_{zz} \sim {\cal J}$]
	{
		\includegraphics[width=0.25\hsize]{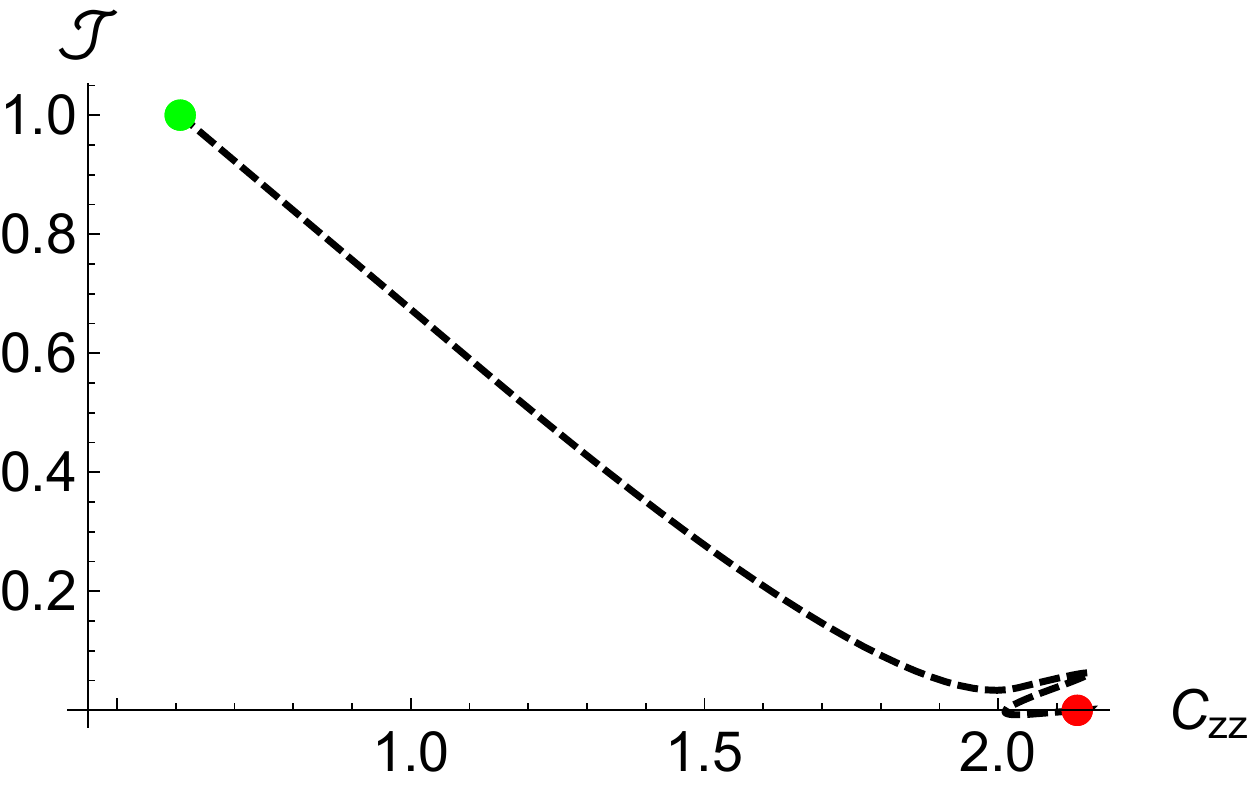}	
	}
	\subfigure[${\cal J} \sim n$]
	{
		\includegraphics[width=0.25\hsize]{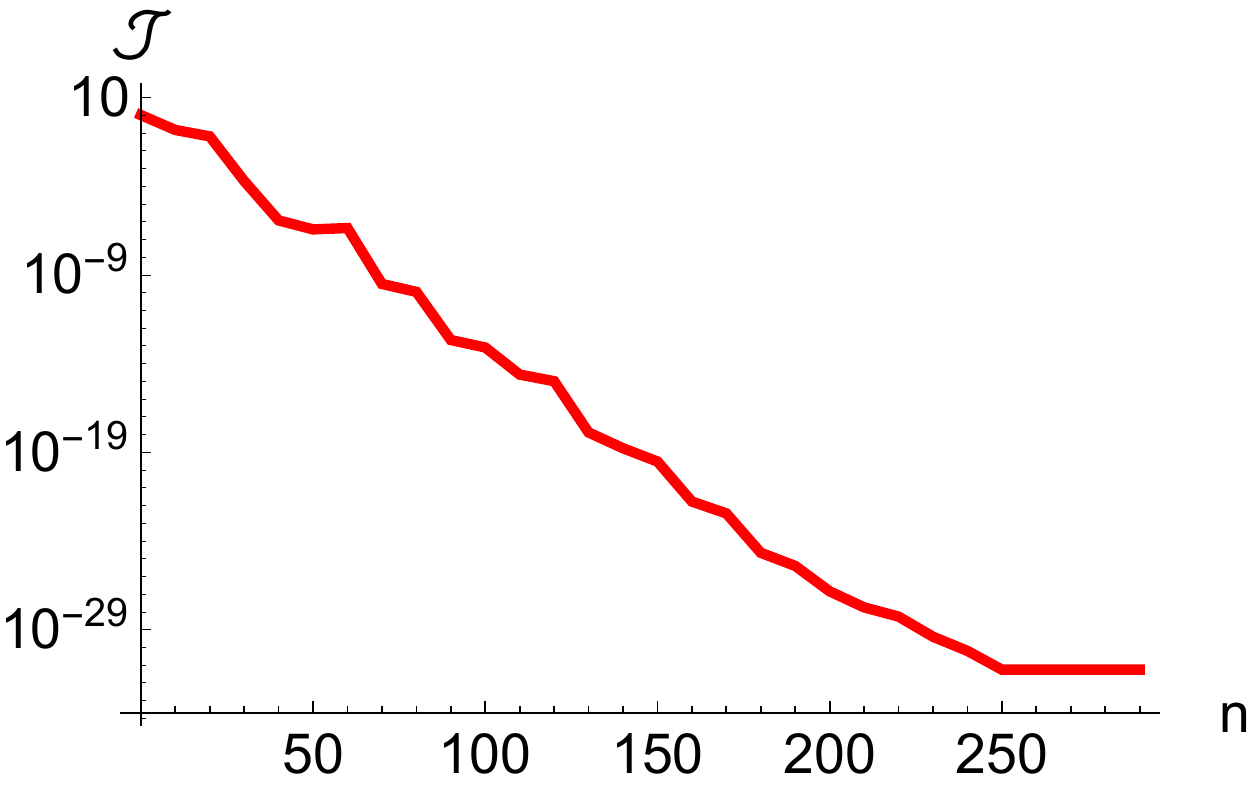}	
	}
\caption{The stable fixed point is found by minimizing ${\cal J}$ via the {\it CG} method for the CE2 approximation of the solar dynamo for $\Omega=1.8$ and $f_{x,y,z}\sim(0, \ 5)$, where (a) and (b) show the path of the cumulants, $C_z$ and $C_{zz}$, as the solution converges to ${\cal J}=0$ and (c) illustrates the convergence of ${\cal J}$ as a function of the number of iterations, $n$. The initial guess and the terminal solution of the minimization are shown as green and red dots in (a) and (b). The optimal solution converges to the stable time-invariant solution of CE2 via the time stepping method.}
\label{solarpt4}
\end{figure}

We first study gradient based optimization methods for computing the fixed point of the CE2 approximations, in the presence of noise. For $\sigma_{x,y,z}^2=5$, the dynamical system can be accurately approximated by timestepping the CE2 equations and the optimal solution of Eq. (\ref{defmis}) is the global minimal of ${\cal J}$. Shown in Fig. (\ref{solarpt4}) 
is the convergence to the optimal solution for the cumulants of the CE2 approximation of the dynamo system in the quasiperiodic state for $\Omega=1.8$ and $f_{x,y,z}\sim {\cal N}(0, \ 5)$ via the {\it CG} method, where ${\cal J}$ is normalized by its value at the first iteration. The initial guess is randomly chosen and satisfies the statistical realisibility criteria. The optimal solution of the CE2 system always converges to the unique stable fixed point that is found by the time stepping method. The misfit, ${\cal J}$, converges to zero exponentially. For ${\cal O}(50)$ iterations, the misfit reduces by a factor of $10^9$. For this case the quasi-newton method performs similarly as {\it CG} in terms of the accuracy and the convergence rate. We note that the time stepping method also converges to the stable solution of CE2 system exponentially  but with a slower rate, i.e., the misfit reduces by a factor of $10^9$ within ${\cal O}(10^3)$ time steps, where the optimal time step, $dt=10^{-2}$, is used for the numerical integration. We note that calculating the "downhill" direction for minimizing the misfit, ${\cal J}$, that is directly computed via the symbolic differentiation is computationally twice to three times as expensive as for evolving the cumulant equation one time step forward. 
However, for this problem, the minimization method is approximately $10$ times faster than the time stepping method to obtain the convergence, ${\cal J} < 10^{-9}$, where ${\cal J}$ is normalised to one at the first iteration/time step.

We find that it is very difficult to find the stable fixed point for CE3 approximations via the gradient based method. By taking the third order cumulant into the consideration, the misfit, ${\cal J}$, is no longer convex. The optimization either converges slowly to unstable or non-realizable fixed points or is trapped by the local minima that are introduced by the third order cumulants. 

\begin{figure}[htp]
	\centering
	\subfigure[$C_{z} \sim {\cal J}$]
	{
		\includegraphics[width=0.25\hsize]{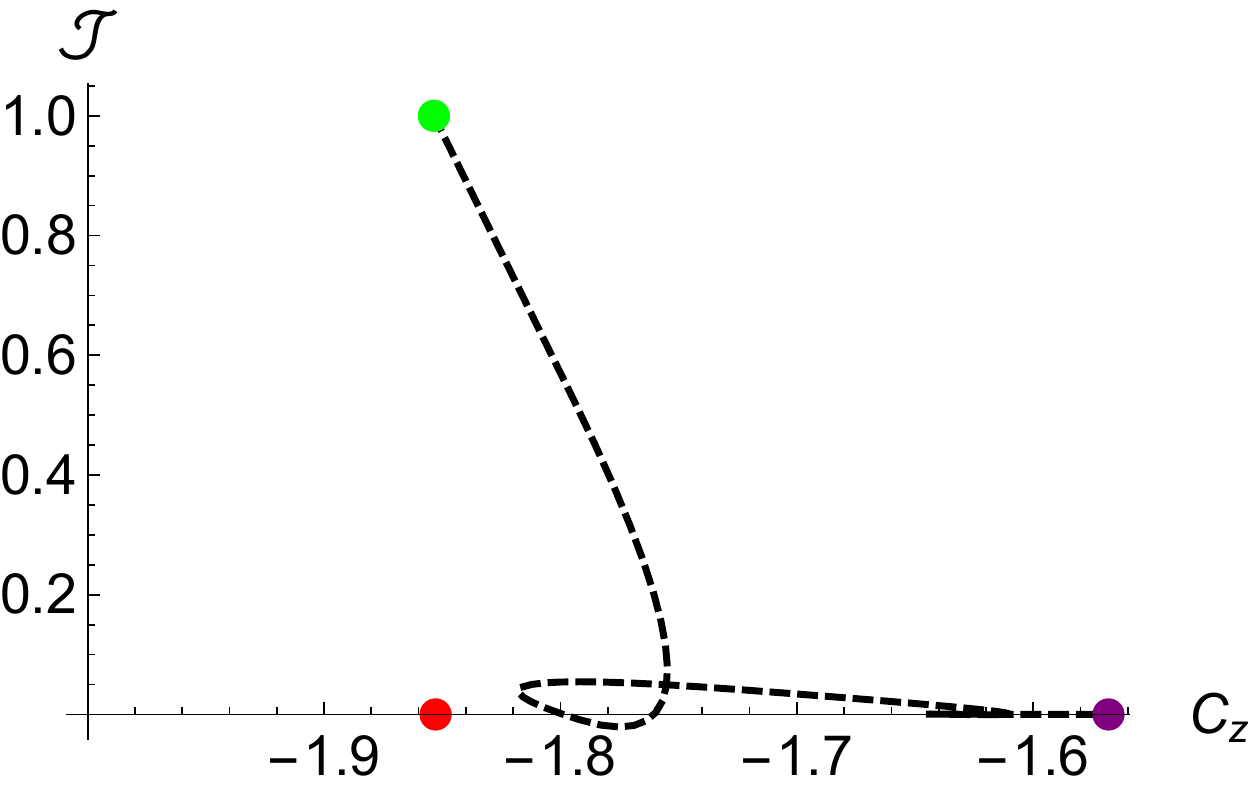}	
	}
	\subfigure[$C_{zz} \sim {\cal J}$]
	{
		\includegraphics[width=0.25\hsize]{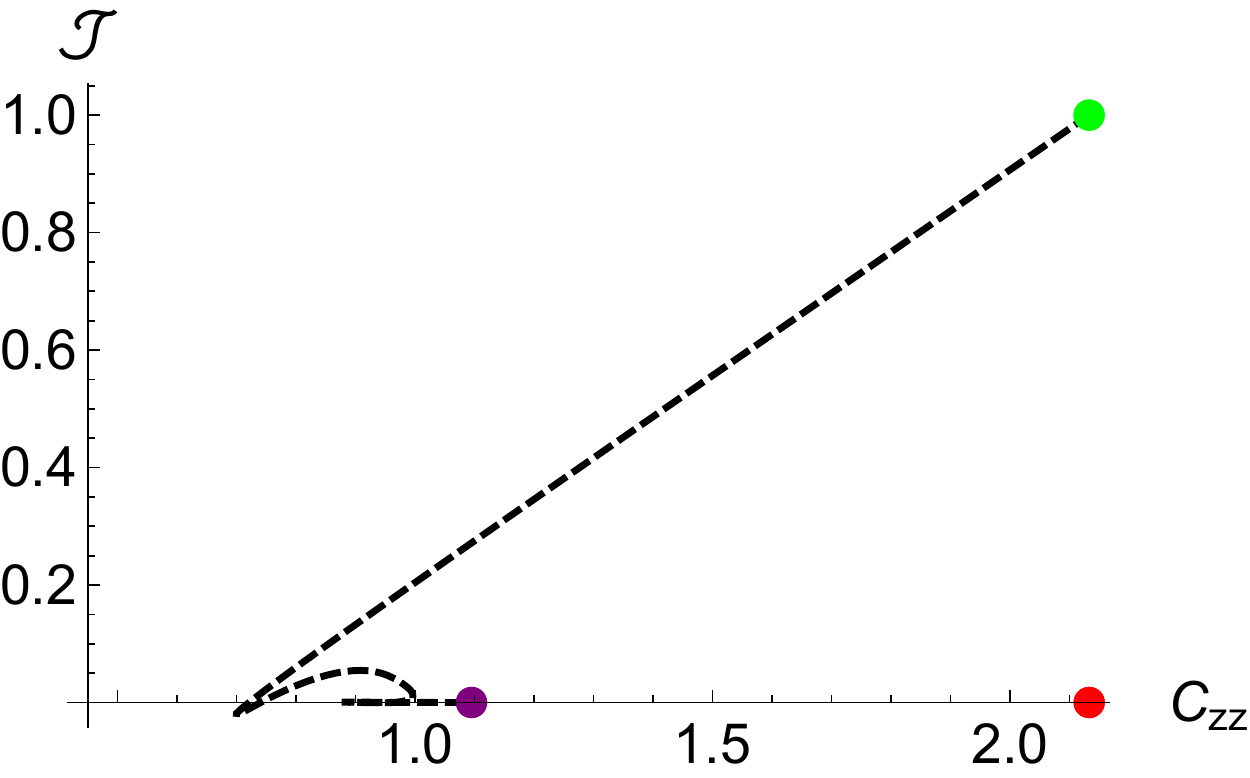}	
	}
	\subfigure[${\cal J} \sim n$]
	{
		\includegraphics[width=0.25\hsize]{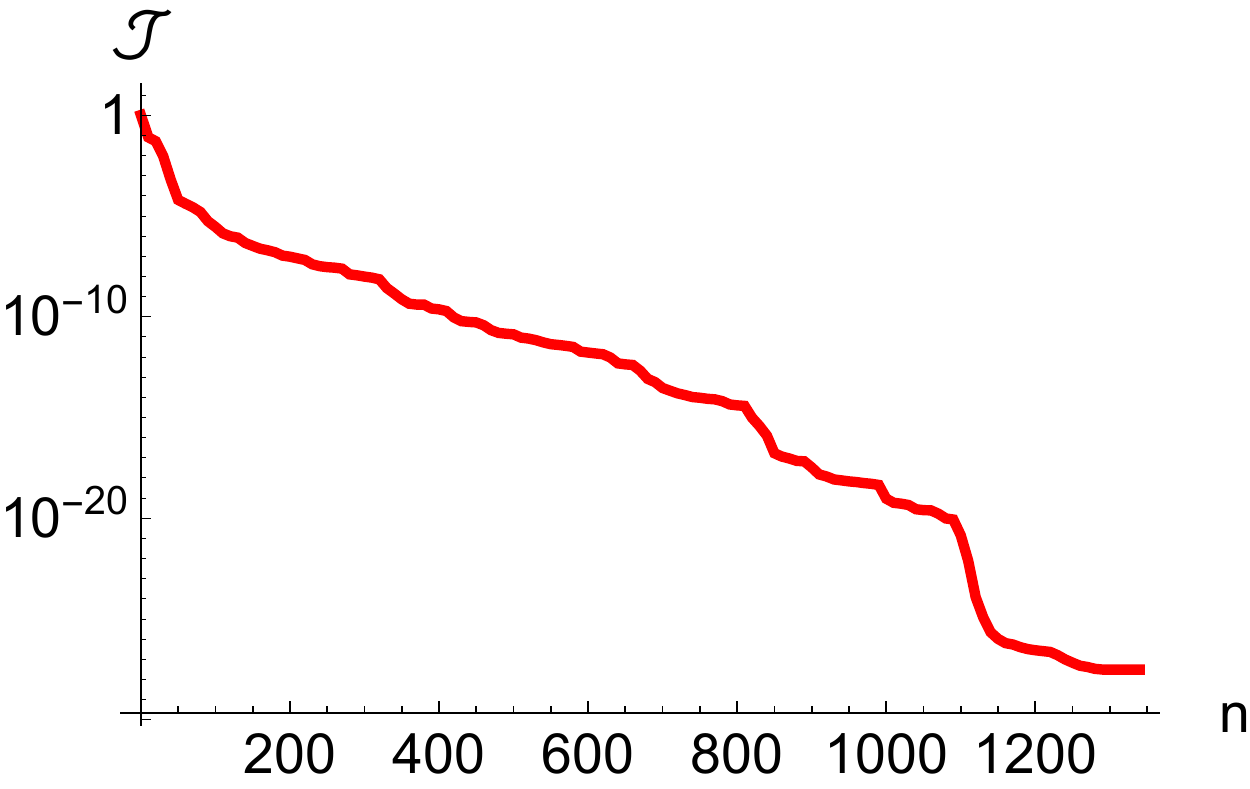}	
	}
\caption{Illustration of the path of the low-order cumulant, $C_z$ and $C_{zz}$, as ${\cal J}$ converges to zero in (a) and (b) and the convergence of ${\cal J}$ as the number of iterations in (c), where, in (a) \& (b), the red and purple dots give the initial guess and terminal solution of the optimization and the red dots are the true solution.}
\label{solarpt6}
\end{figure}

As an example, we use {\it CG} to continue the stable fixed point of the dynamo system from $\Omega=1.8$ (found by timestepping) to $\Omega=1.81$ with $\Delta \Omega=0.01$, for the same stochastic force level ($f_{x,y,z}\sim {\cal N}(0, \ 5)$). The results are in Fig. (\ref{solarpt6})
which shows the path of the low-order cumulant, $C_z$ and $C_{zz}$, as ${\cal J}$ converges to zero and the convergence of ${\cal J}$ as the number of iterations, where the green and purple dots in Figs. (\ref{solarpt6}a \& b) represent the initial guess and terminal solution of the optimization and the red dots are for the true solution found by timestepping. Disappointingly, the optimal solution converges to an unstable fixed point of CE3 system. Similar performance is observed for different $\Omega$ and $\Delta \Omega$ with and without the stochastic force, $f_{x,y,z}$, even for $\Delta\Omega=10^{-3}$.

\begin{figure}[htp]
	\centering
	\subfigure[$C_{z} \sim {\cal J}$ for $f_{x,y,z}\sim{\cal N}(0, \ 5)$]
	{
		\includegraphics[width=0.3\hsize]{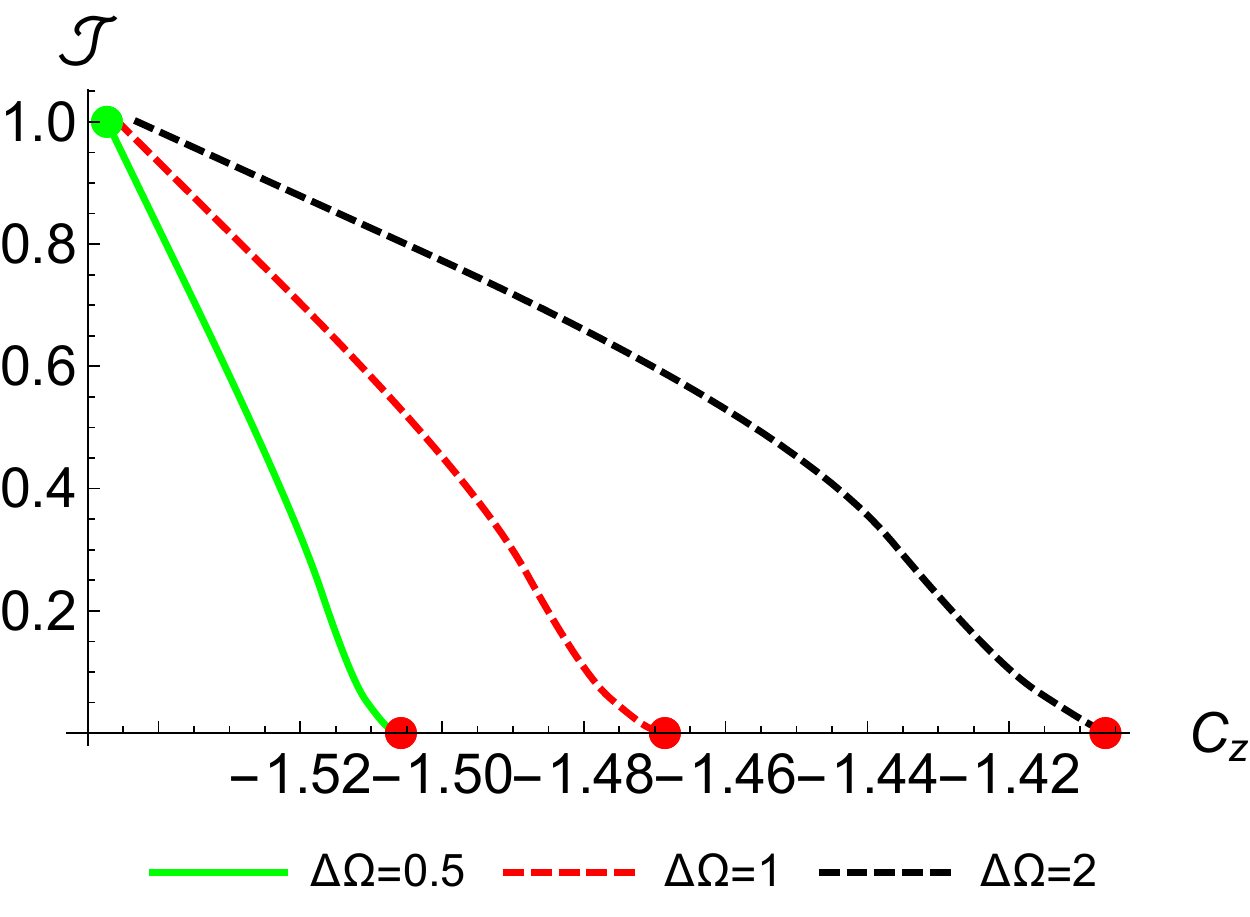}	
	}
	\subfigure[$C_{zz} \sim {\cal J}$ for $f_{x,y,z}\sim{\cal N}(0, \ 5)$]
	{
		\includegraphics[width=0.3\hsize]{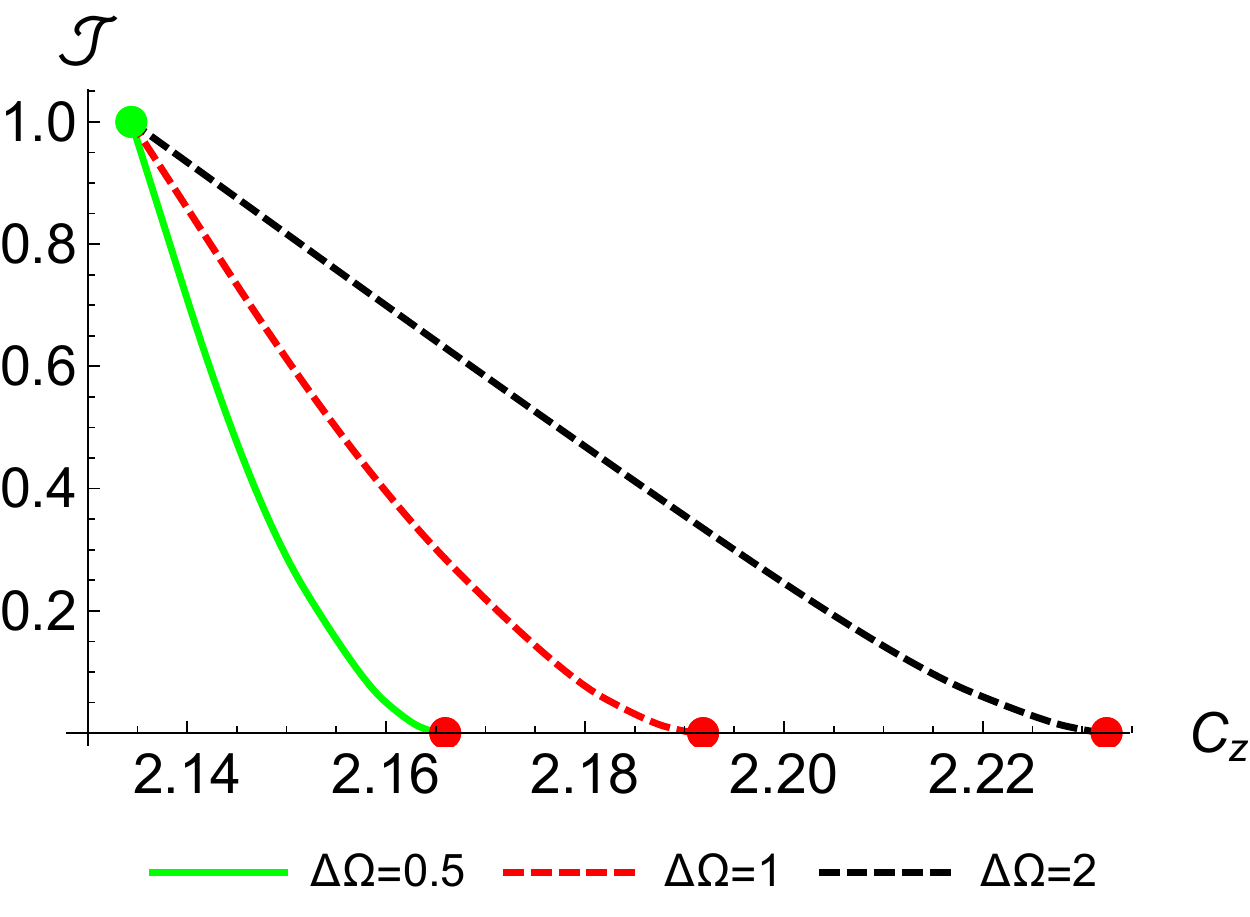}	
	}
	\subfigure[${\cal J}\sim n$ for $f_{x,y,z}\sim{\cal N}(0, \ 5)$]
	{
		\includegraphics[width=0.3\hsize]{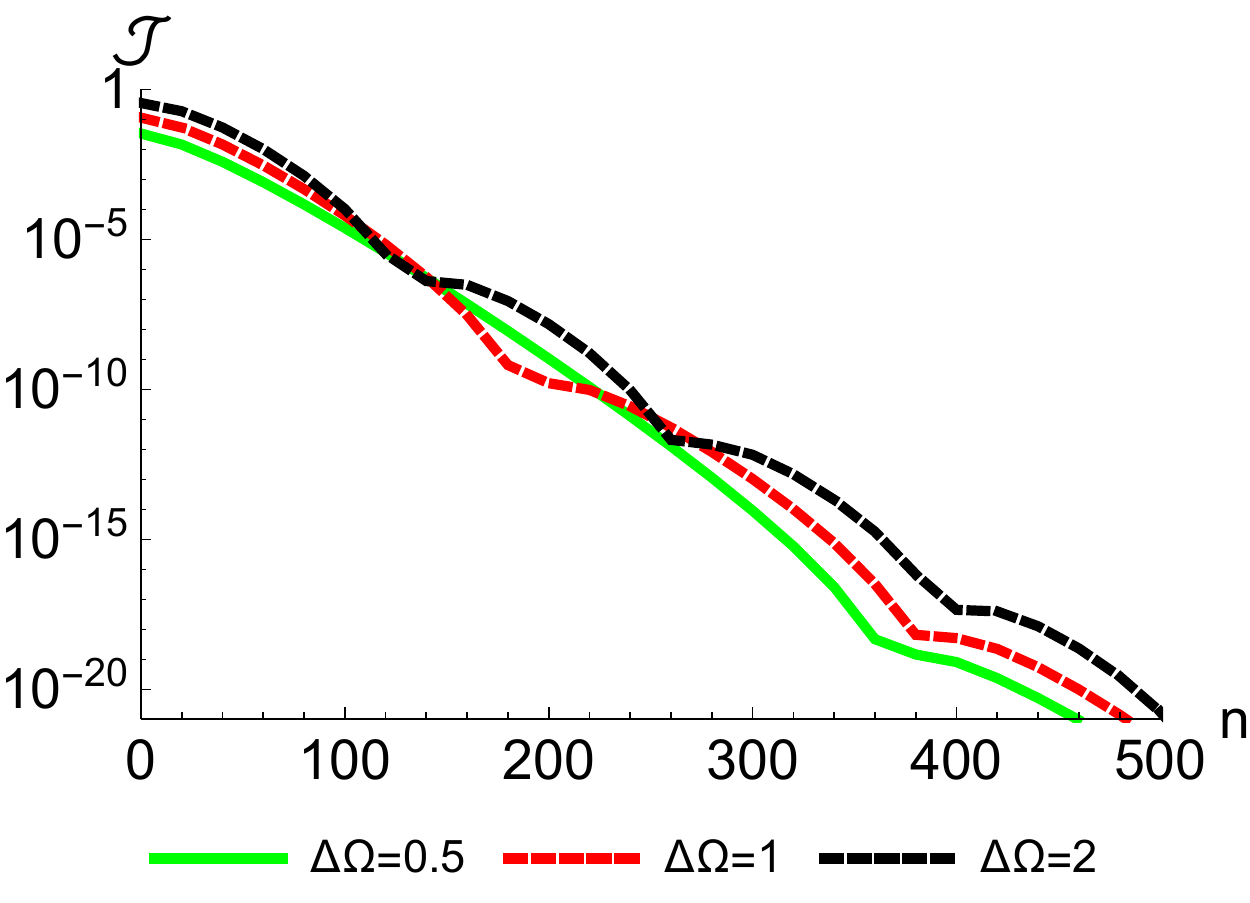}	
	}\\
	\subfigure[$C_{z} \sim {\cal J}$ for $f_{x,y,z}=0$]
	{
		\includegraphics[width=0.3\hsize]{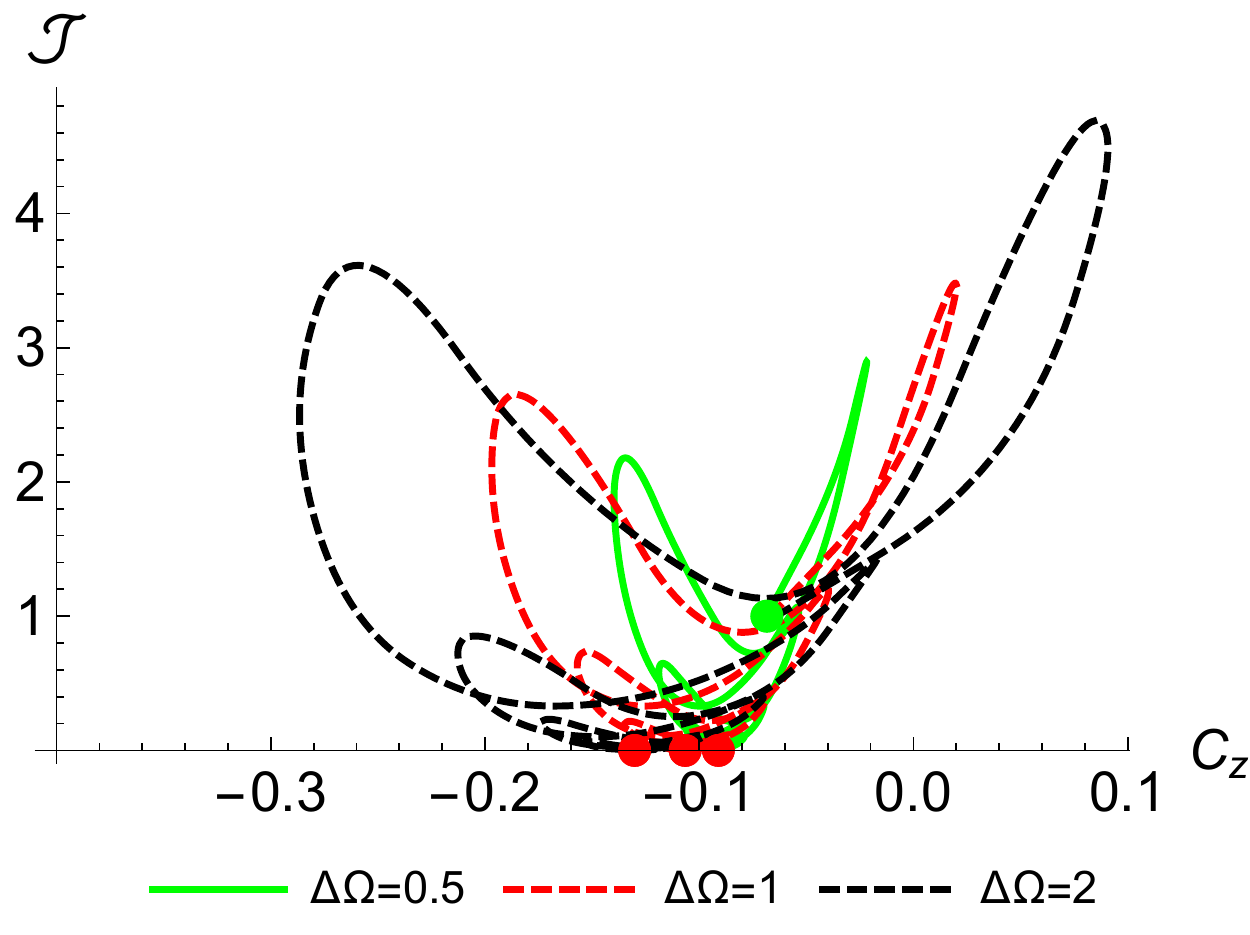}	
	}
	\subfigure[$C_{zz} \sim {\cal J}$ for $f_{x,y,z}=0$]
	{
		\includegraphics[width=0.3\hsize]{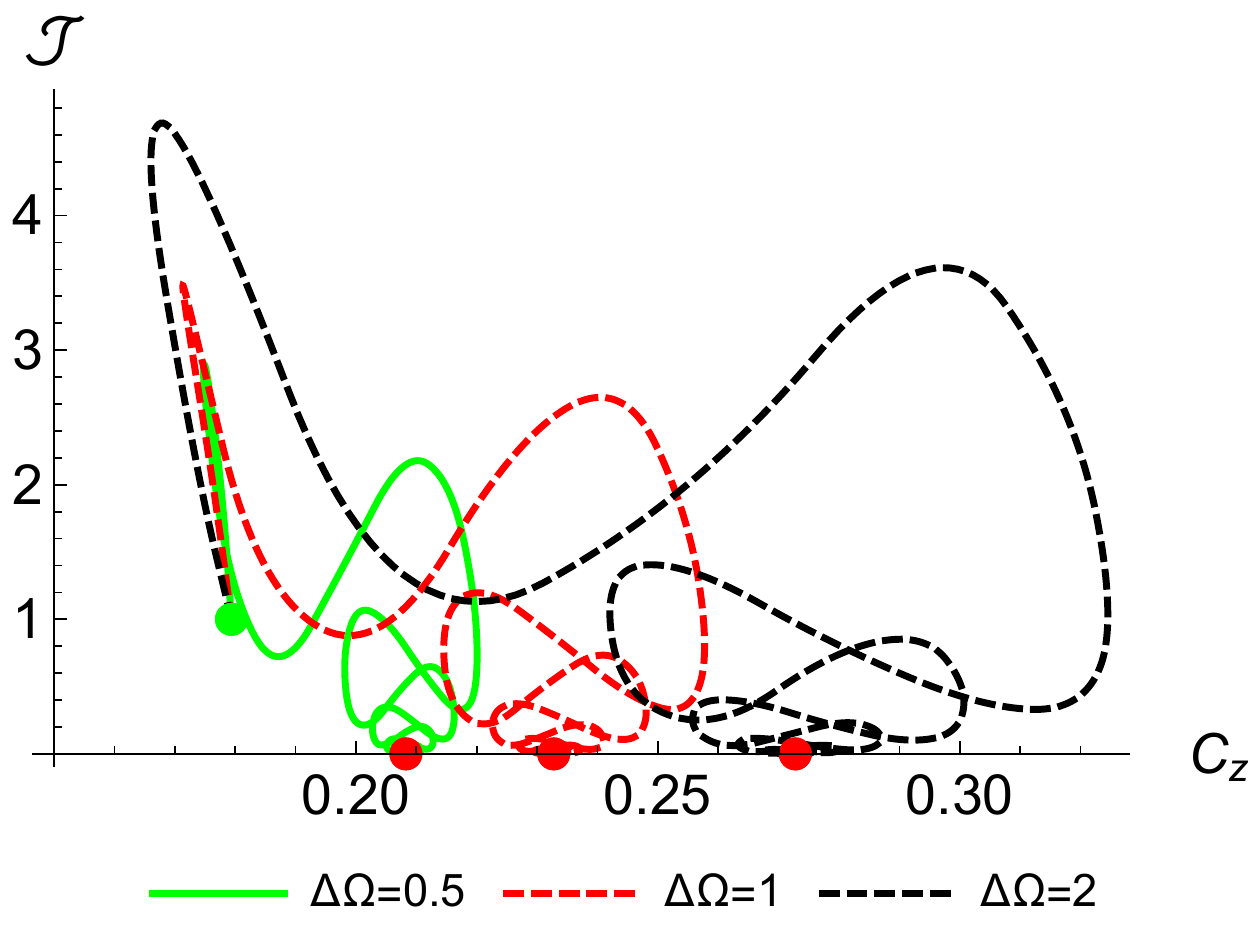}	
	}
	\subfigure[$C_{zz} \sim {\cal J}$ for $f_{x,y,z}=0$]
	{
		\includegraphics[width=0.3\hsize]{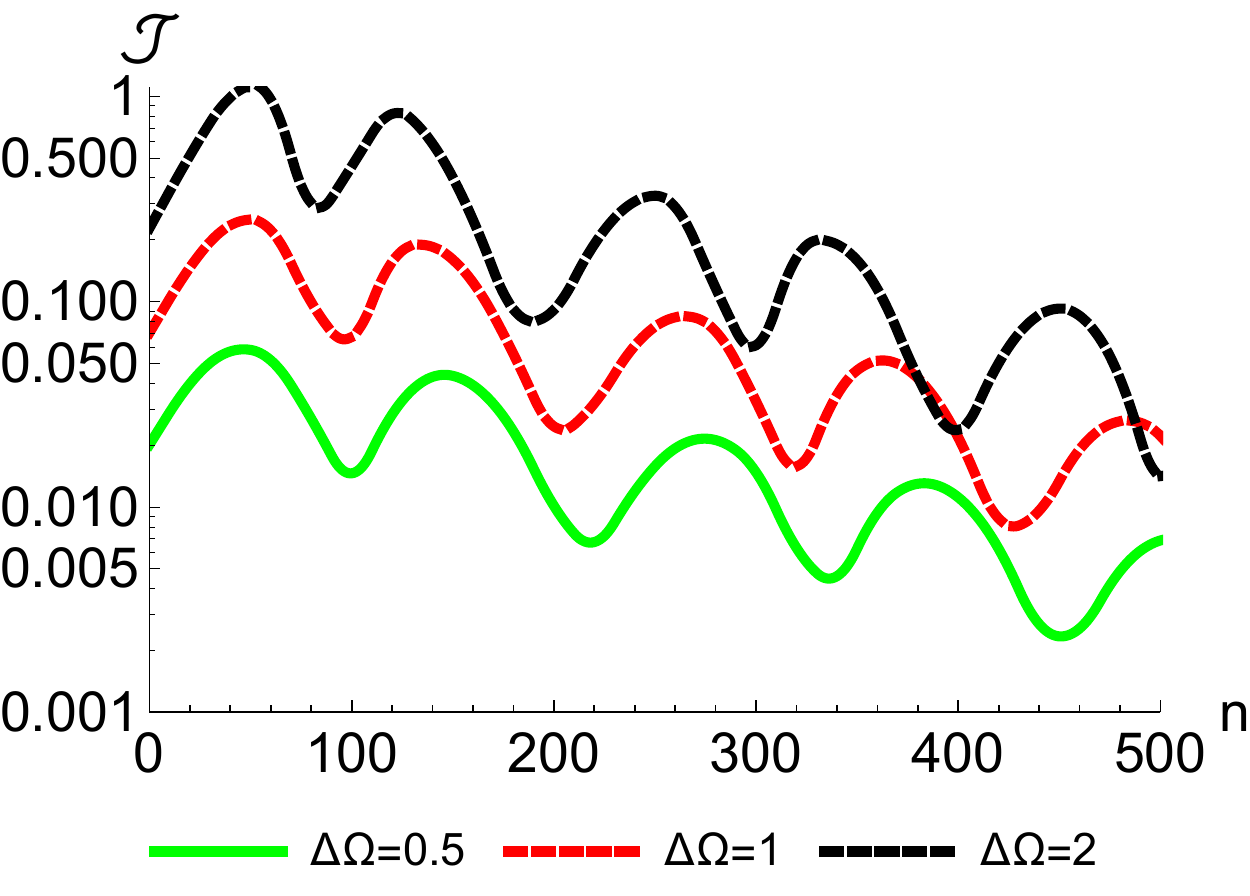}	
	}
\caption{The migration of the stable and time-invariant solution from $\Omega=1.8$ to $1.8+\Delta \Omega$ for the dynamo system for $f_{x,y,z}\sim{\cal N}(0, \ 5)$ and $f_{x,y,z}=0$ via the time stepping method, where (a)-(b) \& (d)-(e) show the path of the low-order statistics, $C_z$ and $C_{zz}$ as the misfit, ${\cal J}$, tends to zero and (c) \& (f) are the illustrations of the exponential convergence of the misfit as the number of iterations.}
\label{solarpt7}
\end{figure}

The stable, time-invariant solution of the {CE3 approximation of the} quasiperiodic dynamo system can always be obtained by the time stepping method. An interesting calculation is to determine the path of approach to a fixed point using a nearby solution as an initial guess. Shown in Fig. (\ref{solarpt7})
are the path of the low-order cumulants, $C_z$ and $C_{zz}$, as ${\cal J}$ approaches to zero and the convergence of ${\cal J}$ as the number of time steps for the CE3 approximation of the dynamo system. Here the initial state is that calculated at $\Omega=1.8$ and we show 3 different cases with  $\Delta\Omega=0.5, 1$ and $2$, where ${\cal J}$ is normalized by its value at the first time step. For the case with  stochastic forcing, the convergence of ${\cal J}$ is accelerated from that for  the dynamo system in the absence of the stochastic force. {Interestingly, the path of the cumulant, e.g., $C_z$ and $C_{zz}$, demonstrates a strong self-similarity as we evolve the CE3 equations in time to reduce ${\cal J}$ for different $\Delta\Omega$. The form of the approach to the fixed point suggests that taking reasonably large values of $\Delta\Omega$ is the best strategy for continuation of solutions. For the cumulant system with non-negligible third order cumulant for $f_{x,y,z}=0$, the misfit, ${\cal J}$, is never found convex as the solution of the cumulant equations converges to the stable fixed point. The same observations have been made for all other test cases. We speculate that this is the reason why the minimization method fails to optimize the CE3 system.}

\subsection{Solar dynamo in the chaotic state}

\begin{figure}[htp]
	\centering
	\subfigure[]
	{
		\includegraphics[width=0.15\hsize]{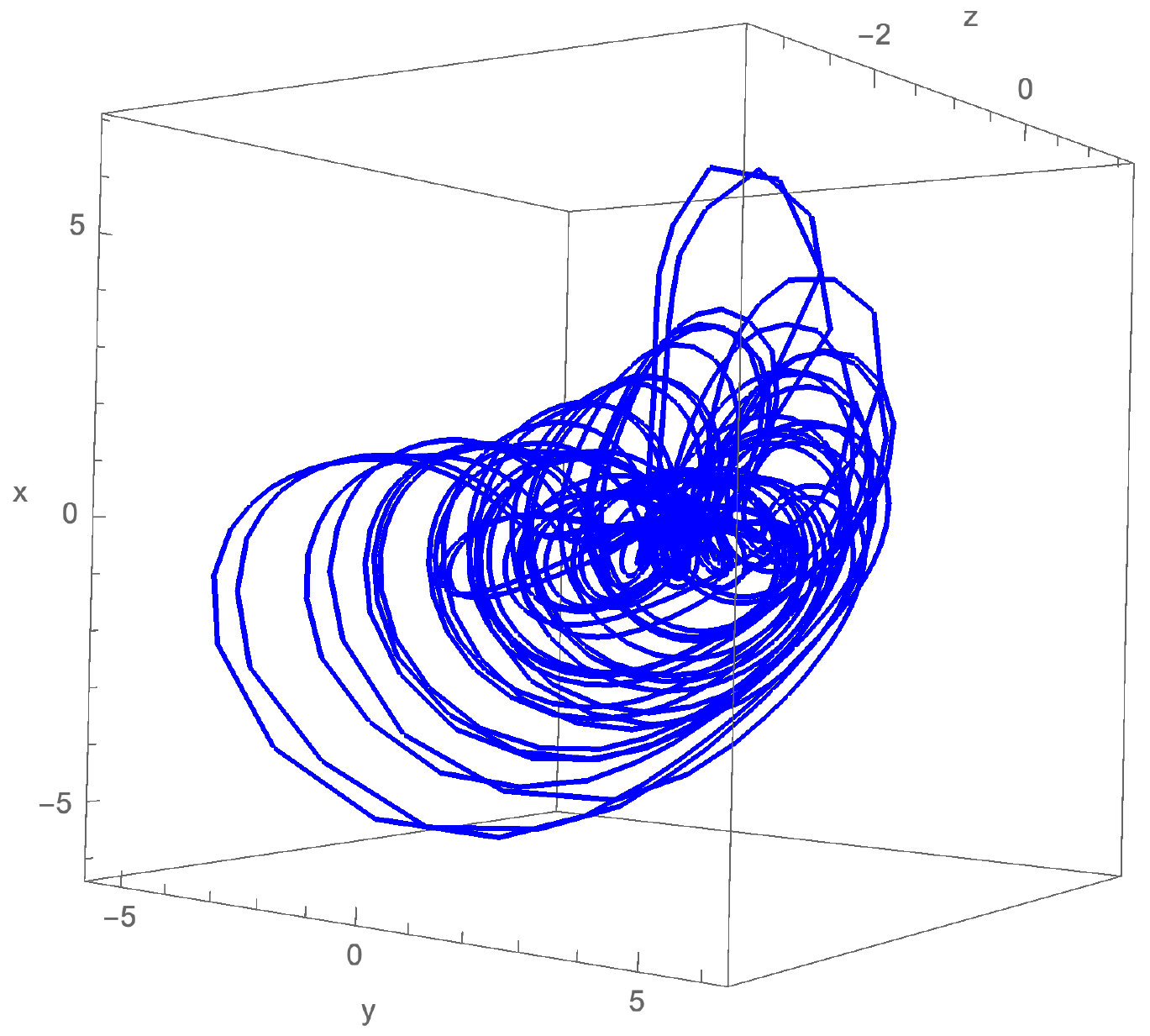}	
	}
	\subfigure[]
	{
		\includegraphics[width=0.15\hsize]{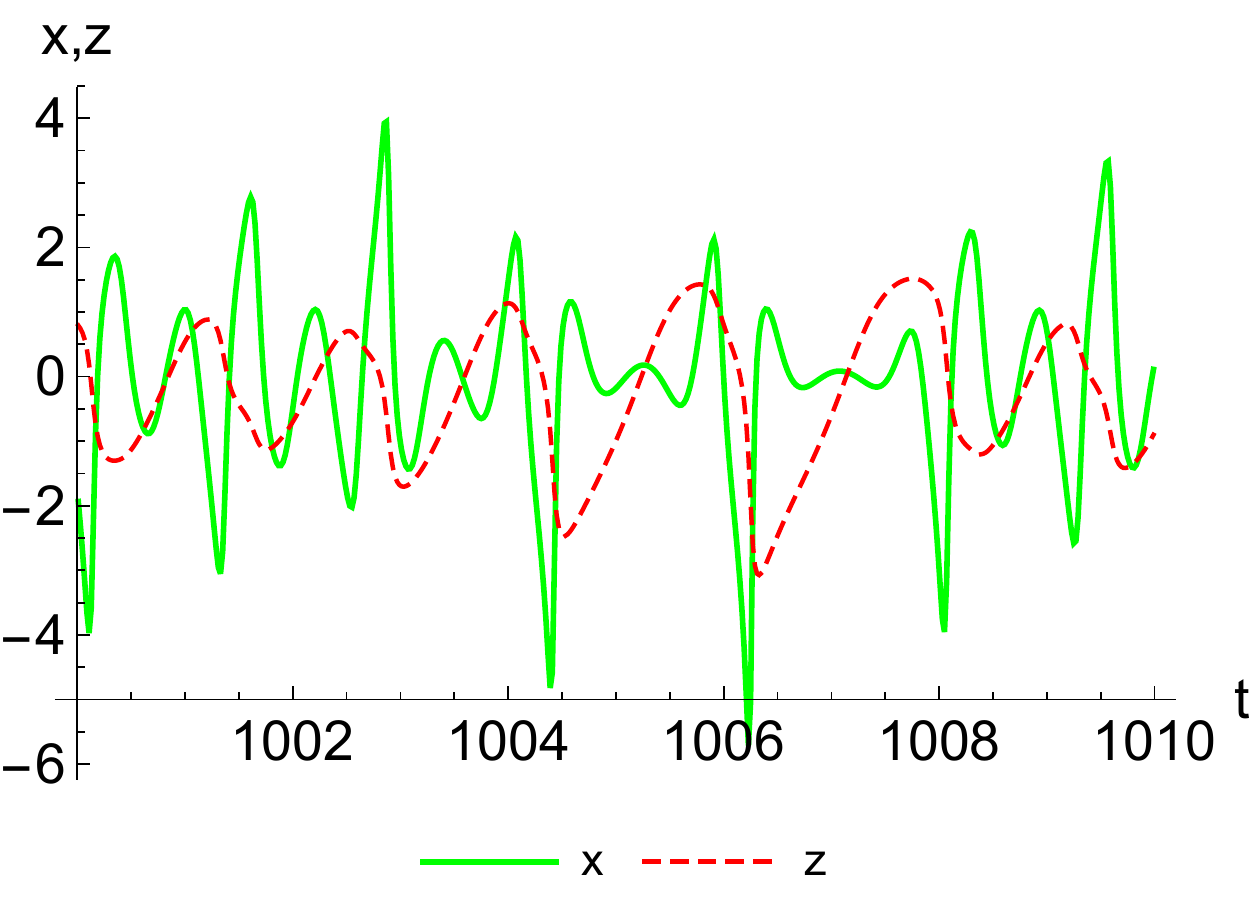}	
	}
	\subfigure[]
	{
		\includegraphics[width=0.15\hsize]{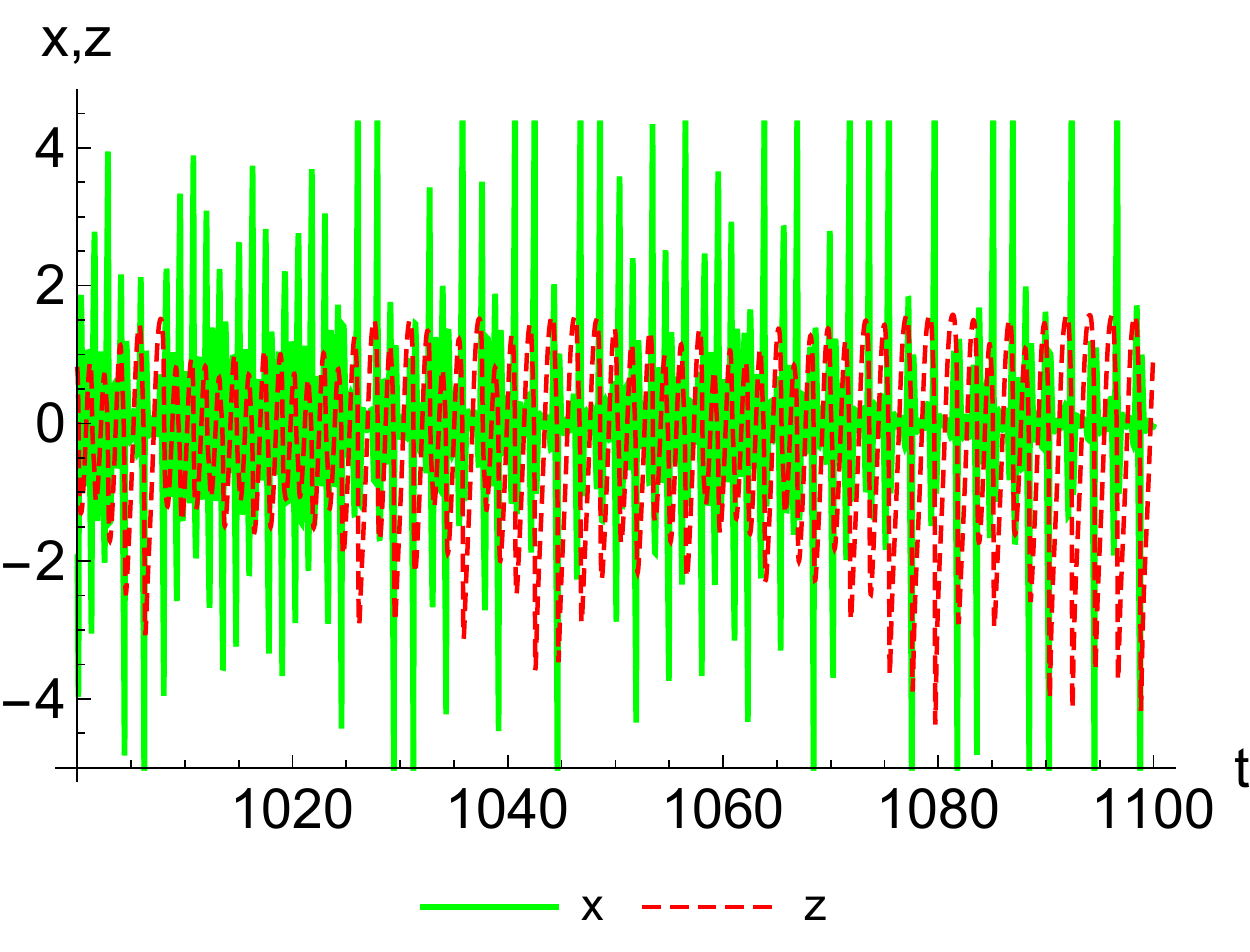}	
	}
	\subfigure[]
	{
		\includegraphics[width=0.15\hsize]{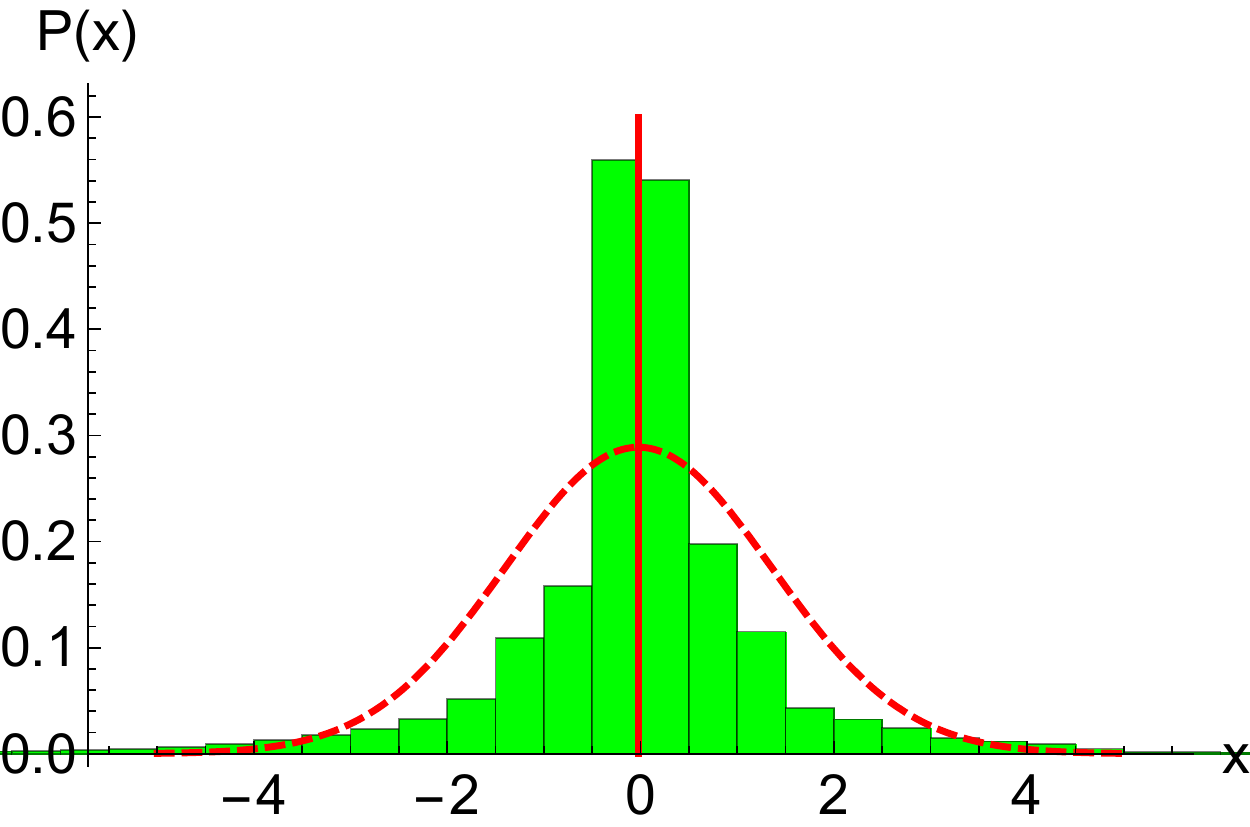}	
	}
	\subfigure[]
	{
		\includegraphics[width=0.15\hsize]{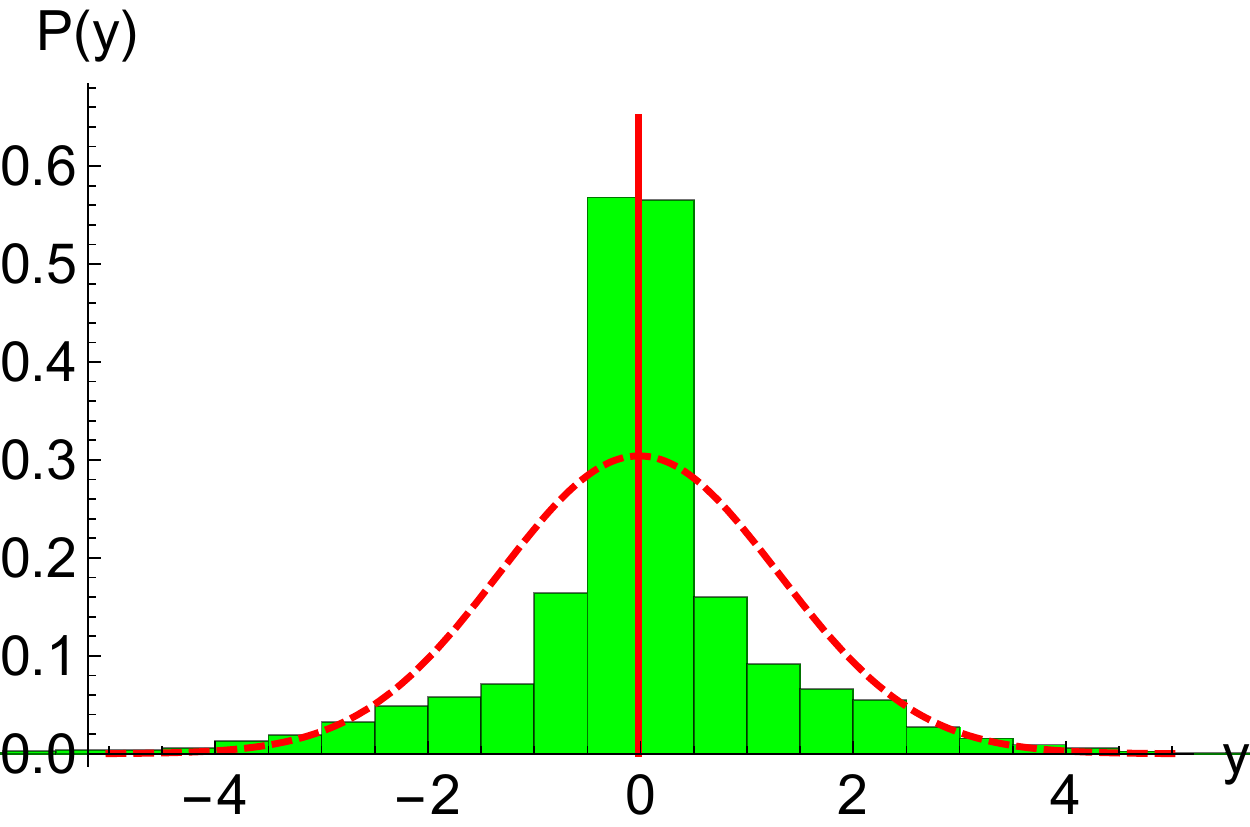}	
	}
	\subfigure[]
	{
		\includegraphics[width=0.15\hsize]{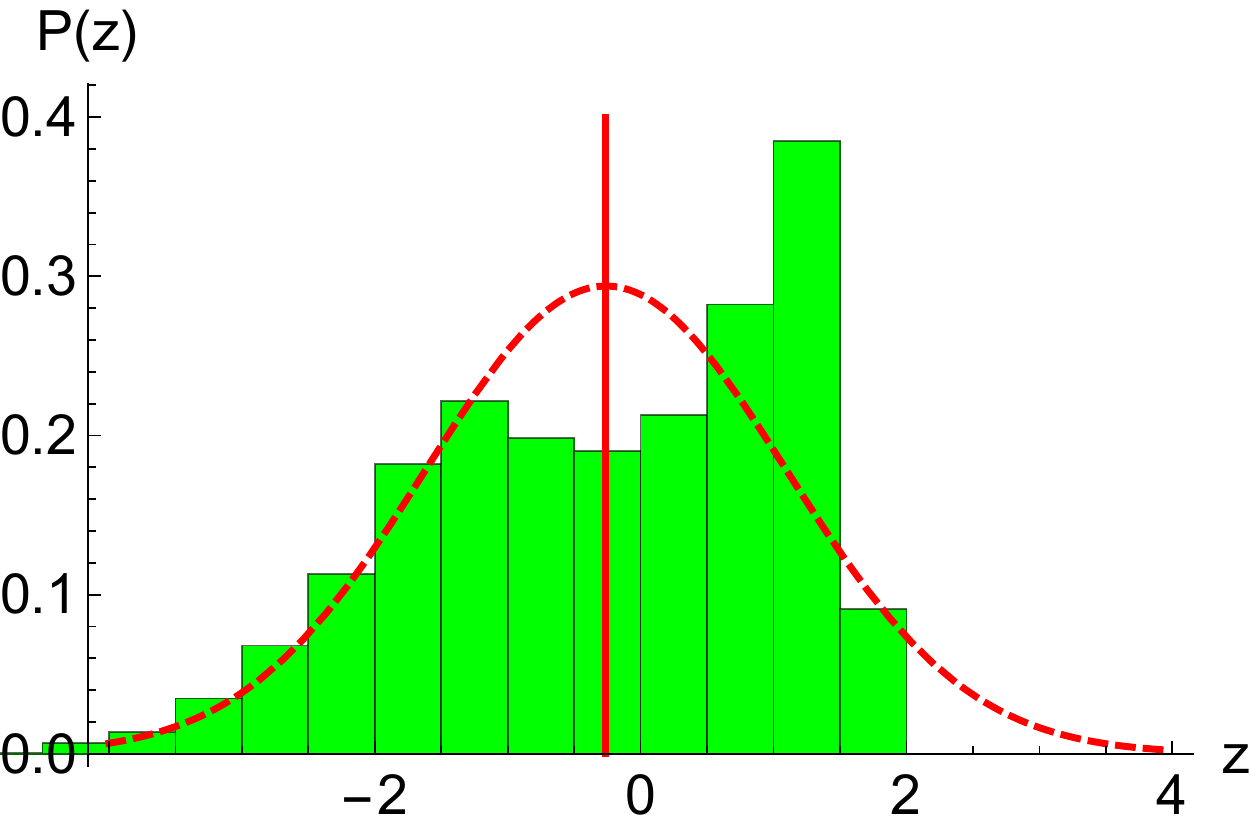}	
	}
\caption{The plots of the trajectory, time series and the PDFs of the solar dynamo in the chaotic state for $\Omega=20$ without the stochastic force, i.e., $f_{x,y,z}=0$, where the statistics obtained via the ensemble average of DNS are shown in green histogram and the Gaussian distribution with the same mean and variance are in dashed red.}
\label{solar2pt1}
\end{figure}

\begin{figure}[htp]
	\centering
	\subfigure[]
	{
		\includegraphics[width=0.15\hsize]{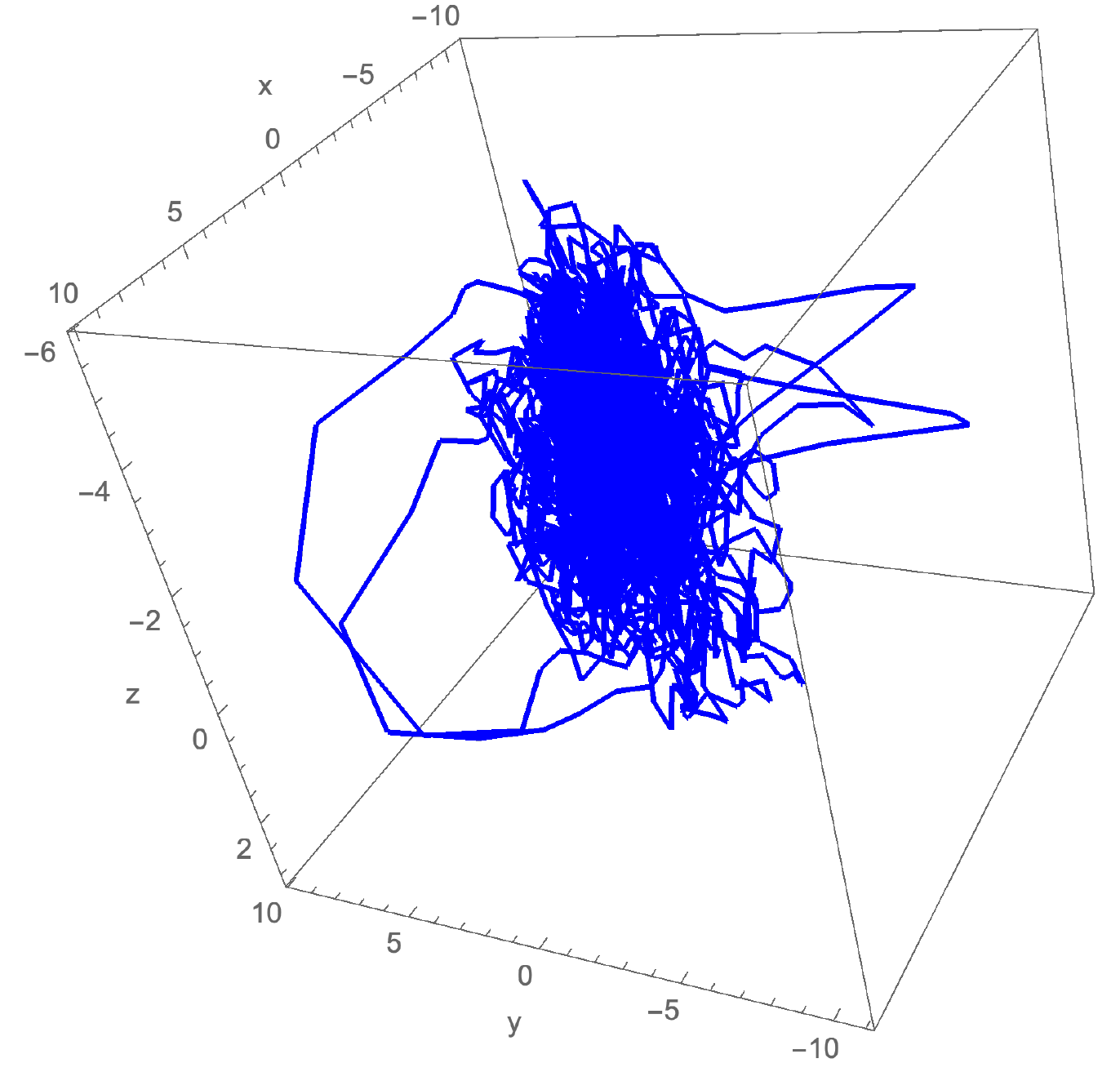}	
	} 
	\subfigure[]
	{
		\includegraphics[width=0.15\hsize]{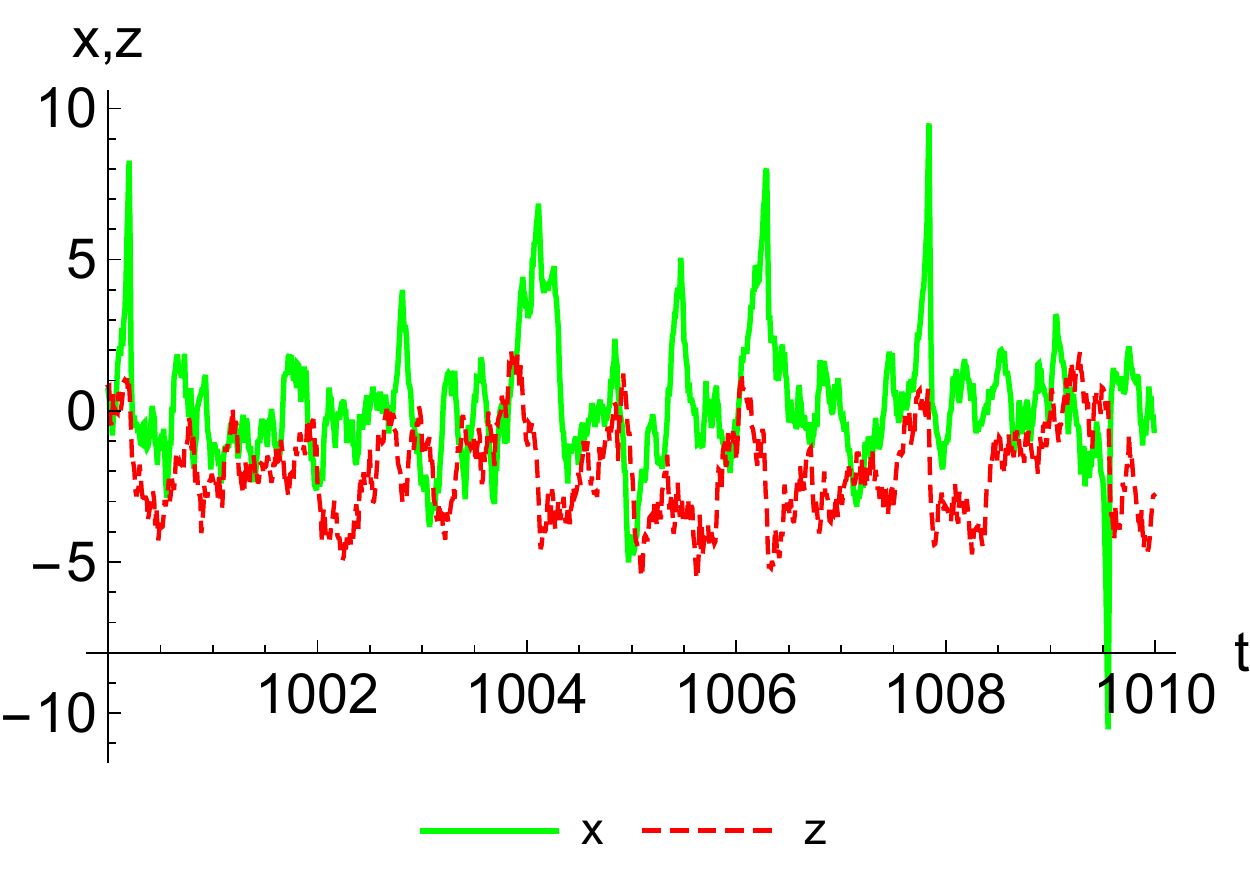}	
	}
	\subfigure[]
	{
		\includegraphics[width=0.15\hsize]{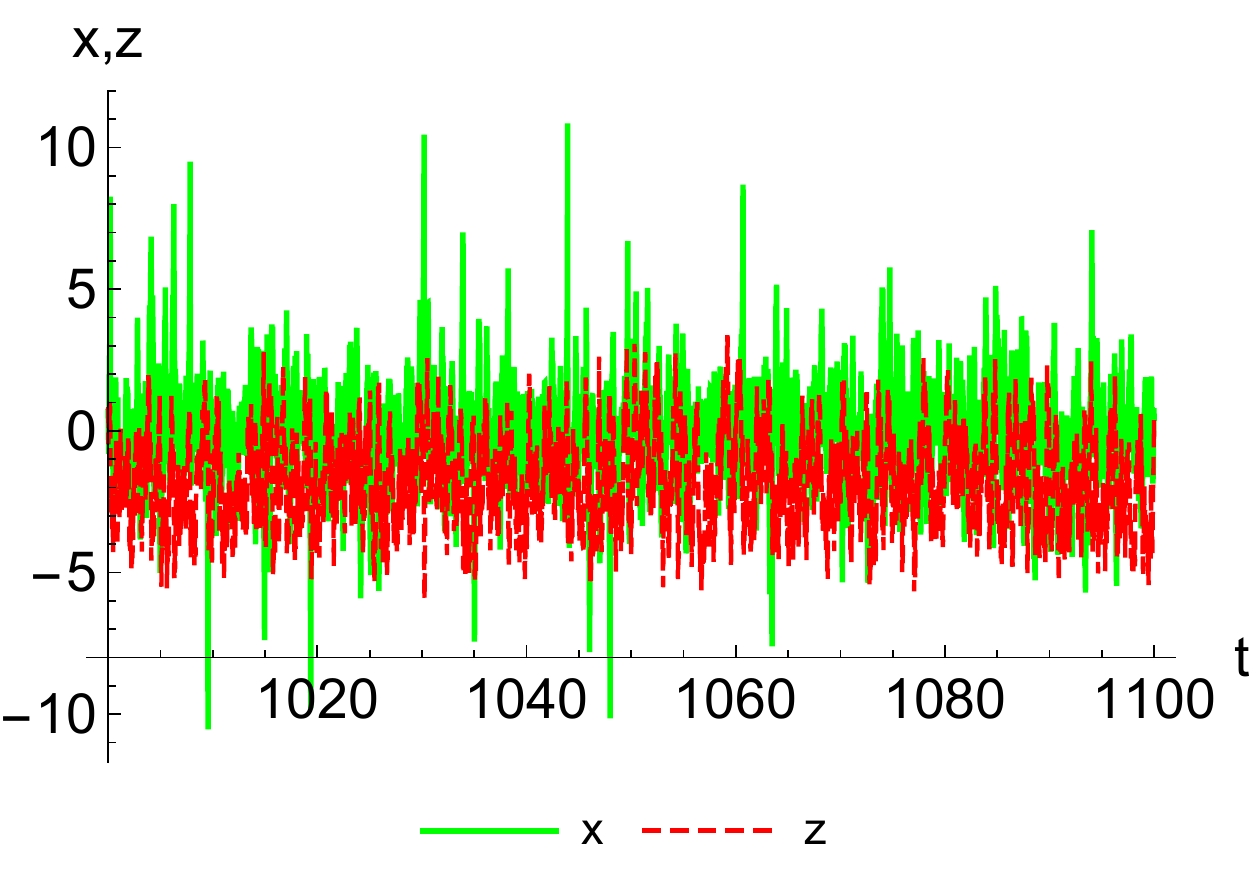}	
	}
	\subfigure[]
	{
		\includegraphics[width=0.15\hsize]{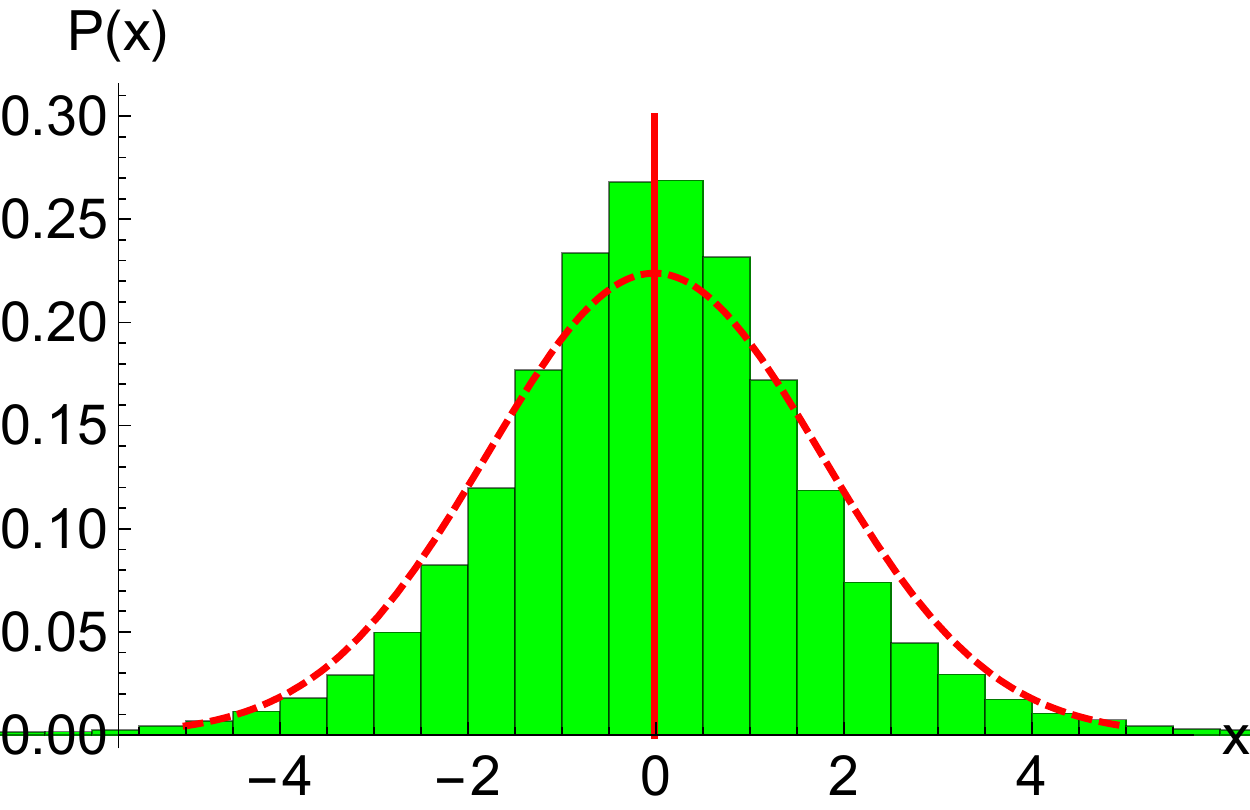}	
	}
	\subfigure[]
	{
		\includegraphics[width=0.15\hsize]{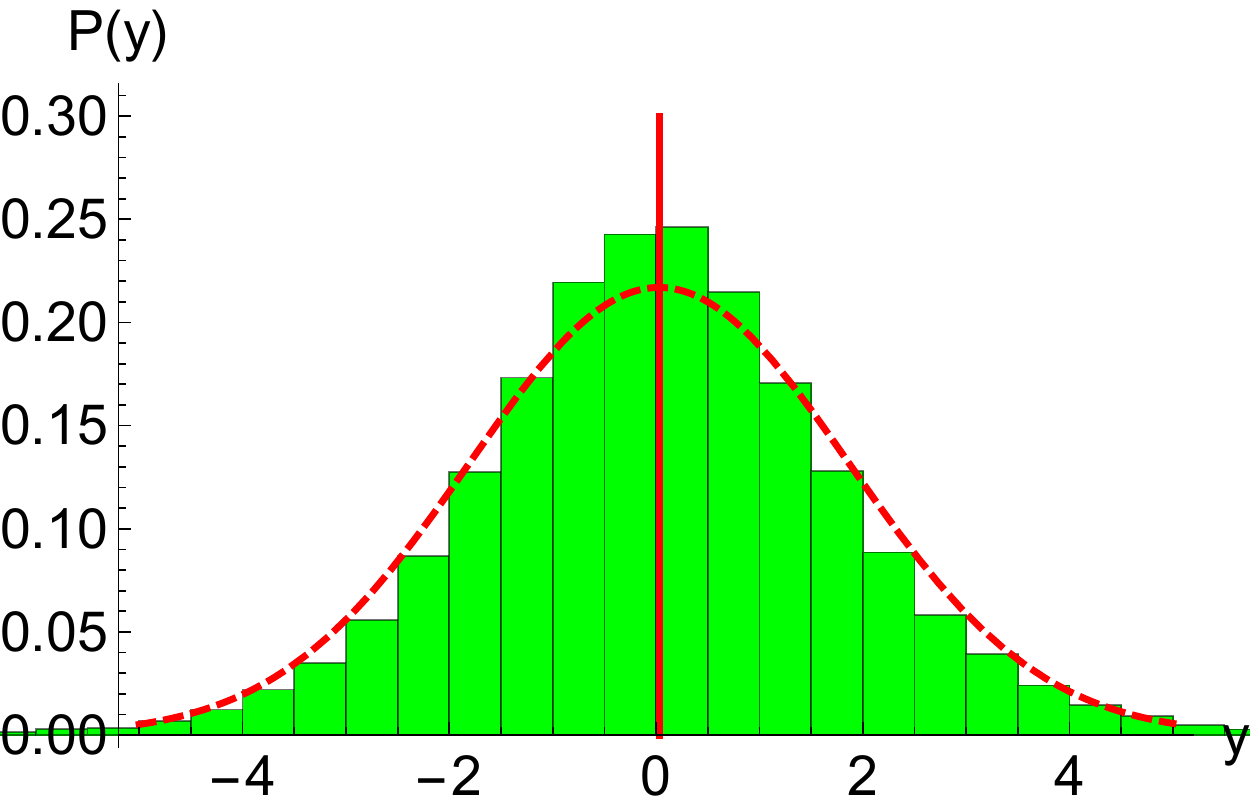}	
	}
	\subfigure[]
	{
		\includegraphics[width=0.15\hsize]{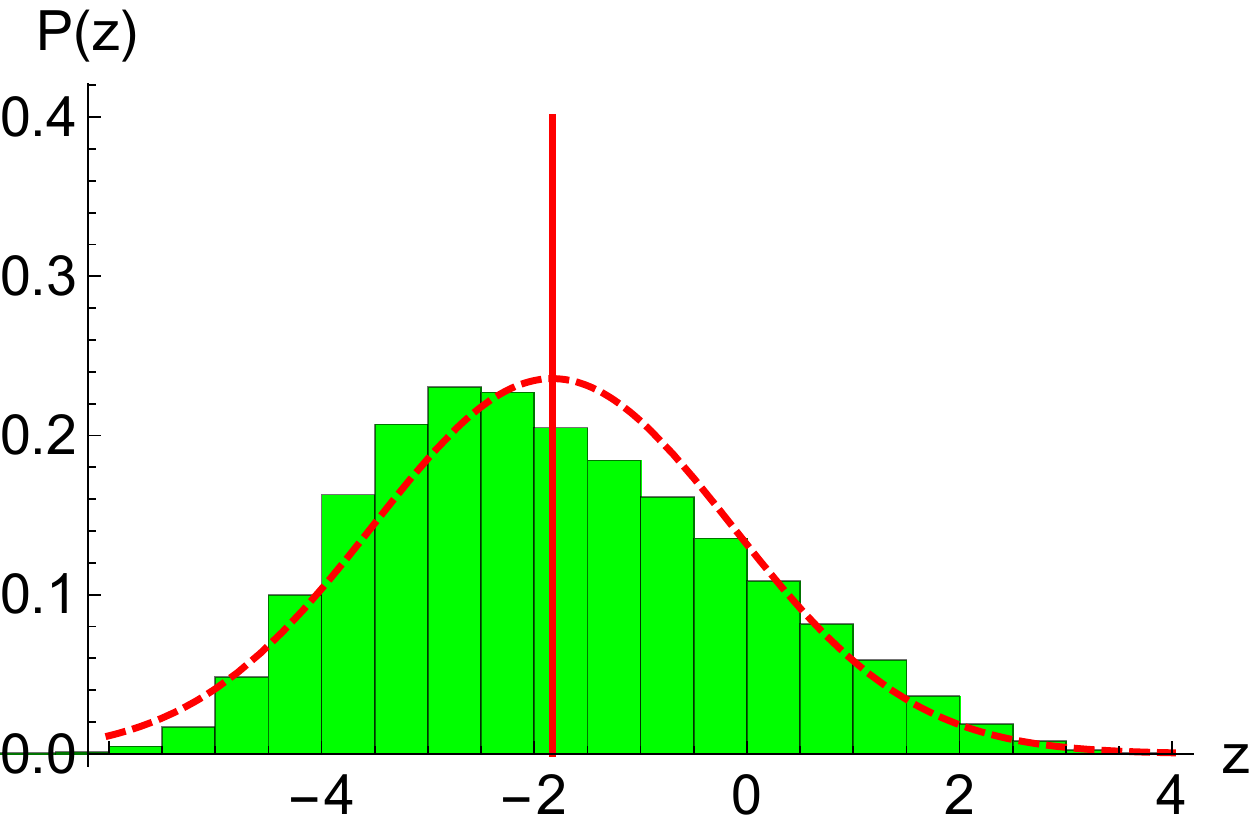}	
	}
\caption{The plots of the trajectory, time series and the PDFs of the solar dynamo in the chaotic state for $\Omega=20$ and $f_{x,y,z}\sim{\cal N}(0, \ 20)$.}
\label{solar2pt2}
\end{figure}

Having investigated the utility of DSS for relatively simple attractors, we move on to a case where the solar dynamo model yields a chaotic attractor. This is achieved by further increasing the control parameter $\Omega$. Fig (\ref{solar2pt1})
illustrates the typical trajectory, time series and PDFs of the dynamo system in the chaotic state for $\Omega=20$ in the absence of stochastic terms, $f_{x,y,z}=0$. Here the PDFs of $x$, $y$ and $z$ that are shown as green histograms that are obtained via the direct numerical simulation and the Gaussian distributions with the same mean and variance are shown in dashed red curves for comparison. The PDFs of the magnetic field, $x$ and $y$, are symmetric and narrowly distributed in the region near zero; and have stronger tails than the Gaussian of the same height. The PDF of the velocity field, $z$, is strongly asymmetric, which indicates the importance of the third order cumulant of $P(z)$. Addition of stochastic terms, $f_{x,y,z}$, increase the randomness of the dynamo action but also acts to regularize the PDFs of $x$,  $y$ and $z$ to be closer to Gaussian, e.g., see Fig. (\ref{solar2pt2}) for $\Omega=20$ and $f_{x,y,z}\sim{\cal N}(0, \ 20)$. Note this is very strong stochastic driving.

\subsection{Direct statistical simulation of the solar dynamo in the chaotic states}

As for the test cases in the quasiperiodic state, the solar dynamo in the chaotic states is accurately approximated by CE3 equations for all test cases (i.e. those both with and without noise), where the eddy damping parameter, $\tau_d$, must be introduced to stabilize the numerical simulation. The most accurate solutions occur for $\tau_d$ within the range from ${\cal O}(10^{-2})$ to ${\cal O}(10^{-1})$. CE2 however does not converge in the absence of noise. 

With the addition of noise, when the PDFs of $x$, $y$ and $z$ are more Gaussian for $\sigma_{x,y,z}^2>1$,  the CE2 equations become stable and accurately represent the statistics of the dynamo action. Detailed results are found in Table (\ref{kindynamo_ces2}).
\begin{table}[htp]
\centering
\begin{tabular}{l|c|c|c|c|c|c|c|c|c|c|c}
    & $F$  &$\tau_d^{-1}$ & $C_x$  & $C_y$  & $C_z$    & $C_{xx}$ & $C_{xy}$ & $C_{xz}$ & $C_{yy}$   & $C_{yz}$ & $C_{zz}$ \\\hline
DNS &   $0$        &              & $0$ & $0$  & $-0.25$ & $1.92$ & $0.11$ & $0$ & $1.76$ & $0$ & $1.72$ \\\hline
CE2.5 &              &    $20$   & $0$ & $0$  & $-0.20$ & $2.02$ & $0$ & $0$ & $2.02$ & $0$ & $0.58$ \\\hline
CE3 &                  &    $15$   & $0$ & $0$  & $-0.19$ & $2.01$ & $0$ & $0$ & $2.01$ & $0$ & $0.60$ \\\hline
CE3 &                  &    $50$   & $0$ & $0$  & $-0.20$ & $2.02$ & $0$ & $0$ & $2.04$ & $0$ & $0.49$ \\\hline\hline
DNS & ${\cal N}(0,\ 5)$ &               & $0$ & $0$  & $-1.09$ & $2.34$ & $-0.31$ & $-0.01$& $2.20$ & $0$ & $1.84$ \\\hline
CE2 &                            &                & $0$ & $0$  & $-1.00$ & $2.17$ & $0$ & $0$& $2.17$ & $0$ & $2.40$ \\\hline
CE2.5 &                          &      $50$  & $0.17$ & $0.04$  & $-0.63$ & $1.73$ & $0.04$&$-0.25$ & $2.16$& $0.57$ & $1.67$ \\\hline\hline
CE3 &                            &      $50$  & $0$ & $0$  & $-1.13$ & $2.24$ & $0$ & $0$& $2.24$ & $0$ & $2.23$ \\\hline\hline
DNS & ${\cal N}(0,\ 20)$ &              & $0$ & $0$  & $-1.85$ & $3.09$ & $-0.60$ & $0$& $3.33$ & $0$ & $2.79$ \\\hline
CE2 &                              &               & $0$ & $0$  & $-1.83$ & $4.15$ & $0$ & $0$& $4.15$ & $0$ & $3.93$ \\\hline
CE2.5 &                           &    $10$   & $-0.34$ & $0.32$  & $-1.21$ & $2.11$ & $0.18$ & $0.43$& $3.70$ & $1.92$ & $2.28$ \\\hline
CE2.5 &                           &    $50$   & $-0.44$ & $0.16$  & $-1.27$ & $2.01$ & $0.22$ & $-0.20$& $3.44$ & $1.95$ & $2.68$ \\\hline
CE3 &                              &    $50$   & $0$ & $0$  & $-2.07$ & $4.20$ & $0$ & $0$& $4.20$ & $0$ & $3.88$ \\\hline
CE3 &                              &    $100$  & $0$ & $0$  & $-1.95$ & $4.18$ & $0$ & $0$& $4.18$ & $0$ & $3.85$ \\\hline
CE3 &                              &    $500$  & $0$ & $0$  & $-1.86$ & $4.16$ & $0$ & $0$& $4.16$ & $0$ & $3.91$ \\\hline
\end{tabular}
\caption{The low-order statistics of the solar dynamo system in the chaotic state for $\Omega=20$ with different stochastic force, $f_{x,y,z}$.}
\label{kindynamo_ces2}
\end{table}


We also observe that the second order cumulant, $C_{zz}$, is least accurately modelled by the CE2/3 approximations with maximum error over $50\%$ for $f_{x,y,z}=0$. By increasing the randomness of the stochastic force, the PDF of $z$ becomes more Gaussian  and the error of $C_{zz}$ is significantly reduced. Again the CE2.5 approximation is found numerically stable for all test cases with the eddy damping parameter, $\tau_d$, within the range from ${\cal O}(10^{-2})$ to ${\cal O}(10^{-1})$. When $\sigma_{x,y,z}^2$ becomes large, the terms that involves the first order cumulants, $C_{x_i}$, are no longer negligible in the third order equations and hence in this limit the CE2.5 approximation becomes less accurate.

\subsection{The fixed points of the solar dynamo in the chaotic state}

\begin{figure}[htp]
\centering
\subfigure[${\cal J} \sim C_z$]
{
	\includegraphics[width=0.3\hsize]{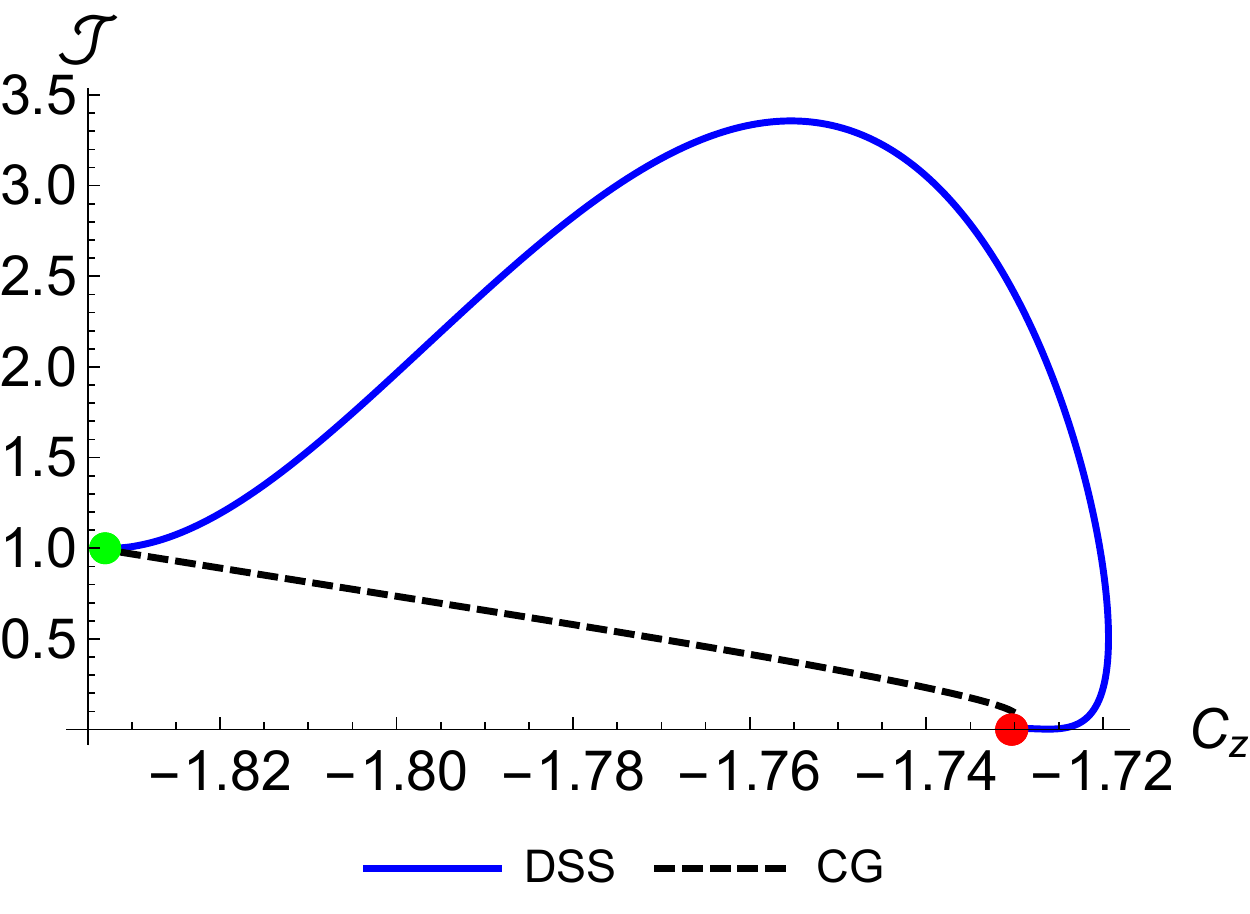}
}
\subfigure[${\cal J} \sim C_{zz}$]
{
	\includegraphics[width=0.3\hsize]{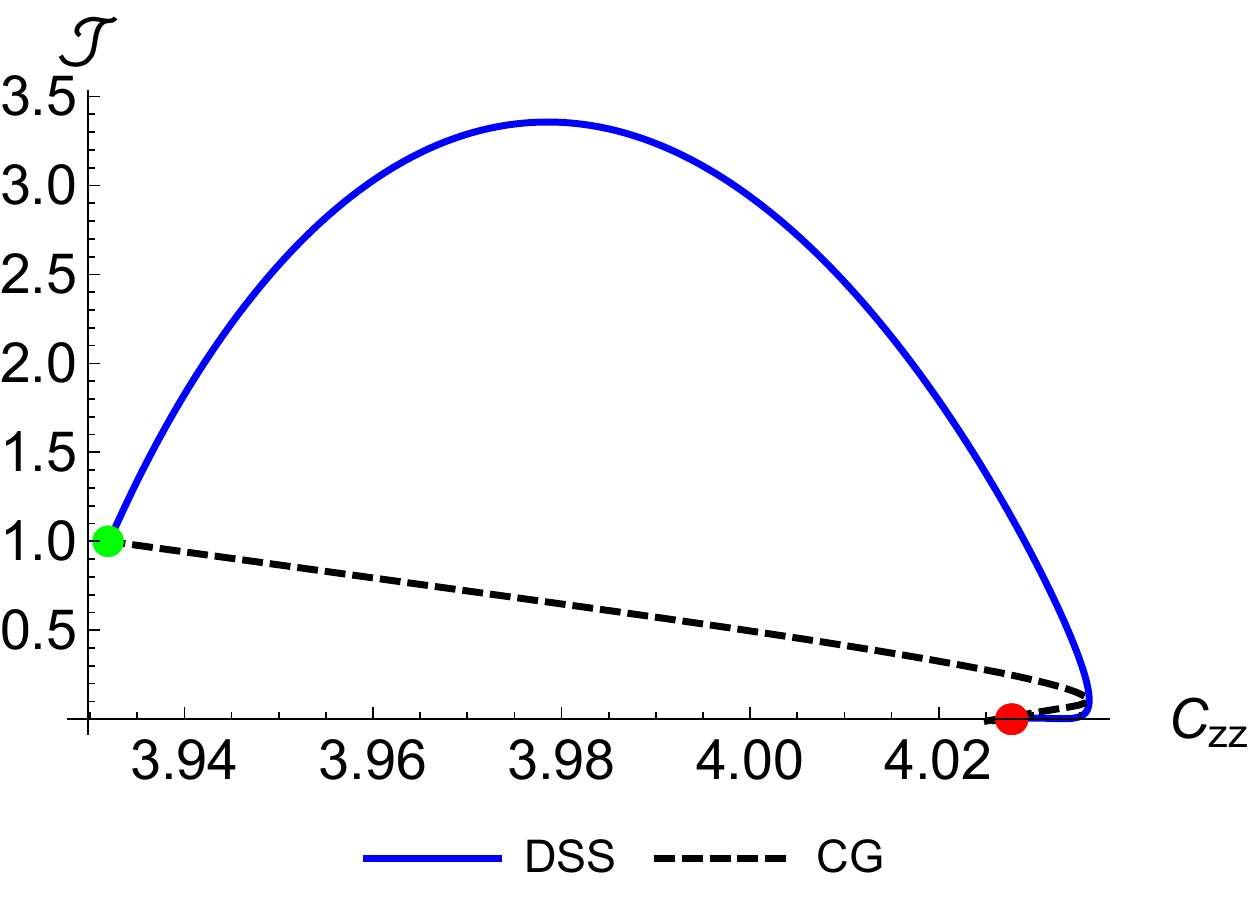}
}
\subfigure[${\cal J} \sim n$]
{
	\includegraphics[width=0.3\hsize]{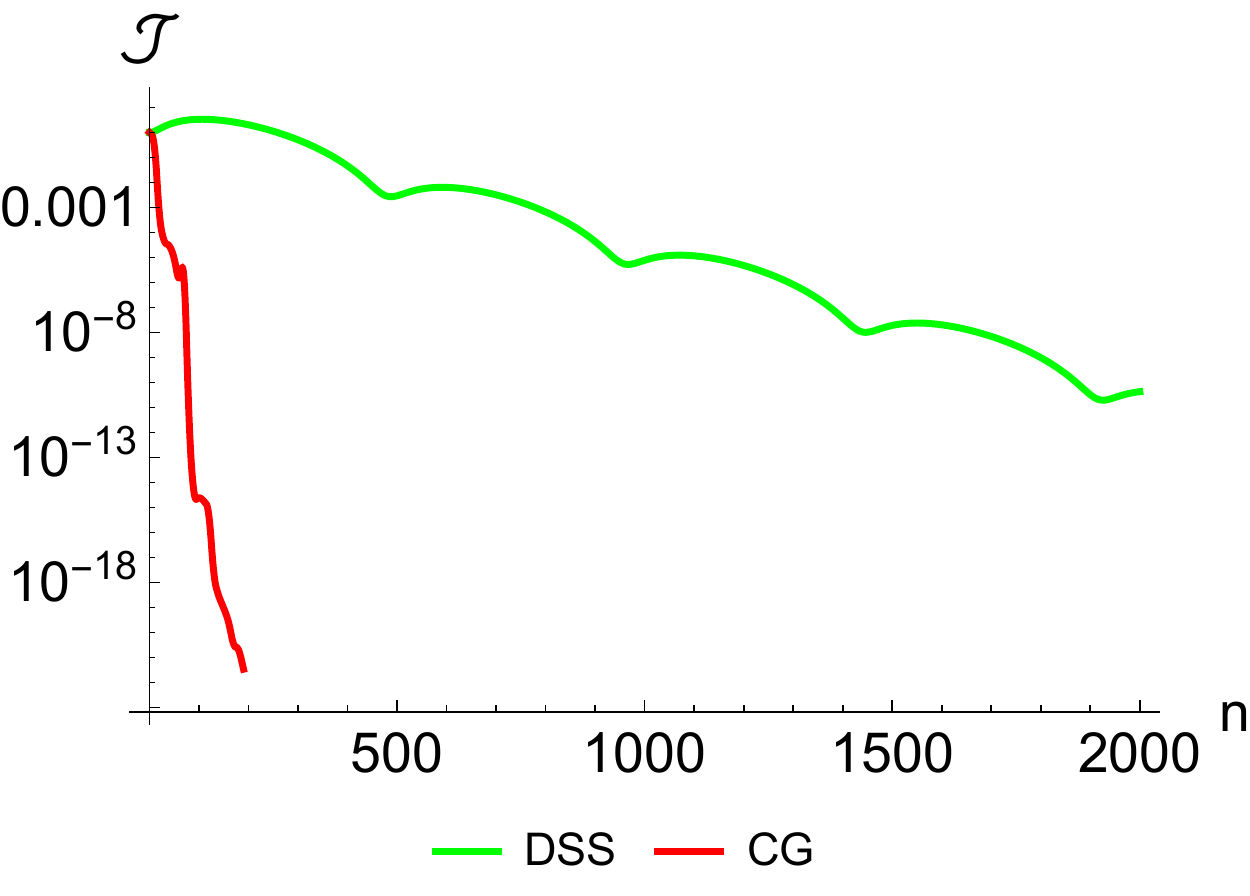}
}
\caption{Illustration of the migration of $C_z$ and $C_{zz}$ from $\Omega=20$ to $30$ via the time stepping and {\it CG} method for CE2 equations, where shown in (a) \& (b) are the path of the low-order statistics as ${\cal J}$ tends to zero and (c) illustrates the exponential convergence of ${\cal J}$ as the number of time steps/iterations, where the stochastic force is $f_{x,y,z}\sim{\cal N}(0, \ 20)$. The green dots in (a) and (b) stand for the initial condition/guess for the time stepping and {\it CG} method and the terminal solutions of {\it CG} shown in red dots that agrees perfectly with the time-invariant solution that obtained via the time stepping method.}
\label{solarMig2}
\end{figure}

The gradient based methods have demonstrated a great efficiency and accuracy for computing the stable fixed point for CE2 approximation as compared with those obtained via the time stepping method. Fig. (\ref{solarMig2}) 
illustrate the typical example for computing of the fixed point for the solar dynamo in the chaotic state for $\Omega=30$ and $f_{x,y,z}\sim{\cal N}(0, \ 20)$ via the {\it CG} and time stepping method, where the initial condition/guess that are shown in green dots in Fig. (\ref{solarMig2} a, b) are taken from the stable fixed point of $\Omega=20$ and the terminal solution shown in red dots agrees perfectly with the time stepping method. For both methods, the misfit, ${\cal J}$, converges to zero exponentially, however the {\it CG} method is approximately $10$ to $10^2$ times faster than the time stepping method to achieve the same numerical accuracy.

\begin{figure}[htp]
\centering
\subfigure[${\cal J} \sim C_z$]
{
	\includegraphics[width=0.3\hsize]{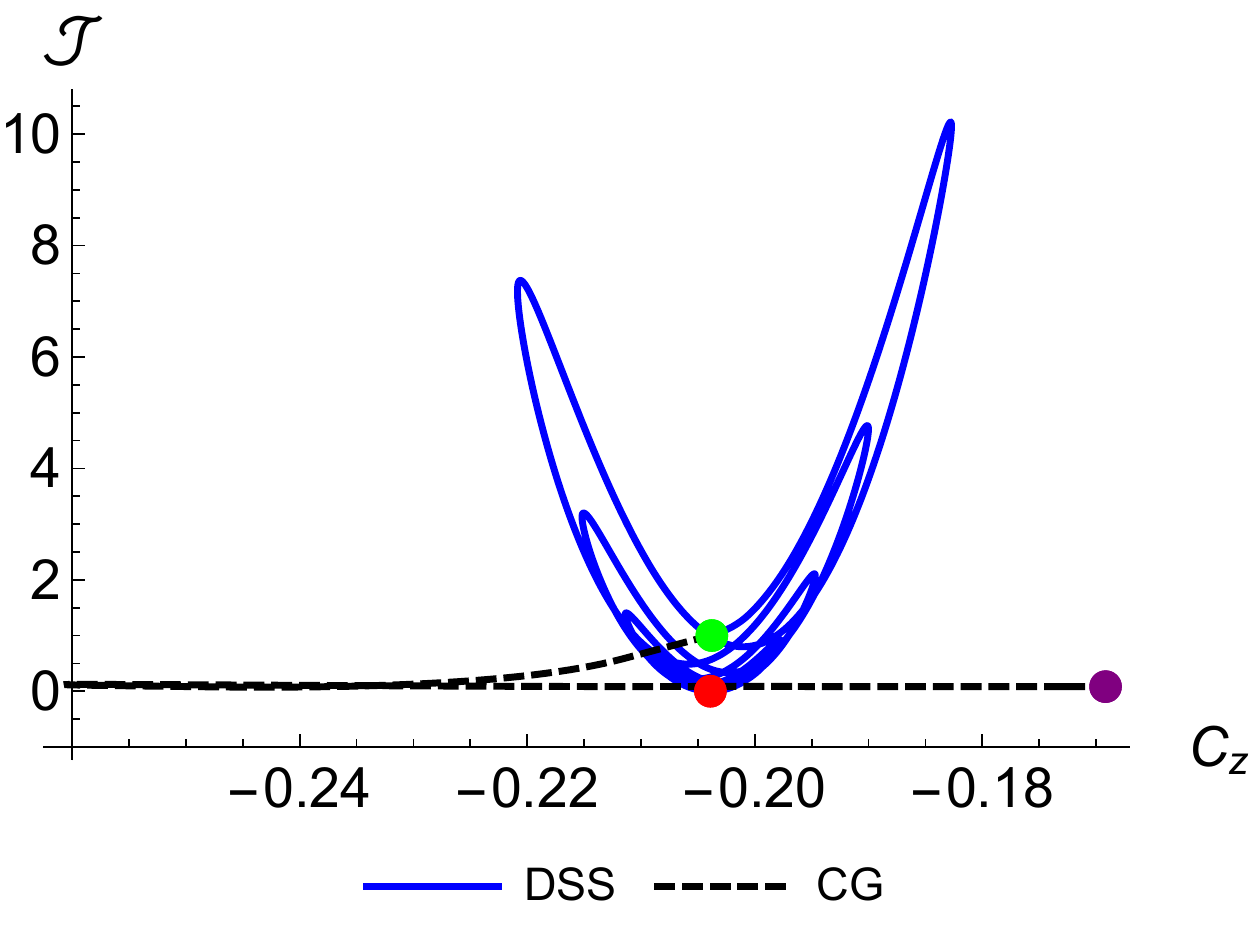}
}
\subfigure[${\cal J} \sim C_{zz}$]
{
	\includegraphics[width=0.3\hsize]{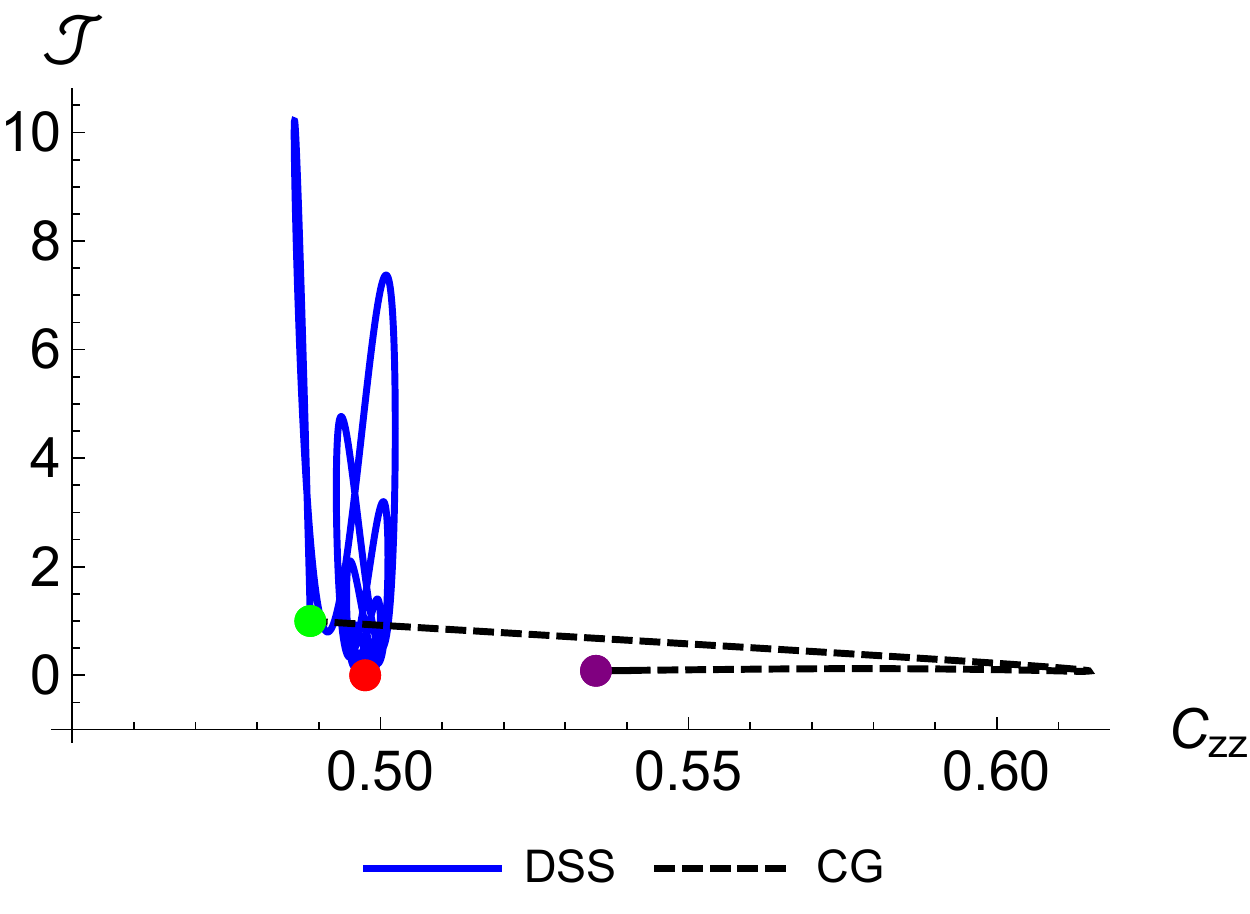}
}
\subfigure[${\cal J} \sim n$]
{
	\includegraphics[width=0.3\hsize]{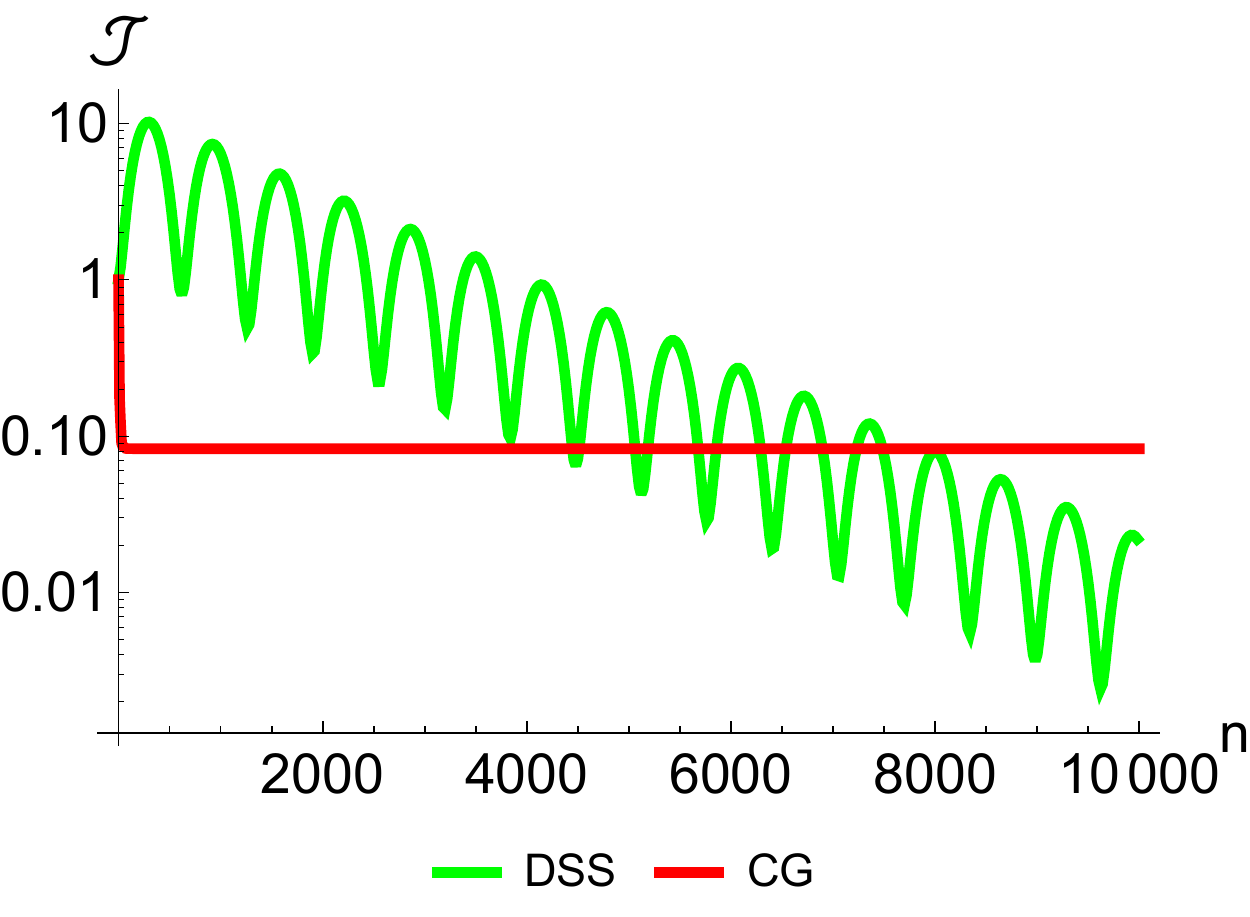}
}
\caption{The migration of $C_z$ and $C_{zz}$ from $\Omega=20$ to $21$ via the time stepping and {\it CG} method for CE3 equations in (a) \& (b) and the convergence of ${\cal J}$ as the number of time steps/iterations in (c), where the stochastic force is zero, i.e., $f_{x,y,z}=0$, the green dots in (a) and (b) stand for the initial condition/guess for the time stepping/{\it CG} method, the red dots are for the time invariant solution found by time stepping method and the purple ones are for the terminal solution of {\it CG}.}
\label{solarMig4}
\end{figure}

As before, for the CE3 approximations the solution of the gradient based optimization method either converges to other (unstable) fixed point or gets stuck in a local minimum. Fig. (\ref{solarMig4})
shows the comparison of {\it CG} method and time stepping method for computing the fixed point of CE3 equations of the dynamo system in the chaotic state for $\Omega=21$ and $f_{x,y,z}=0$, where the initial condition (shown as the green dots in Fig. (\ref{solarMig4}a \& b)), are taken from the stable fixed point for $\Omega=20$. Here the stable fixed point obtained by the time stepping method is shown as red dots and purple dots show  the terminal solution for the {\it CG} method. In this case, after a few iterations, the solution of {\it CG} is trapped by the local minima, while the time stepping method successfully finds the stable fixed points. 

\begin{figure}[htp]
\centering
\subfigure[${\cal J} \sim C_z$ for $f_{x,y,z}=0$]
{
	\includegraphics[width=0.3\hsize]{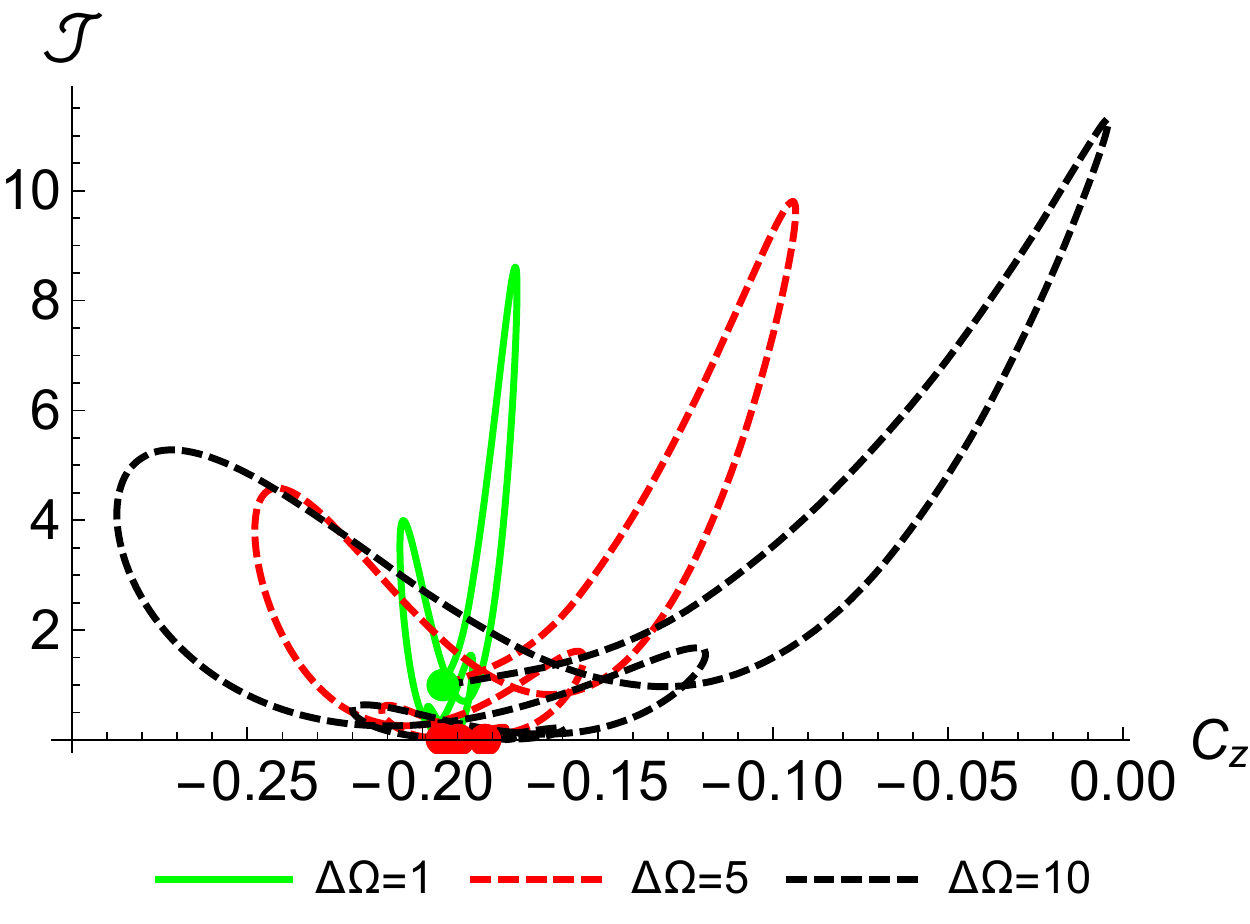}
}
\subfigure[${\cal J} \sim C_{zz}$ for $f_{x,y,z}=0$]
{
	\includegraphics[width=0.3\hsize]{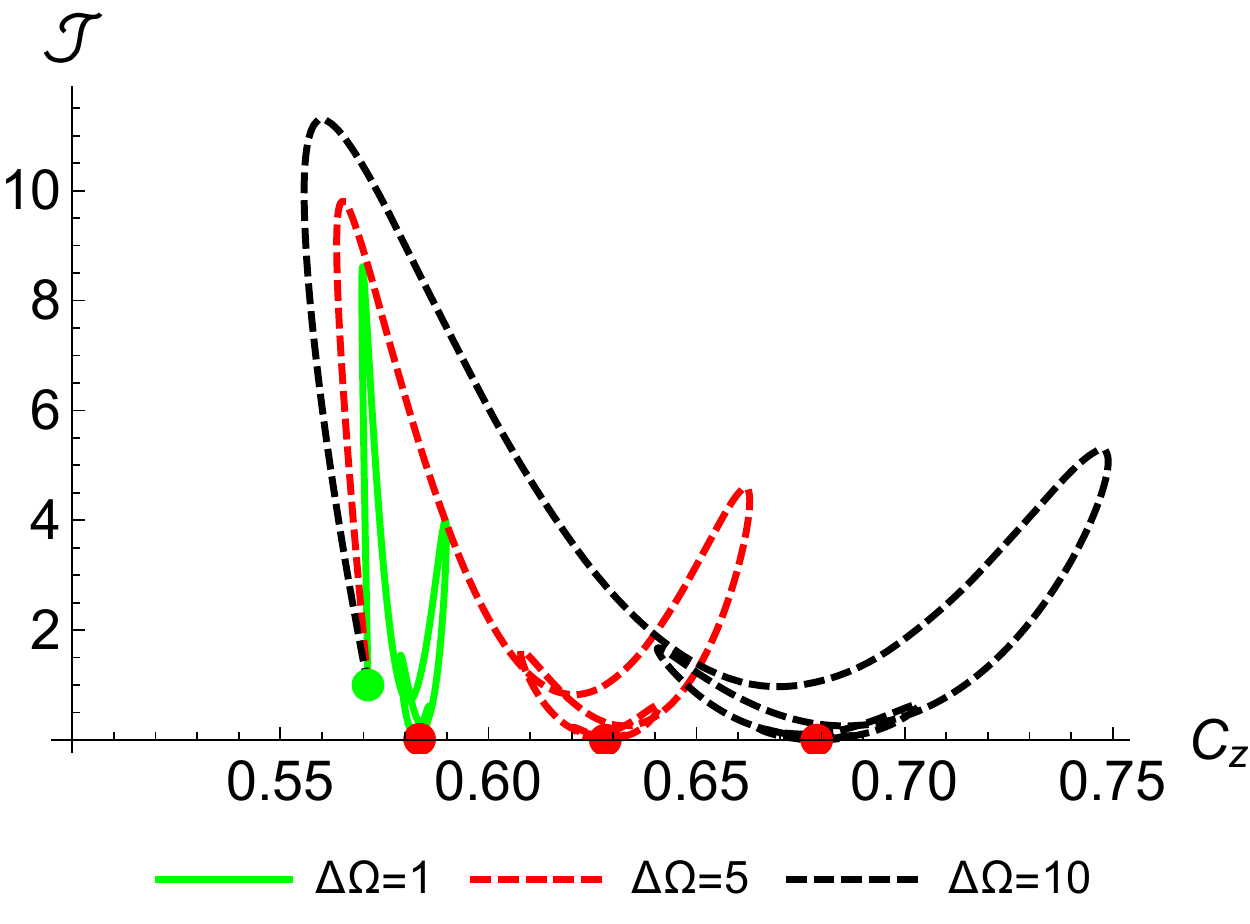}
}
\subfigure[${\cal J} \sim n$ for $f_{x,y,z}=0$]
{
	\includegraphics[width=0.3\hsize]{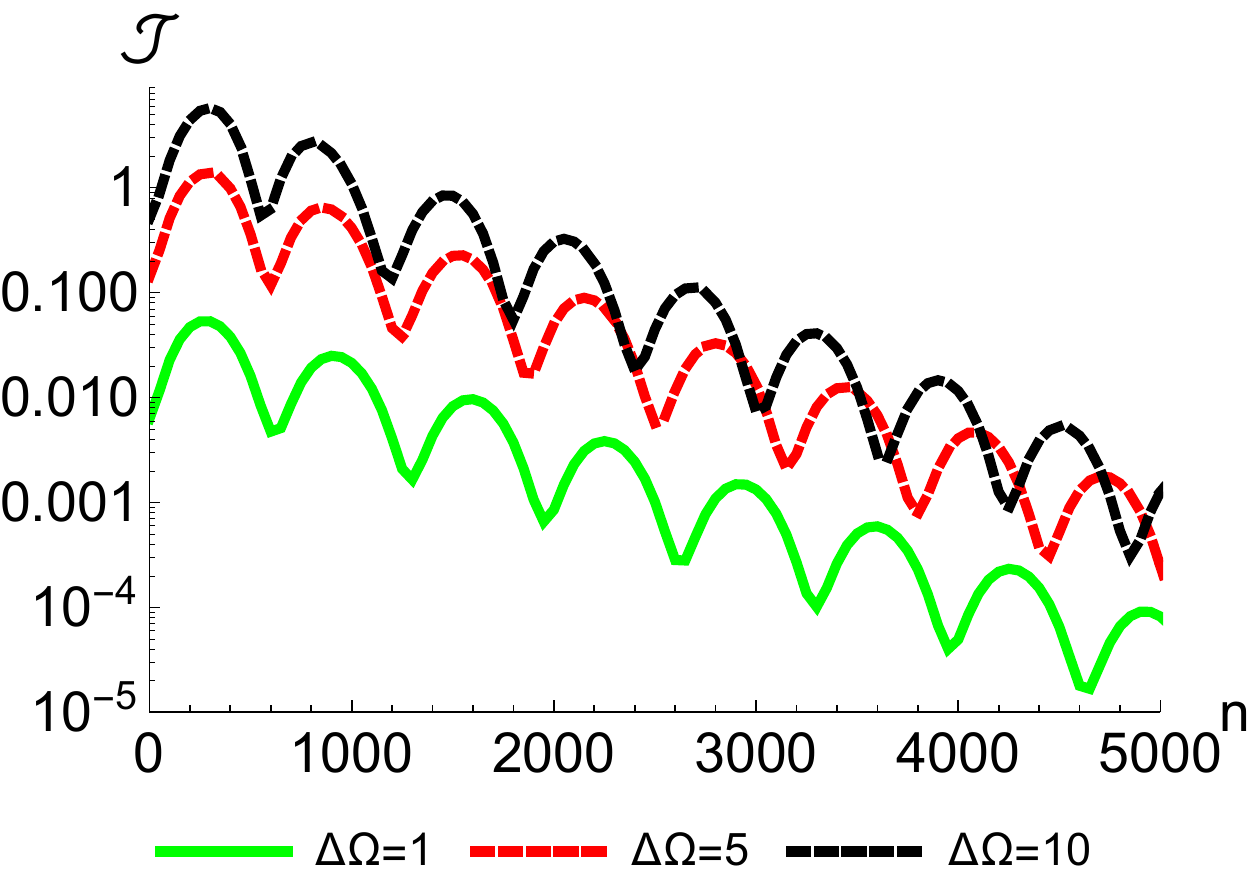}
}\\
\subfigure[${\cal J} \sim C_z$ for $f_{x,y,z}\sim{\cal N}(0, \ 5)$]
{
	\includegraphics[width=0.3\hsize]{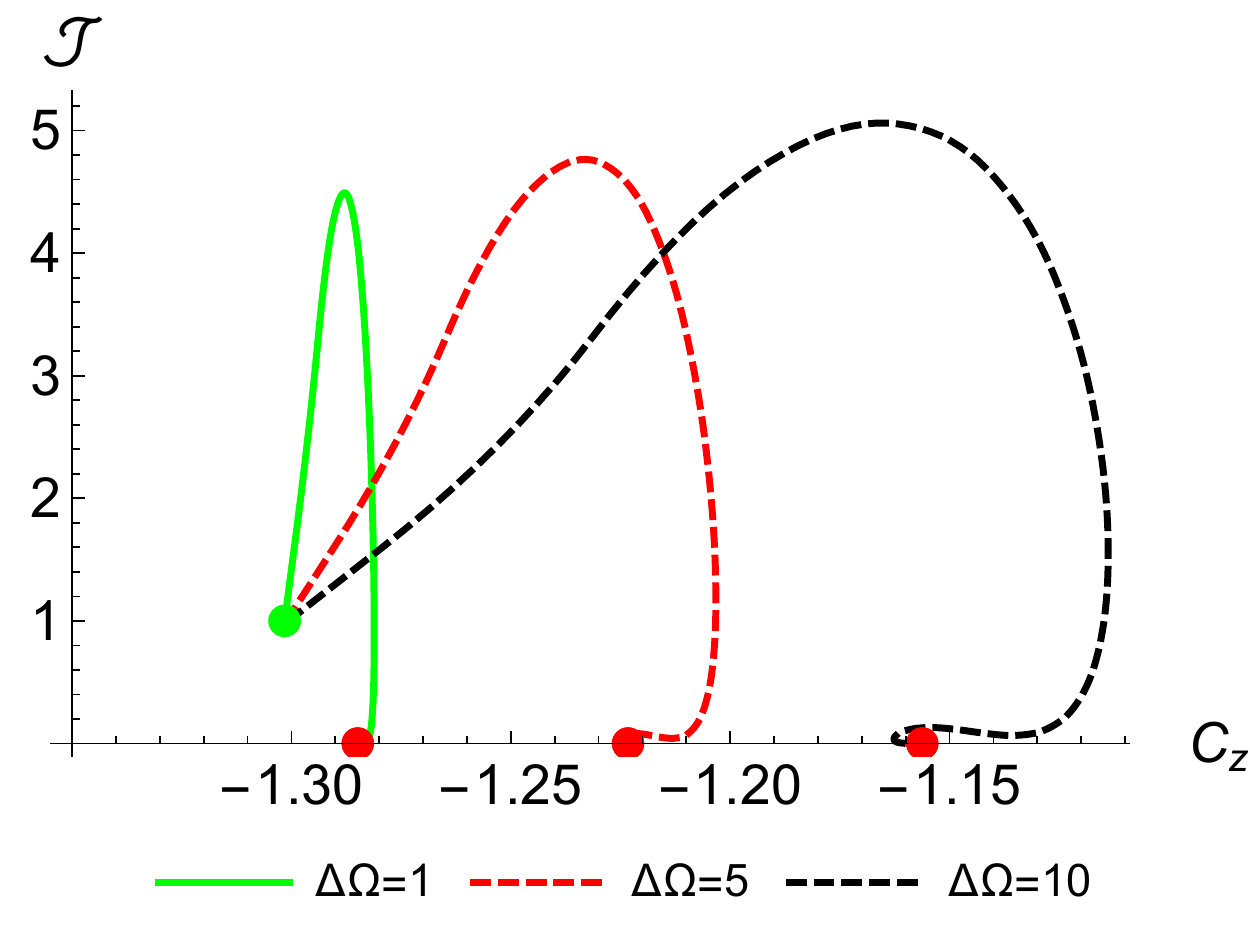}
}
\subfigure[${\cal J} \sim C_{zz}$ for $f_{x,y,z}\sim{\cal N}(0, \ 5)$]
{
	\includegraphics[width=0.3\hsize]{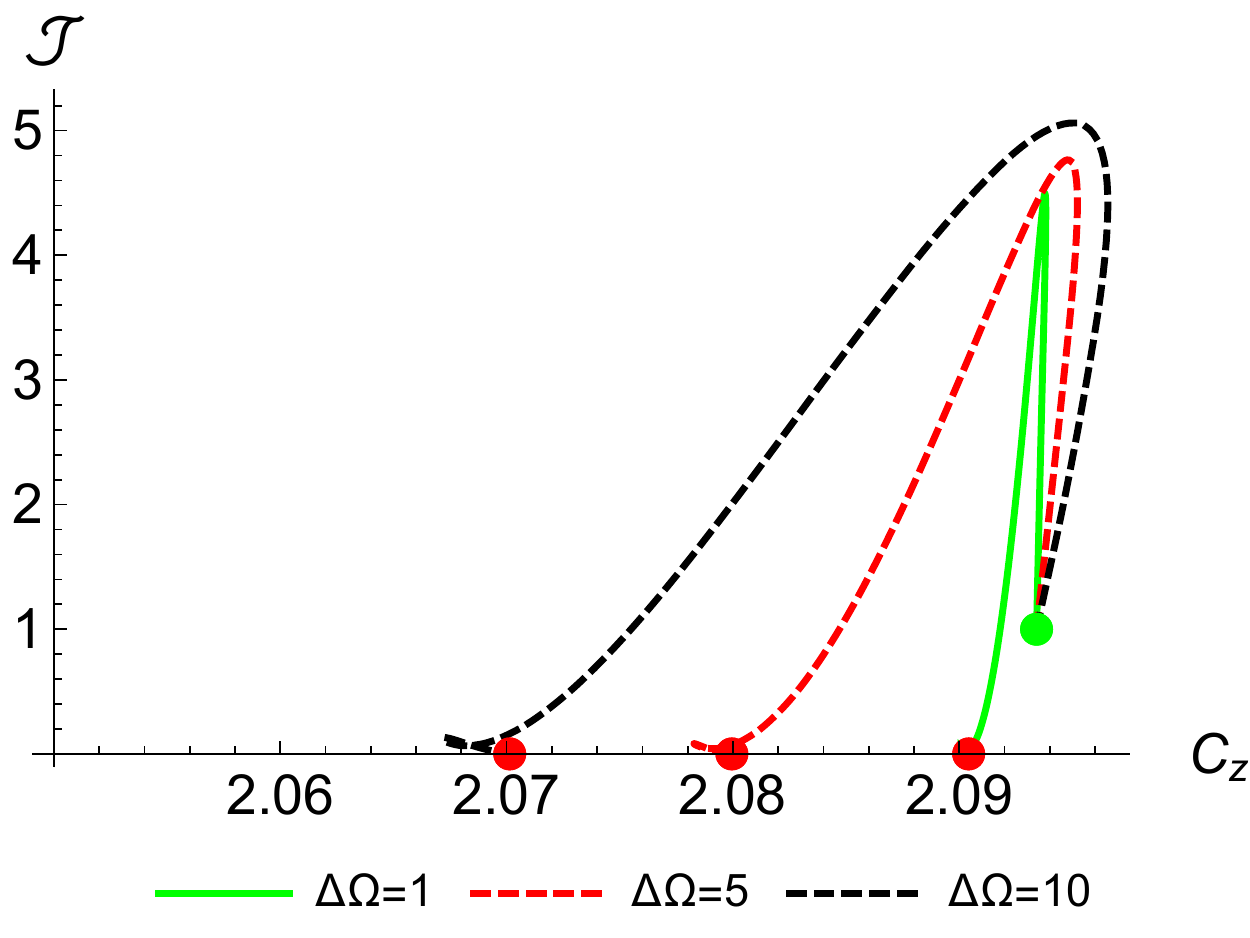}
}
\subfigure[${\cal J} \sim n$ for $f_{x,y,z}\sim{\cal N}(0, \ 5)$]
{
	\includegraphics[width=0.3\hsize]{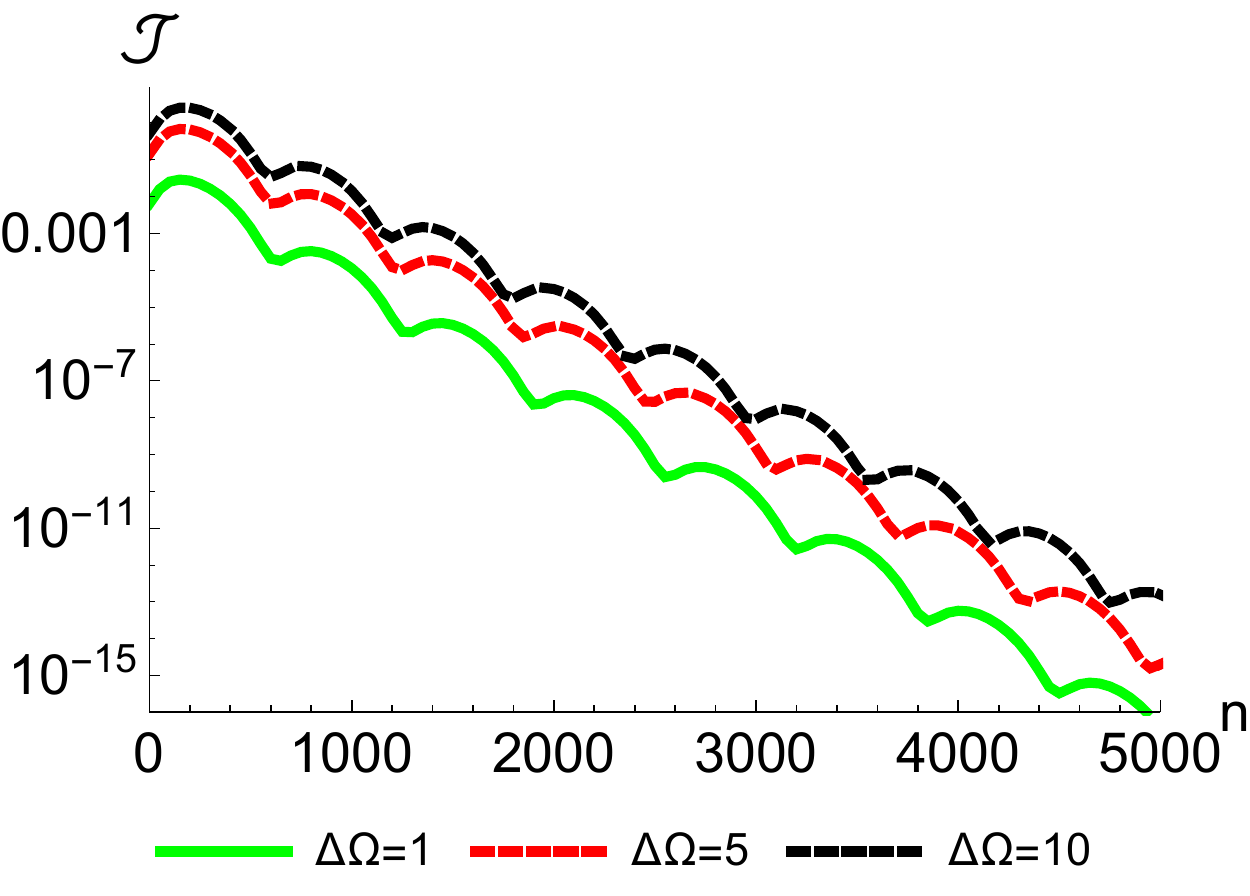}
}
\caption{Illustration of the fixed points of the CE3 approximation of the solar dynamo system for different $\Omega$ found by the time stepping method, where (a) \& (b) show the path of $C_z$ and $C_{zz}$ as ${\cal J}$ tends to zero and (c) shows the exponential convergence of ${\cal J}$ as the number of iterations.}
\label{solarMig5}
\end{figure}

Hence, for CE3, the stable fixed point can be accurately computed by the time stepping method. Fig. (\ref{solarMig5})
illustrates the fixed points of the CE3 equations of the dynamo system found by the time stepping method for $\Omega=21,25$ and $30$ with and without the stochastic force, where the initial condition is taken from the fixed point of $\Omega=20$. The convergence rate of ${\cal J}$ is exponential for all cases. The effect of noise is to increase the convergence rate of the timestepping method over that for the case with no noise.

\clearpage

\section{The self-excited disc dynamo}
\label{disc}

We now turn our attention to a low-order model that has very different dynamics. This model was proposed by \citet{cm1993} to study the chaotic nature of the self-excited dynamo system and replicate some of the dynamics of the geodynamo with irregular reversals on a long timescale. The dynamo system comprises two mutually induced Faraday (Bullard) disks that are powered by the thermally driven convecting flow. The governing equation set is defined as
\begin{eqnarray}
     \left(d_t + \alpha \eta \right) x  &=& \alpha \omega  y  z, \nonumber \\
     \left(d_t +  \eta  \right)  y &=& \omega  x  z, \nonumber \\
     \left(d_t + \kappa  \right) z &=& \kappa ( u - x  y ), \nonumber \\
     \left(d_t + 1 \right) u &=& \xi  z - v  z, \nonumber \\
     \left(d_t +  1 \right) v &=& u  z.
\label{disc_dynamo}
\end{eqnarray}
Here the dimensionless functions, $x$ and $y$ represents the magnetic field, $z$ is the velocity field and $u$ and $v$ represent the temperature field. The set of control parameters is given by $\alpha$, the magnetic inductance, $\omega$, a geometric factors, whilst $\kappa$ and $\eta$ measure the diffusivity of the temperature and magnetic field and $\xi$ measures the thermal driving. In the absence of magnetic field (for $x=y=0$), the low-order disc dynamo system is identical to the Lorenz63 system \citep{Lz_63}. The magnetic field can itself display extremely complicated dynamics --- including reversals. For this section, we omit the stochastic driving. 

As in the previous section, we shall use the cumulant equations to study the long-term evolution of the statistics of the disc dynamo and compare these with those obtained from DNS. We use the symbolic package to derive the cumulant expansions of the governing dynamics defined in Eqs. (\ref{disc_dynamo}) up to the third order. The first order cumulant equations read as follows,
\begin{eqnarray}
(d_t + \alpha \eta) C_x &=& \alpha \omega (C_y C_z + C_{yz}),  \nonumber\\
(d_t +  \eta) C_y &=& \omega(C_xC_z + C_{xz}), \nonumber\\
(d_t + \kappa) C_z &=& \kappa(C_u - C_xC_y - C_{xy}), \nonumber\\
(d_t + 1) C_u &=& -C_v C_z - C_{zv} + \xi C_z, \nonumber\\
(d_t + 1) C_v &=& C_u C_z + C_{zu}. 
\end{eqnarray}
The second order expansion consists of fifteen equations and are listed in Appendix \ref{appdisc}. The cumulant equations may then be further truncated according to CE2/2.5/3 truncation rules, respectively.

We focus our studies on the chaotic state of the dynamo system, i.e., $\xi > 40$, where the other control parameters remain the same as those in \citet{cm1993}, i.e., $\omega=\kappa=1$ and $\eta=4$. We vary the magnetic inductance, $\alpha$, to control the dynamo process. For $\alpha>1$, the magnetic field oscillates chaotically but with an unique polarity ($y>0$) for all time; reducing the inductance to $0<\alpha<1$, the magnetic field reverses directions aperiodically.

\begin{figure}
\centering
\subfigure[$x-y-z$]
{
	\includegraphics[width=0.22\hsize]{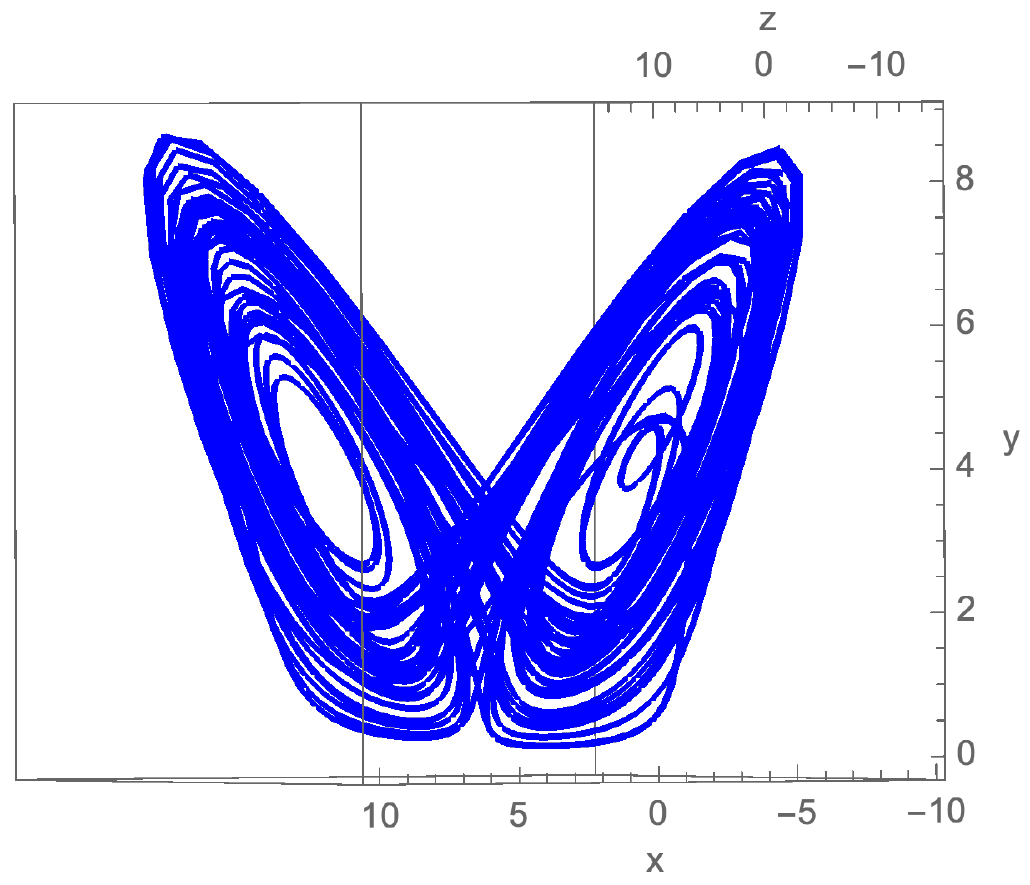}
}
\subfigure[$z-u-v$]
{
	\includegraphics[width=0.22\hsize]{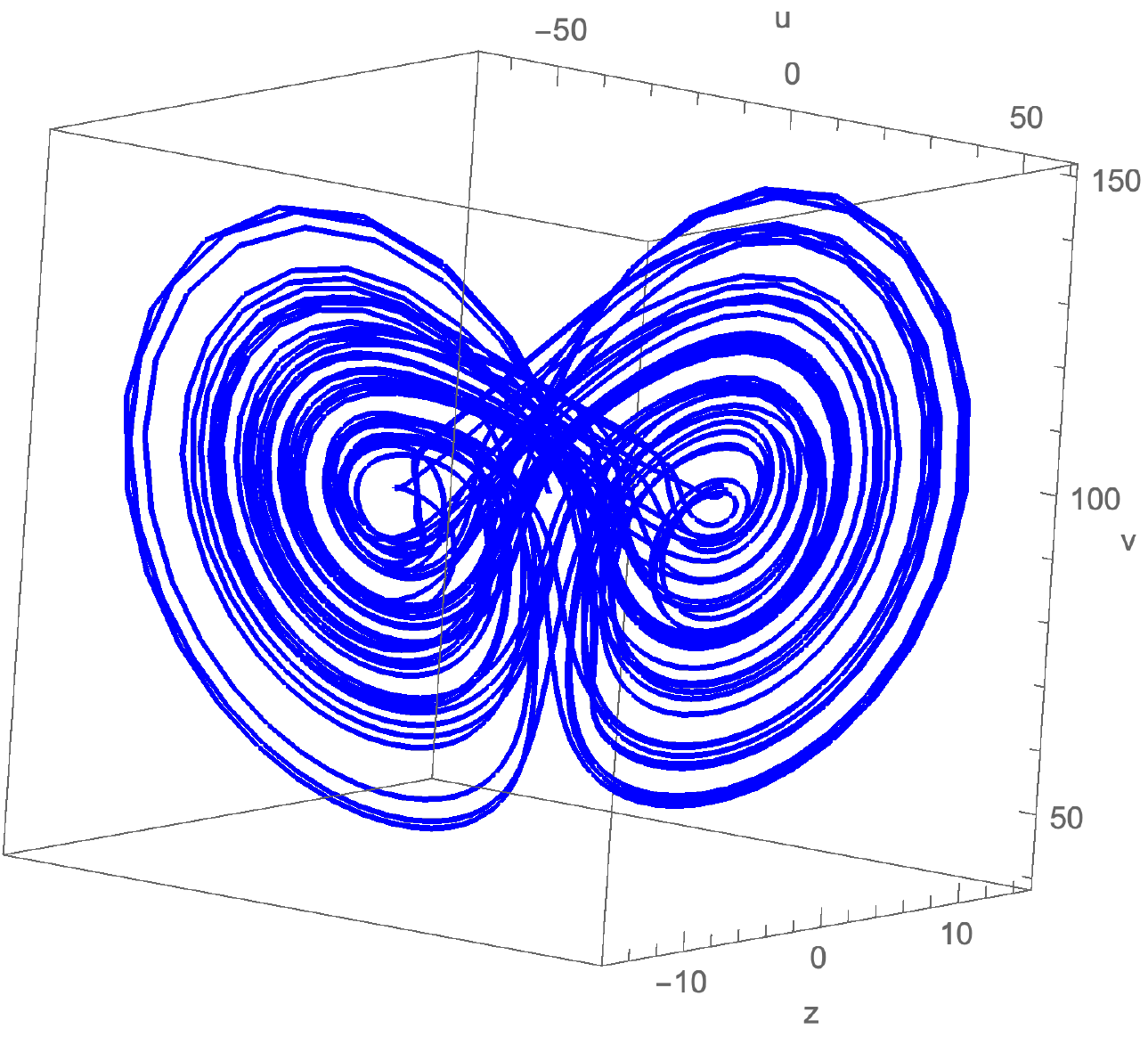}
}
\subfigure[magnetic induction]
{
	\includegraphics[width=0.22\hsize]{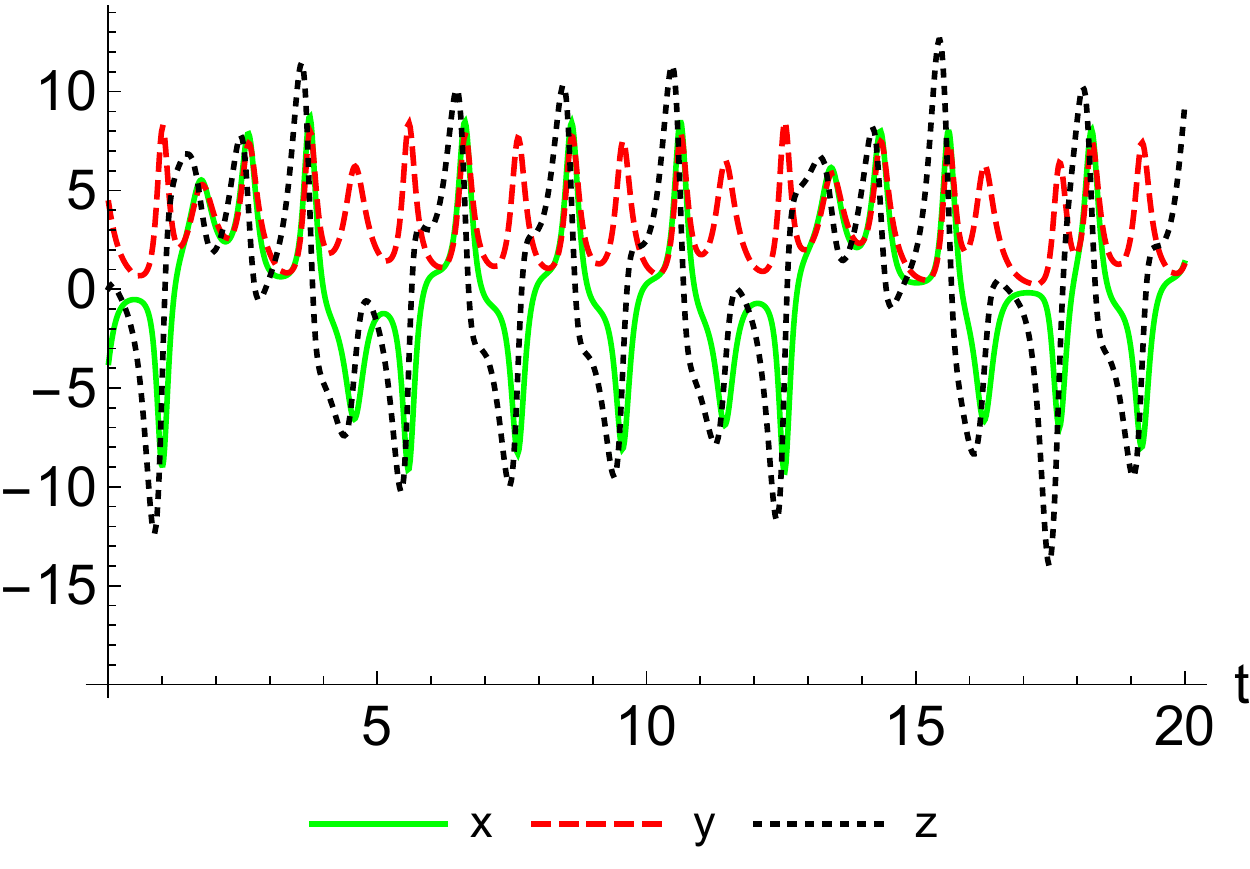}
}
\subfigure[thermal convection]
{
	\includegraphics[width=0.22\hsize]{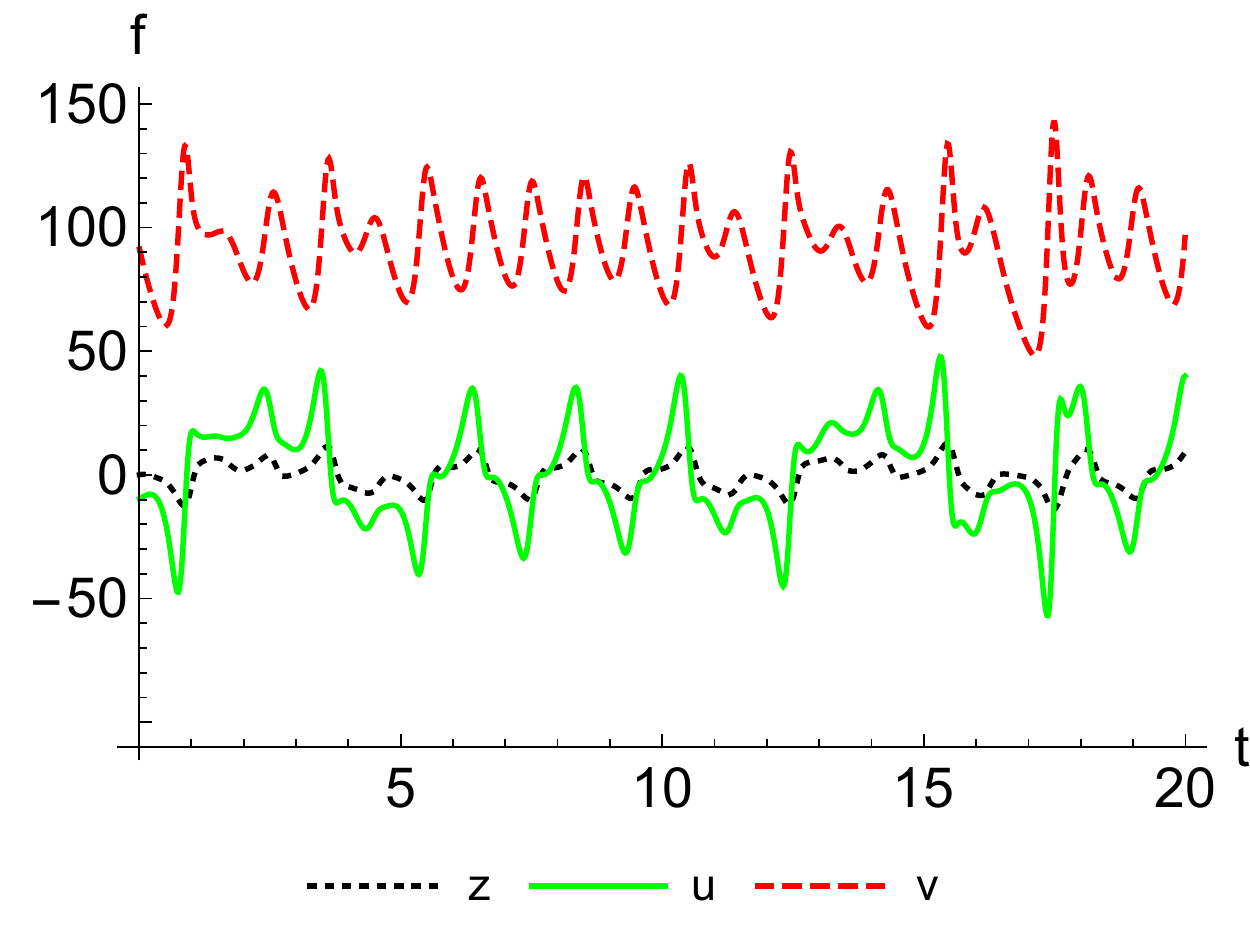}
}\\
\subfigure[$P(x)$]
{
	\includegraphics[width=0.18\hsize]{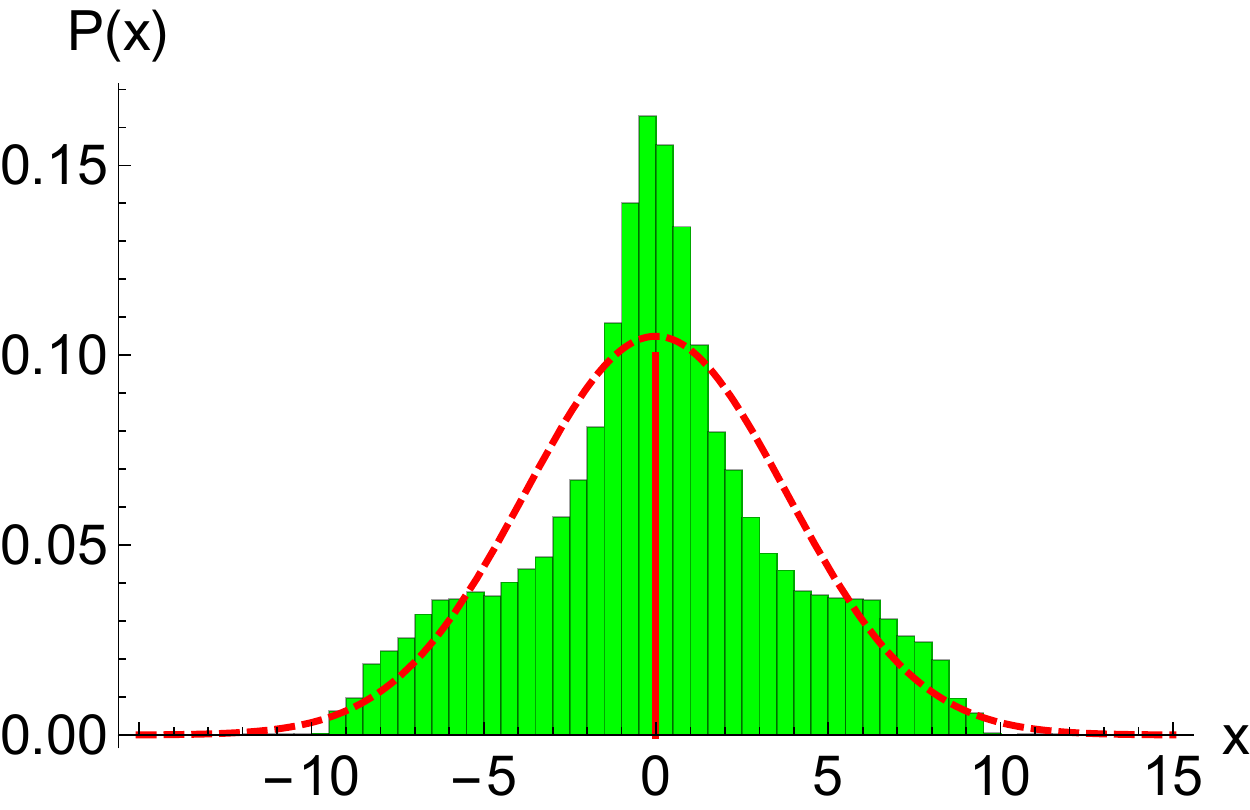}
}
\subfigure[$P(y)$]
{
	\includegraphics[width=0.18\hsize]{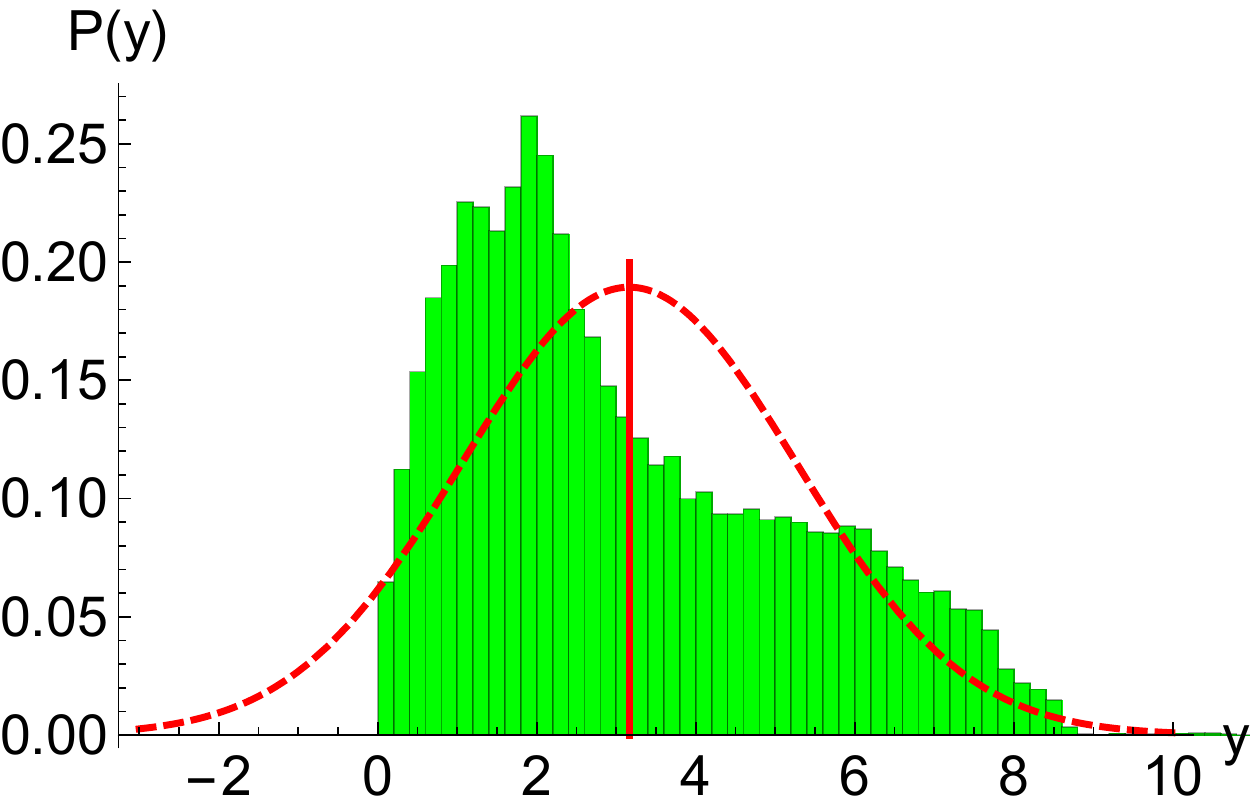}
}
\subfigure[$P(z)$]
{
	\includegraphics[width=0.18\hsize]{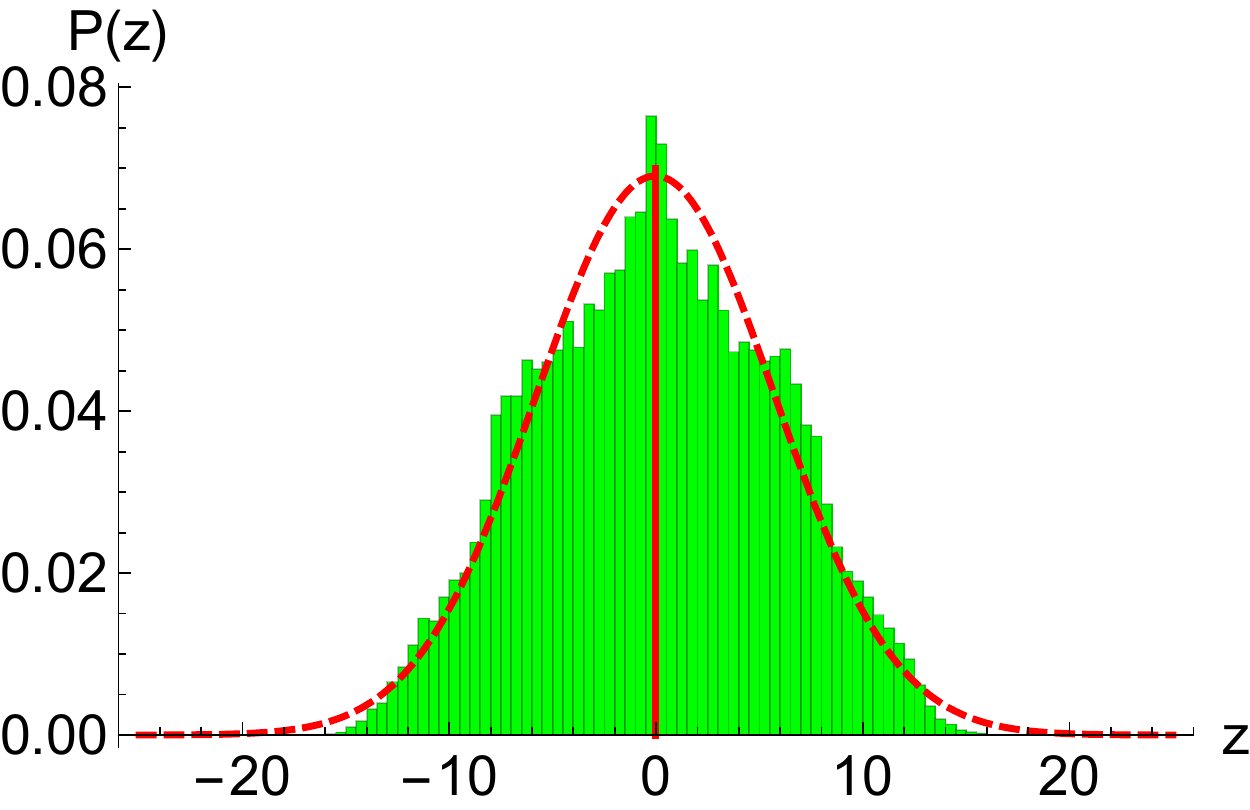}
}
\subfigure[$P(u)$]
{
	\includegraphics[width=0.18\hsize]{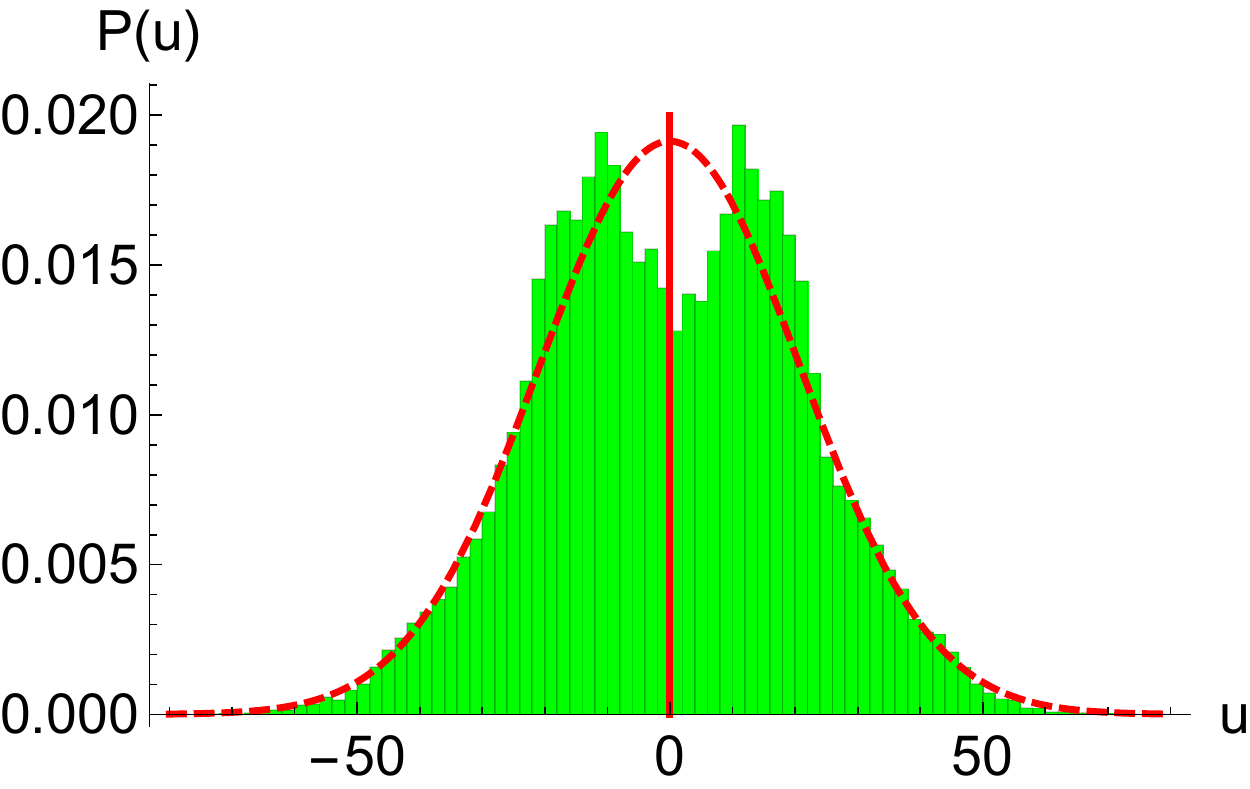}
}
\subfigure[$P(v)$]
{
	\includegraphics[width=0.18\hsize]{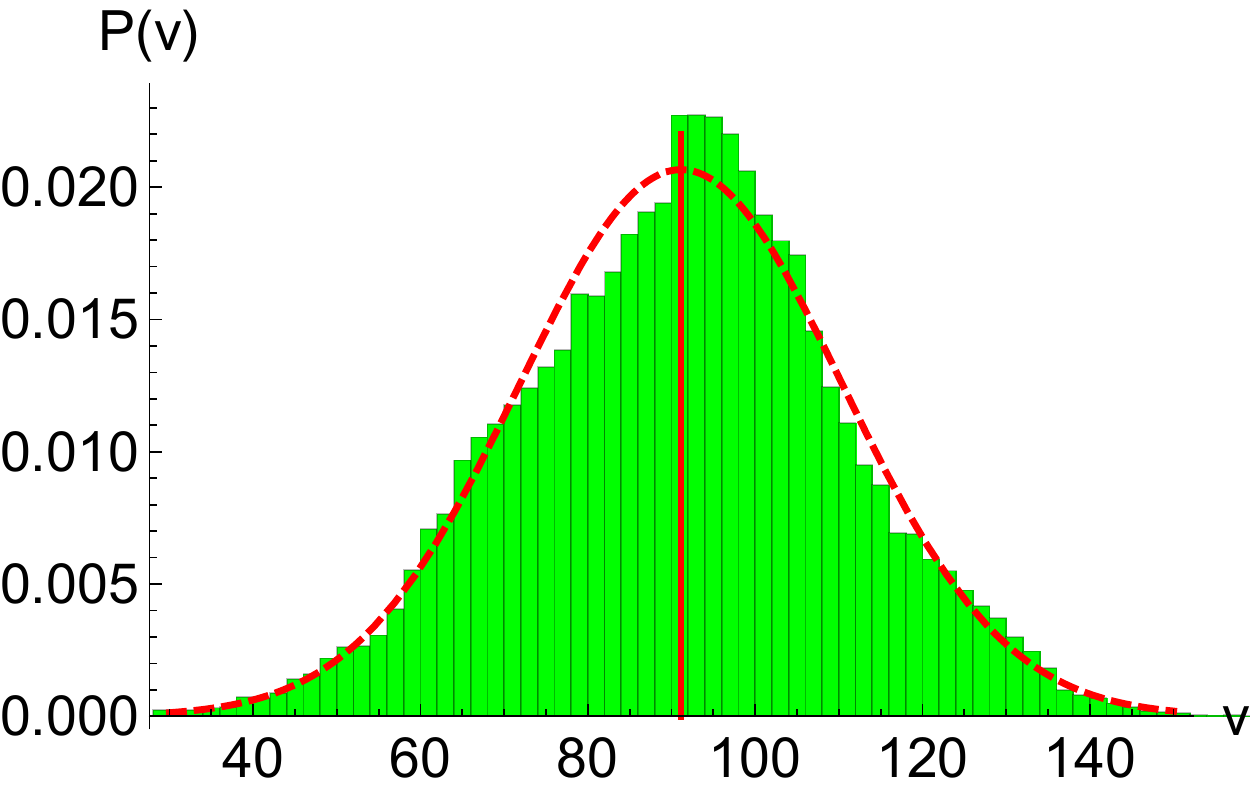}
}
\caption{Illustration  of the self-excited disc dynamo system in the chaotic state with the unique magnetic polarization for $\alpha=1.5$ and $\xi=100$. The projections of the trajectory in $x-y-z$ and $z-u-v$ subspace are shown in (a) and (b), the time series of the magnetic induction and the thermal convection are in (c)-(d) and the PDFs obtained via the ensemble average of DNS of the dynamo system are plotted in (e)-(i), where the red curves stand for the Gaussian distribution with the same mean and variance as those obtained in DNS in green histograms.}
\label{discpt1}
\end{figure}

Illustrated in Fig. (\ref{discpt1})
is a typical solution of the self-excited disc dynamo system in the chaotic state with the unique magnetic polarity. Here  $\alpha=1.5$ and the thermal force is chosen to be $\xi=100$. The trajectory of the thermal convection in the subspace $z-u-v$ is almost identical to a typical solution of the Lorenz63 system in the chaotic state (Fig. \ref{discpt1}b) and the magnetic field follows a similar patterns (Fig. \ref{discpt1}a). The time series shown in Fig. (\ref{discpt1}c \& d) appear to be similar to Lorenz63 in the chaotic state 
\citep{li2_2021}; The system consists of two sets of strange attractors.  

\begin{figure}
\centering
\subfigure[$x-y-z$]
{
	\includegraphics[width=0.22\hsize]{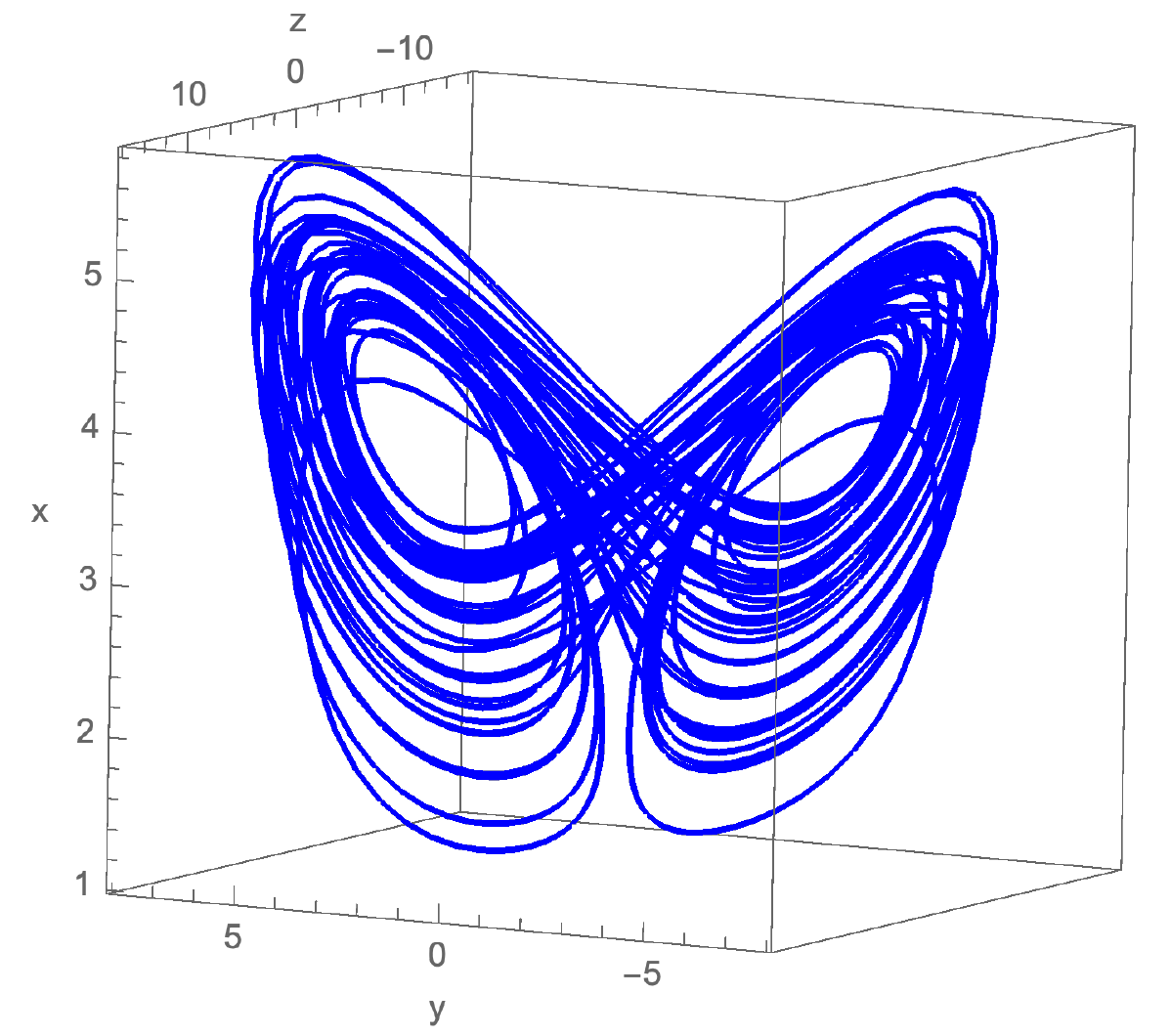}
}
\subfigure[$z-u-v$]
{
	\includegraphics[width=0.22\hsize]{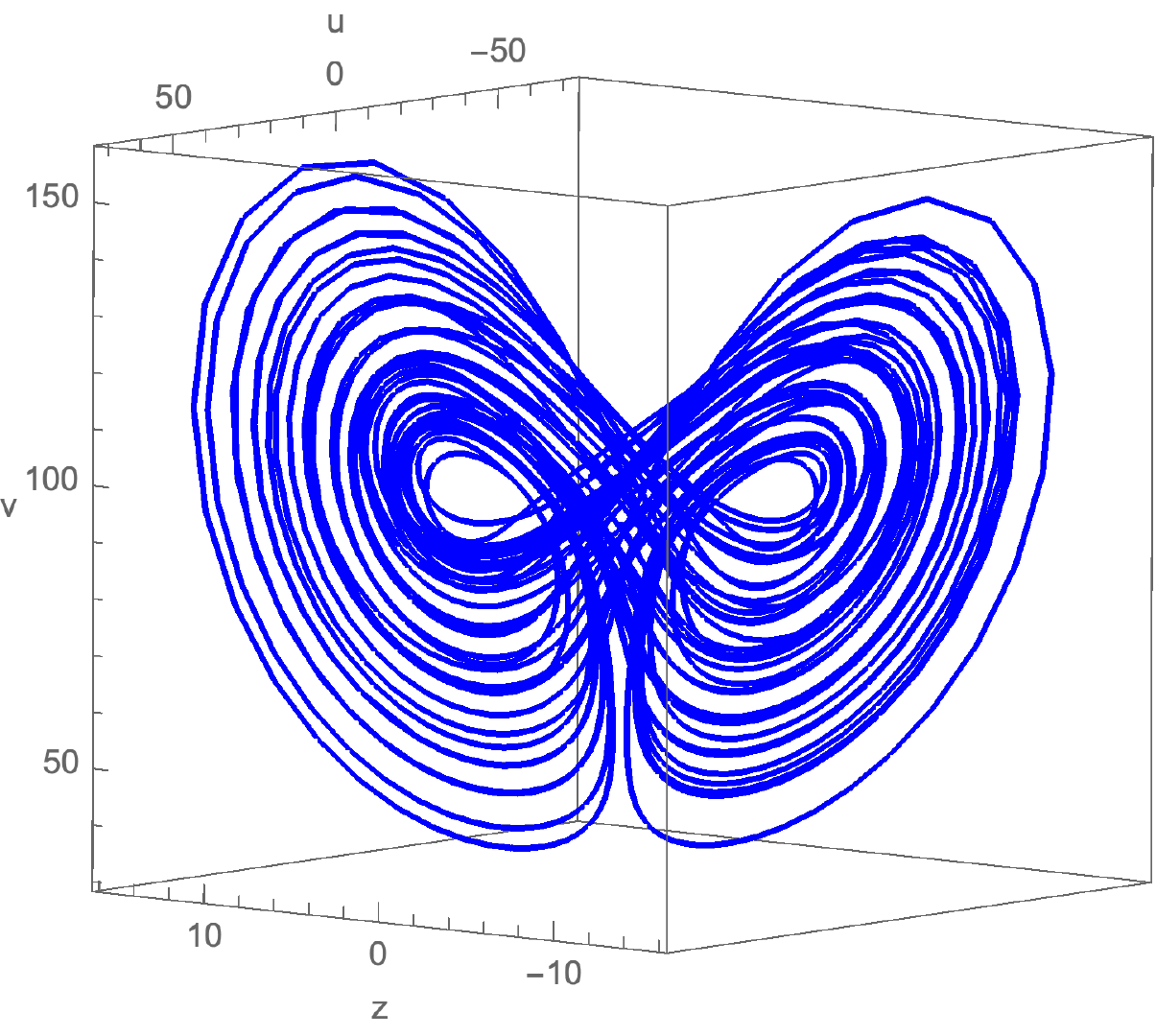}
}
\subfigure[magnetic induction]
{
	\includegraphics[width=0.22\hsize]{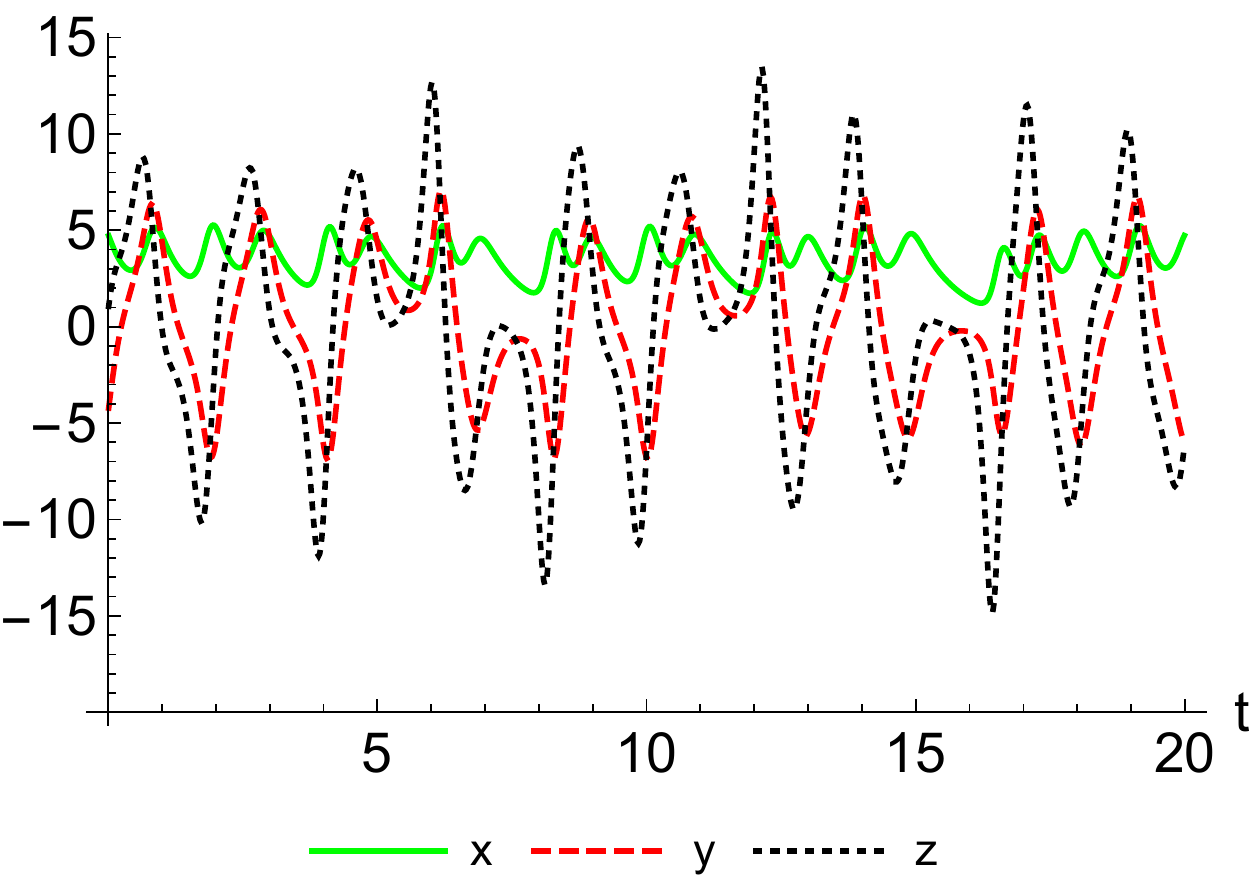}
}
\subfigure[thermal convection]
{
	\includegraphics[width=0.22\hsize]{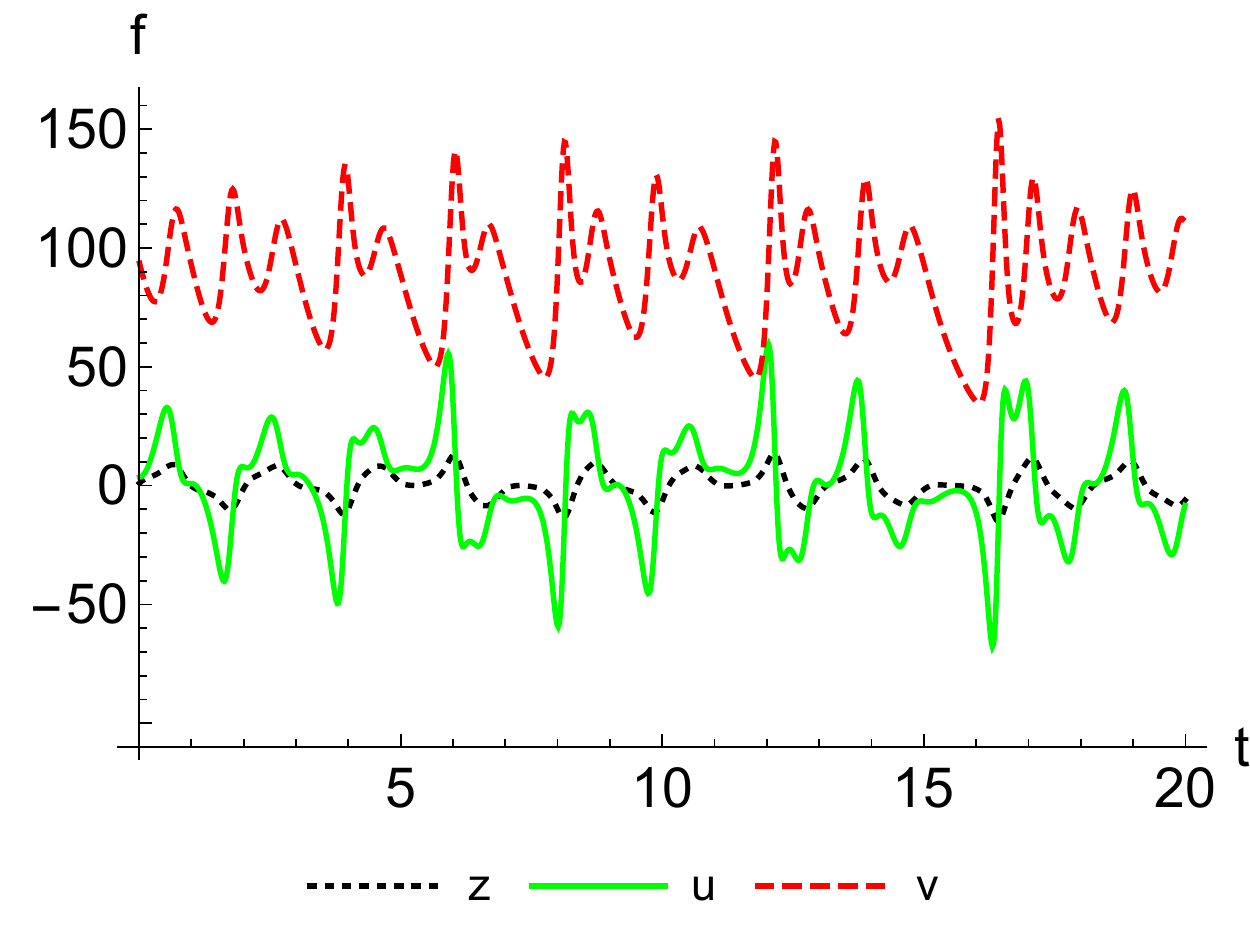}
}\\
\subfigure[$P(x)$]
{
	\includegraphics[width=0.18\hsize]{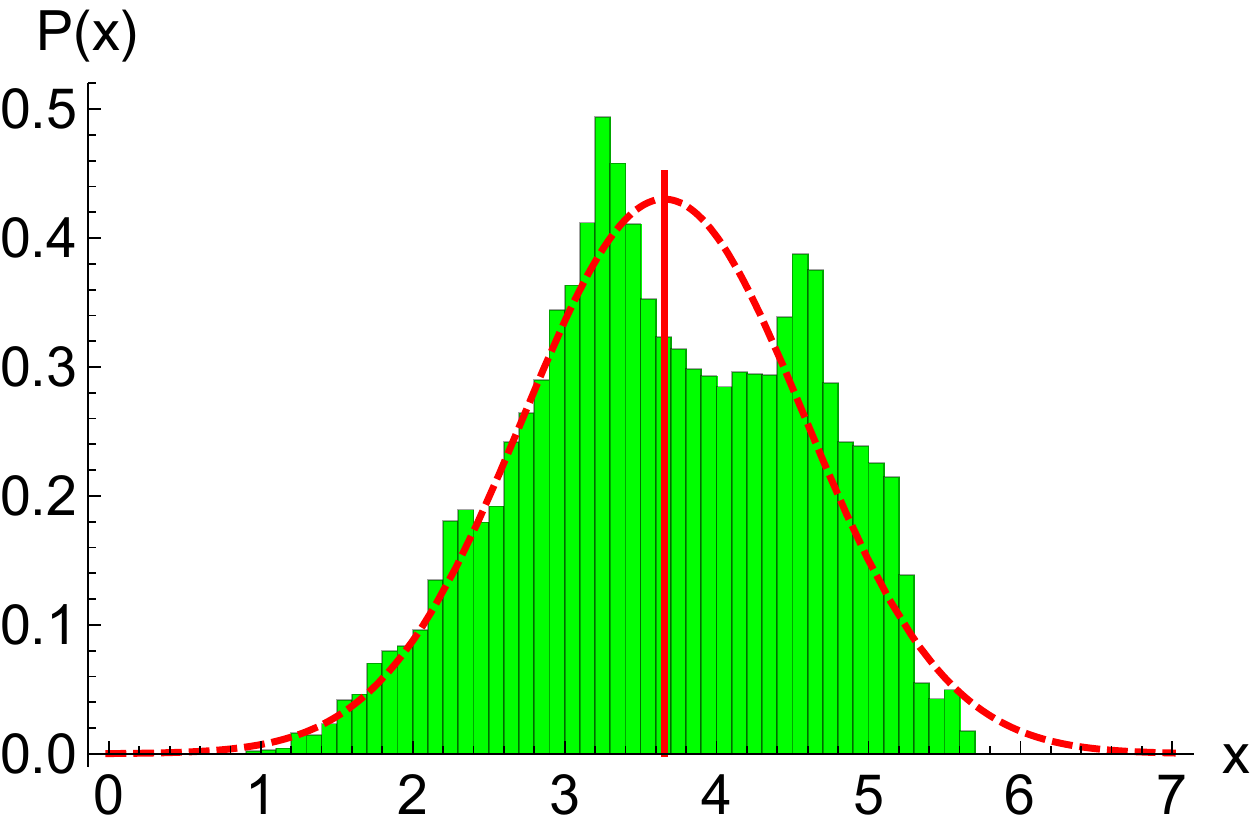}
}
\subfigure[$P(y)$]
{
	\includegraphics[width=0.18\hsize]{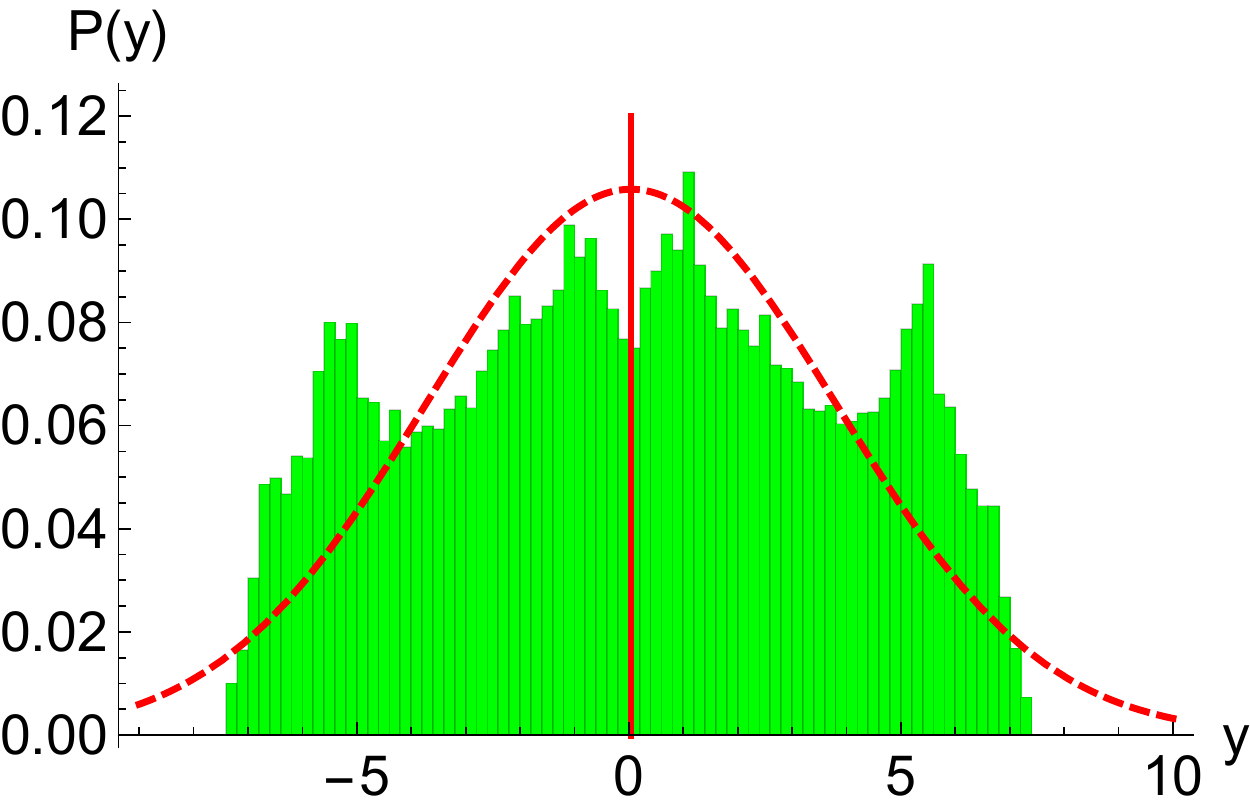}
}
\subfigure[$P(z)$]
{
	\includegraphics[width=0.18\hsize]{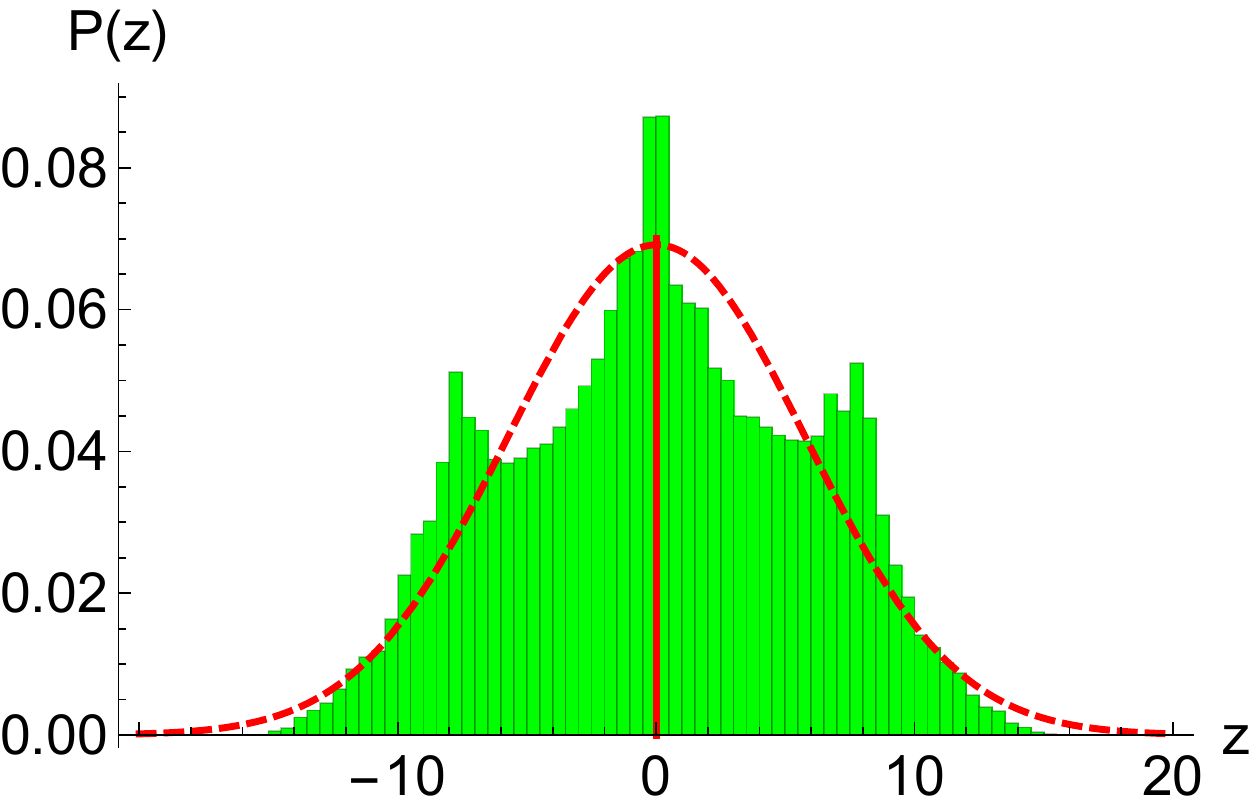}
}
\subfigure[$P(u)$]
{
	\includegraphics[width=0.18\hsize]{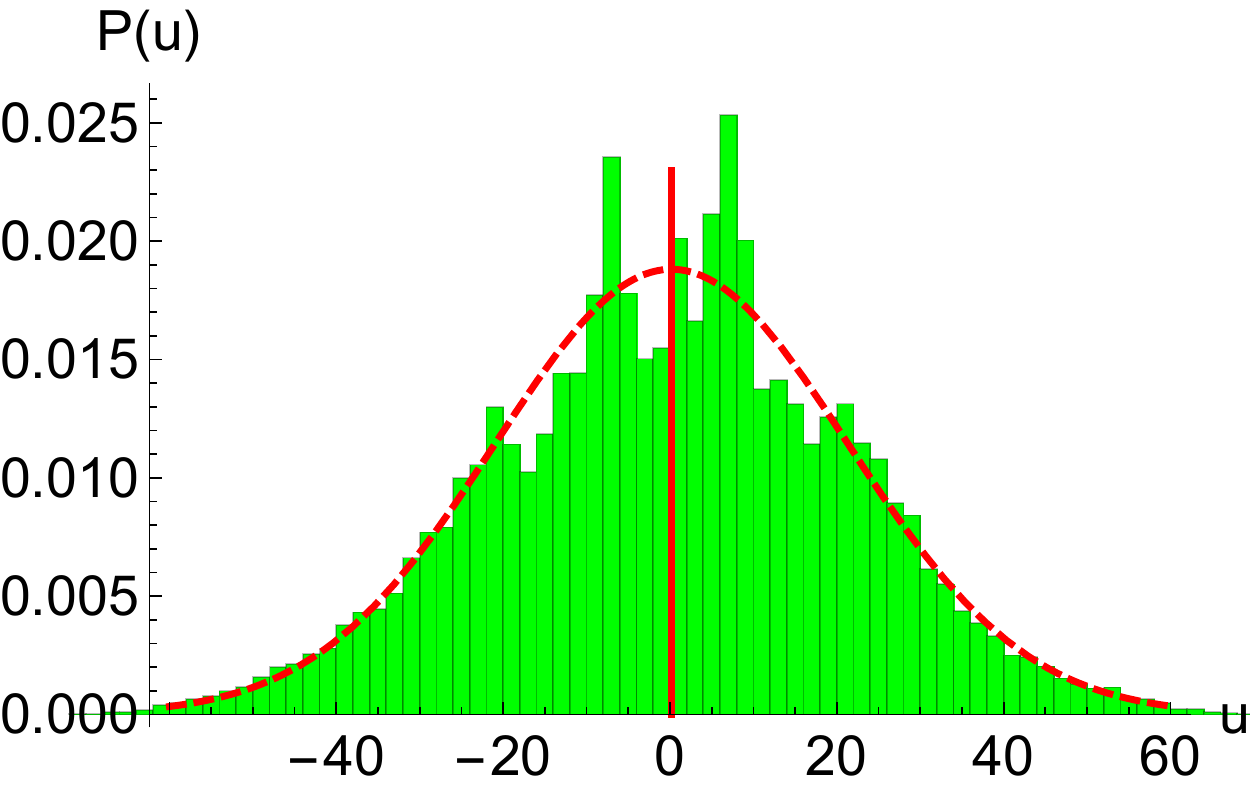}
}
\subfigure[$P(v)$]
{
	\includegraphics[width=0.18\hsize]{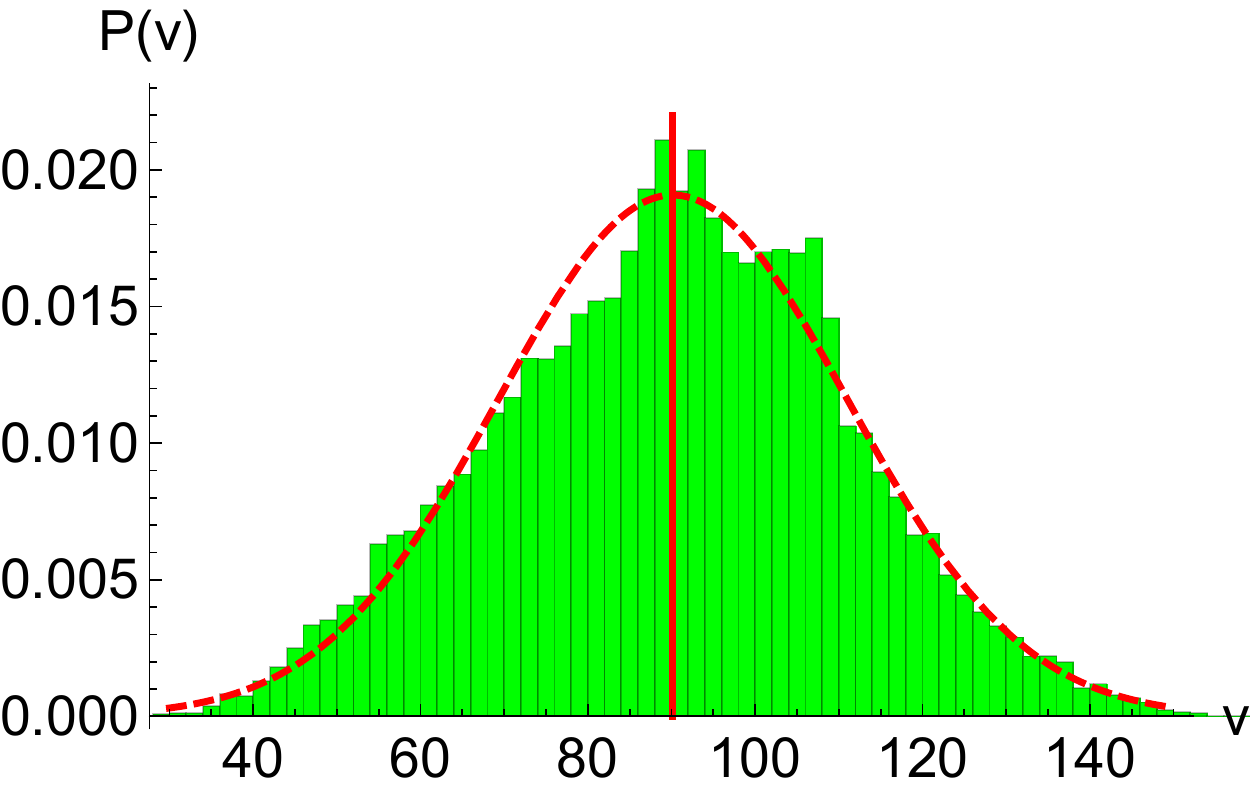}
}
\caption{Illustration of the self-excited disc dynamo system in the chaotic state with the magnetic reversal for $\alpha=0.3$ and $\xi=100$. The projections of the trajectory in $x-y-z$ and $z-u-v$ subspace are shown in (a) and (b), the time series of the magnetic induction and thermal convection are in (c)-(d) and the PDFs obtained via the ensemble average of DNS of the dynamo system are plotted in (e)-(i), where the red curves stand for the Gaussian distribution with the same mean and variance as those obtained in DNS in green histograms.}
\label{discpt2}
\end{figure}

The PDFs of the magnetic field, $y$, and the temperature field, $v$ are bimodal. In Fig. (\ref{discpt1}e-i), the red curves represent the Gaussian distribution with the same mean and variance as those obtained in DNS in green histograms. The dynamo system behaves similarly for $0<\alpha<1$, except for the magnetic field, $y$, that changes sign irregularly in time, see Fig, (\ref{discpt2}) for details.


\subsection{Direct statistical simulation of the disc dynamo in the chaotic state}

We timestep the CE2.5/3 equations for the disc dynamo system to obtain the stable time-invariant solutions for different thermal forcings, $\xi$, and magnetic inductance, $\alpha$. For all the test cases, we find that both the CE2.5 and CE3 approximations converge in time. For both approximations, the eddy damping parameter, $\tau_d$, must be applied to stabilize the numerical computation, where the most accurate solution occurs for $\tau_d$ in the range of ${\cal O}(10^{-2})$ to ${\cal O}(10^{-1})$ (see Table~\ref{disctb1}). In the absence of noise, we find that the CE2 approximation is instantaneously unstable as we forward evolve CE2 equations in time for all $\tau_d$ (as for the solar dynamo case).

In this system the dynamics is such  that the induced magnetic field is highly sensitive to the value of the second order cumulants, $C_{xz}$ and $C_{yz}$. Owing to the truncation of the cumulant hierarchy for the non-Gaussian distribution, it is therefore the mean  of the magnetic field that is sensitive to  $C_{xz}$ and $C_{yz}$ that is least accurately approximated by CE2.5/3; $C_y$ is over estimated by $15$ to $20$ percent for $\xi=100$ and $\alpha=1.5$. However, increase of the thermal force, $\xi$, leads to a significant reduction of this error. For $\xi=150$ and $\alpha=1.5$, all components of the mean trajectory of the magnetic, velocity and temperature field are accurately approximated with the maximum error about $5\%$. Listed in Table (\ref{disctb1})  
\begin{table}[htp]
\centering
\begin{tabular}{l|c|c|c|c|c|c|c|c}
    	    & $\alpha$ &$\xi$ & $\tau_d^{-1}$ & $\|C_{x}\|$ & $\|C_{y}\|$ & $C_{z}$ & $C_{u}$ & $C_{v}$ \\\hline
DNS & $1.5$    &$100$ &               & $0$     & $3.15$     & $0$     & $0$     & $91.11$    \\\hline
CE2.5 &          &  &  $20$    & $0$     & $3.88$     & $0$     & $0$     & $88.86$    \\\hline
CE3 &             & &   $20$    & $0$     & $3.65$     & $0$     & $0$     & $90.34$    \\\hline\hline
DNS & $1.5$    &$150$ &               & $0$     & $4.34$     & $0$     & $0$     & $138.37$    \\\hline
CE2.5&    & &  $30$   & $0$     & $5.13$     & $0$     & $0$     & $135.96$    \\\hline
CE3   &   & &    $20$   & $0$     & $4.56$     & $0$     & $0$     & $136.99$    \\\hline\hline
DNS & $0.3$ & $50$&              & $1.61$     & $0$     & $0$     & $0$     & $36.56$    \\\hline
CE2.5 &         & & $20$       & $2.62$     & $0$     & $0$     & $0$     & $44.69$    \\\hline
CE3    &         & & $20$       & $2.51$     & $0$     & $0$     & $0$     & $44.74$    \\\hline\hline
DNS & $0.3$ & $100$&              & $3.62$     & $0$     & $0$     & $0$     & $89.86$    \\\hline
CE2.5 &         & & $20$       & $4.15$     & $0$     & $0$     & $0$     & $89.44$    \\\hline
CE3 &            &  & $20$       & $3.99$     & $0$     & $0$     & $0$     & $91.04$    \\\hline
\end{tabular}
\caption{The first order cumulants of the self-excited disc dynamo system for different thermal forcing, $\xi$, and magnetic inductance, $\alpha$, where the magnetic field of the dynamo has an unique polarization for $\alpha>1$ and reverses randomly for $\alpha<1$.}
\label{disctb1}
\end{table}
are the first order cumulants of the disc dynamo for different thermal forcing, $\xi$ and magnetic inductance, $\alpha$.
Note that because of the symmetry in the magnetic field ---  under the transformation $x\rightarrow -x$ and $y\rightarrow -y$, the solution of the velocity and temperature field remains invariant ---  it is convenient to report the absolute value, $\|C_x\|$ and $\|C_y\|$ in Table (\ref{disctb1}).

\begin{figure}
\centering
\subfigure[$\alpha=1.5$ and $\tau_d^{-1}=20$]
{
	\includegraphics[width=0.4\hsize]{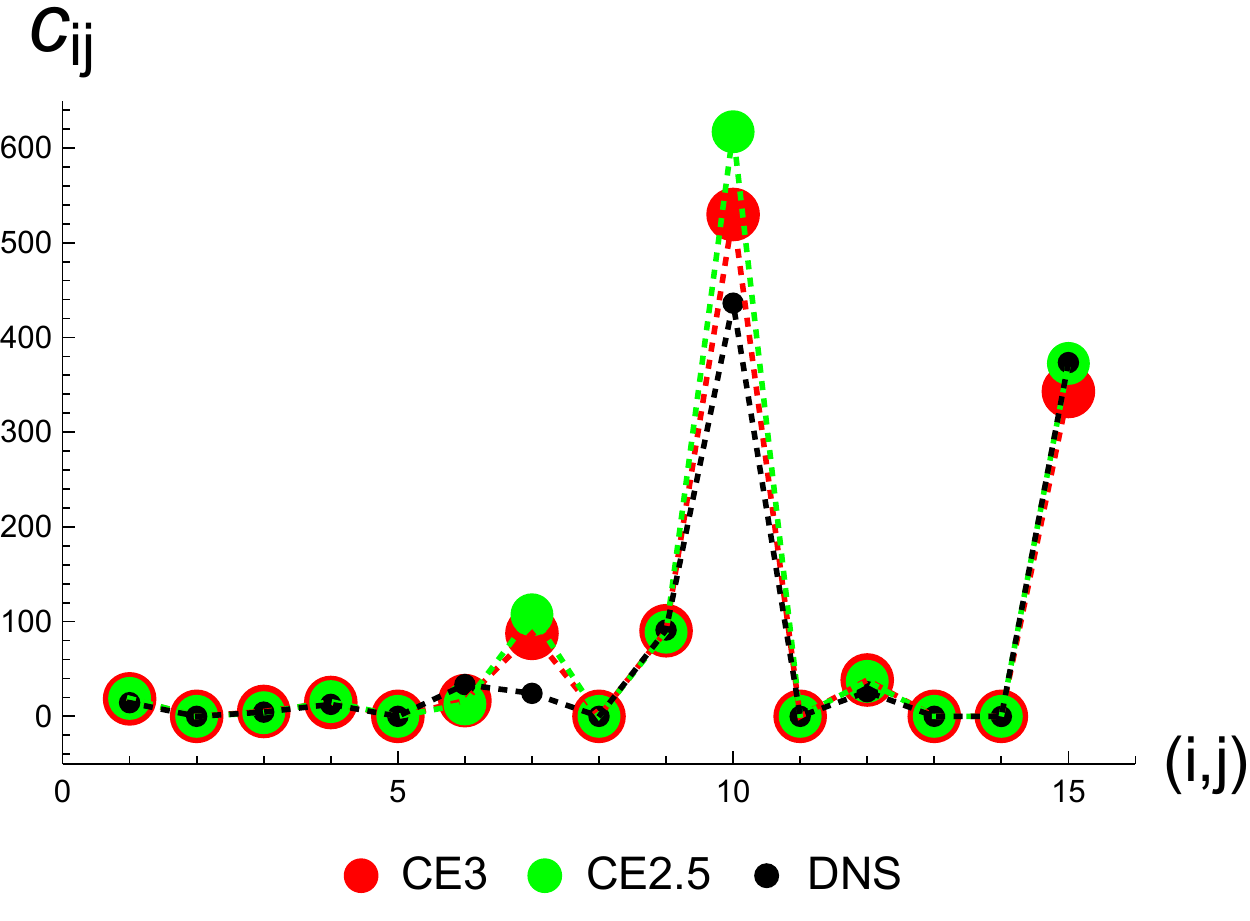}
}
\subfigure[$\alpha=0.3$ and $\tau_d^{-1}=20$]
{
	\includegraphics[width=0.4\hsize]{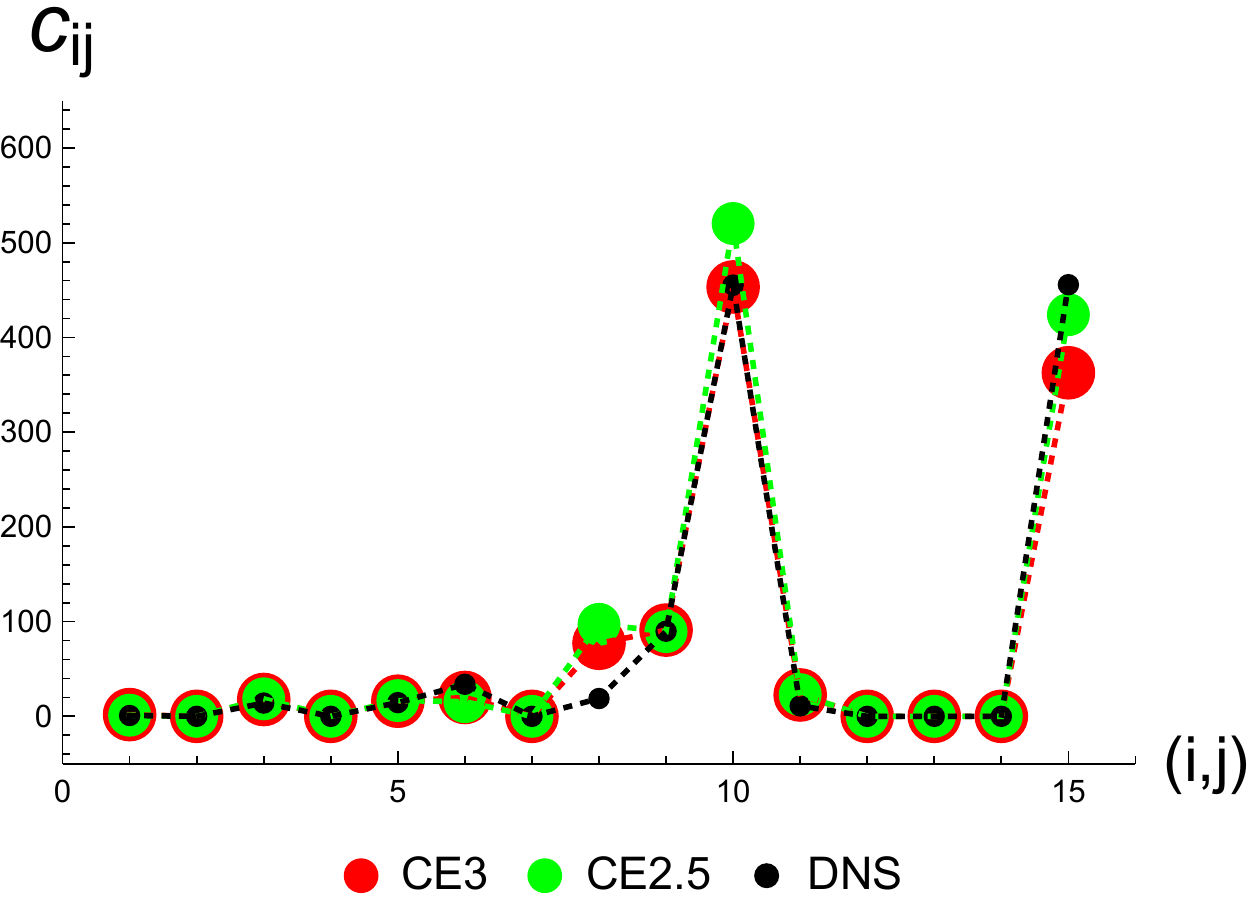}
}
\caption{Second order cumulants obtained in DNS and CE2.5/CE3 without the magnetic reversal in (a) for $\alpha=1.5$ and with the reversal for $\alpha=0.3$ in (b) for $\xi=100$ with the eddy dumping parameter, $\tau_d=1/20$, where the horizontal labels are the collective index of $(x_i, x_j)$ for $i\le j$.}
\label{disdynamoC2}
\end{figure}

The second order cumulants are plotted in Fig. (\ref{disdynamoC2}),
where the index $(i,j)$ are the collective index of $(x_i, x_j)$ for $i \le j$, e.g., the $1$st point in Fig. (\ref{disdynamoC2}a \& b) represents the term, $C_{xx}$, and the $10$th element is for $C_{uu}$. CE2.5 and CE3 provides good estimates of the second cumulants, though we note that the two variances, $C_{uu}$ and $C_{vv}$, that are approximately $5$ to $10$ percent overestimated, are the most inaccurate terms among all second order terms. 

\subsection{The fixed points of the disc dynamo in the chaotic state}

\begin{figure}[htp]
\centering
\subfigure[${\cal J} \sim C_v$ in CE2.5]
{
	\includegraphics[width=0.3\hsize]{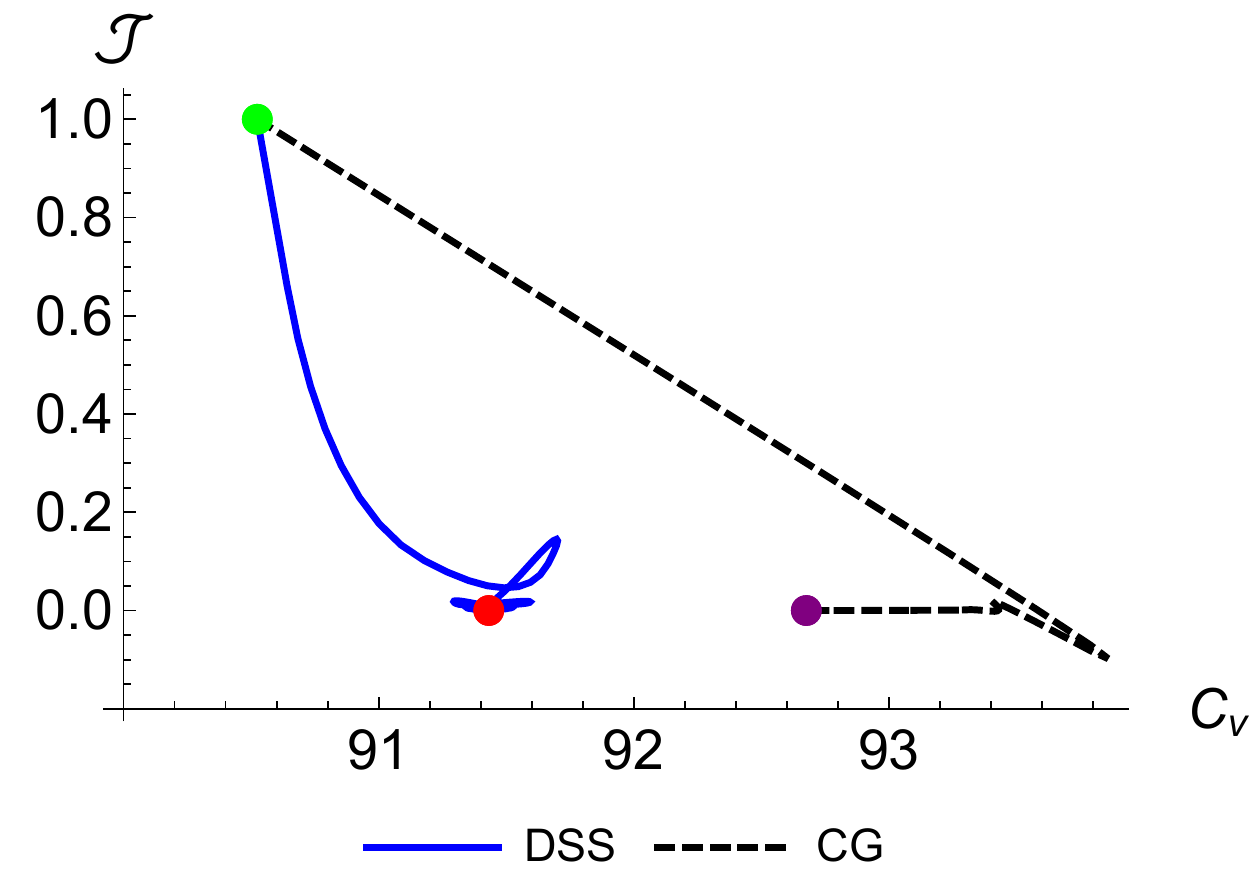}
}
\subfigure[${\cal J} \sim C_{vv}$ in CE2.5]
{
	\includegraphics[width=0.3\hsize]{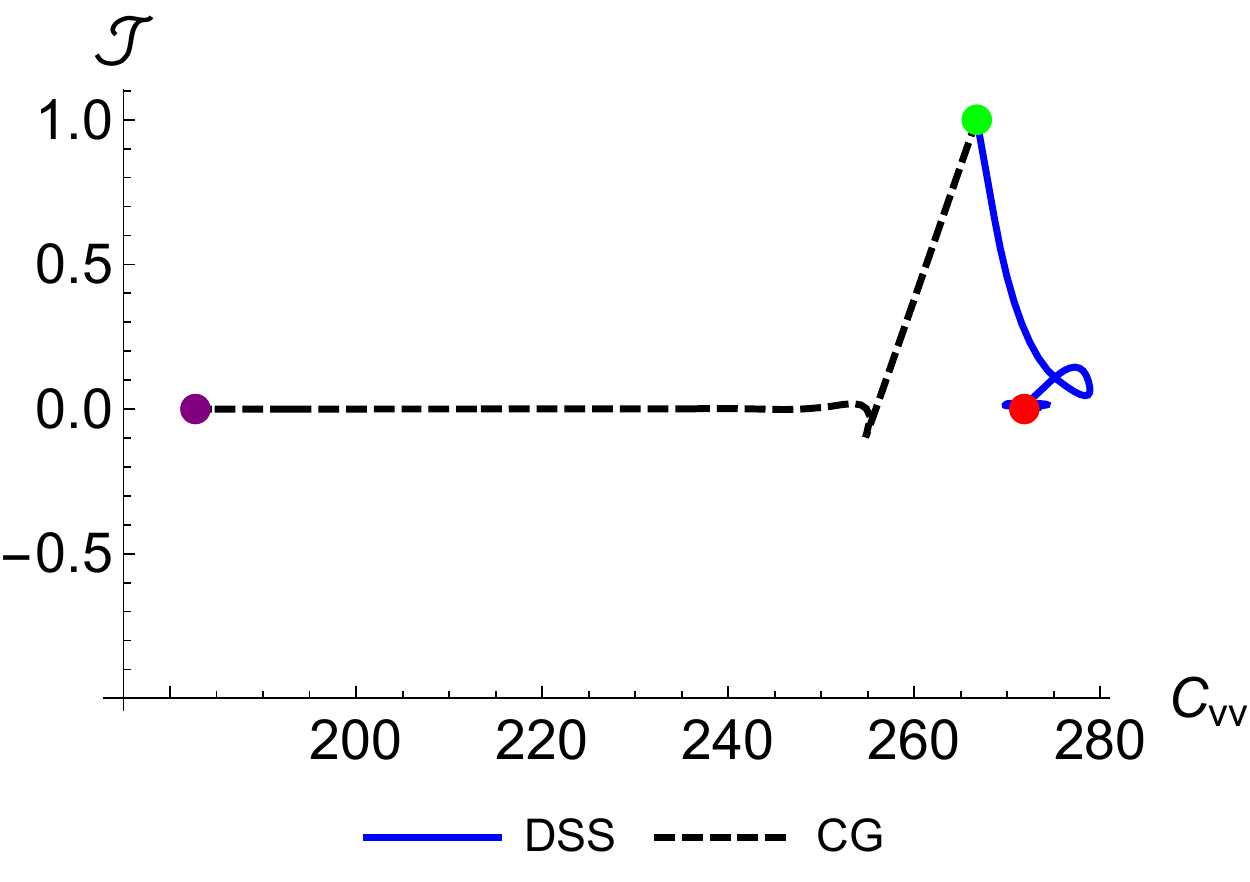}
}
\subfigure[${\cal J} \sim n$ for in CE2.5]
{
	\includegraphics[width=0.3\hsize]{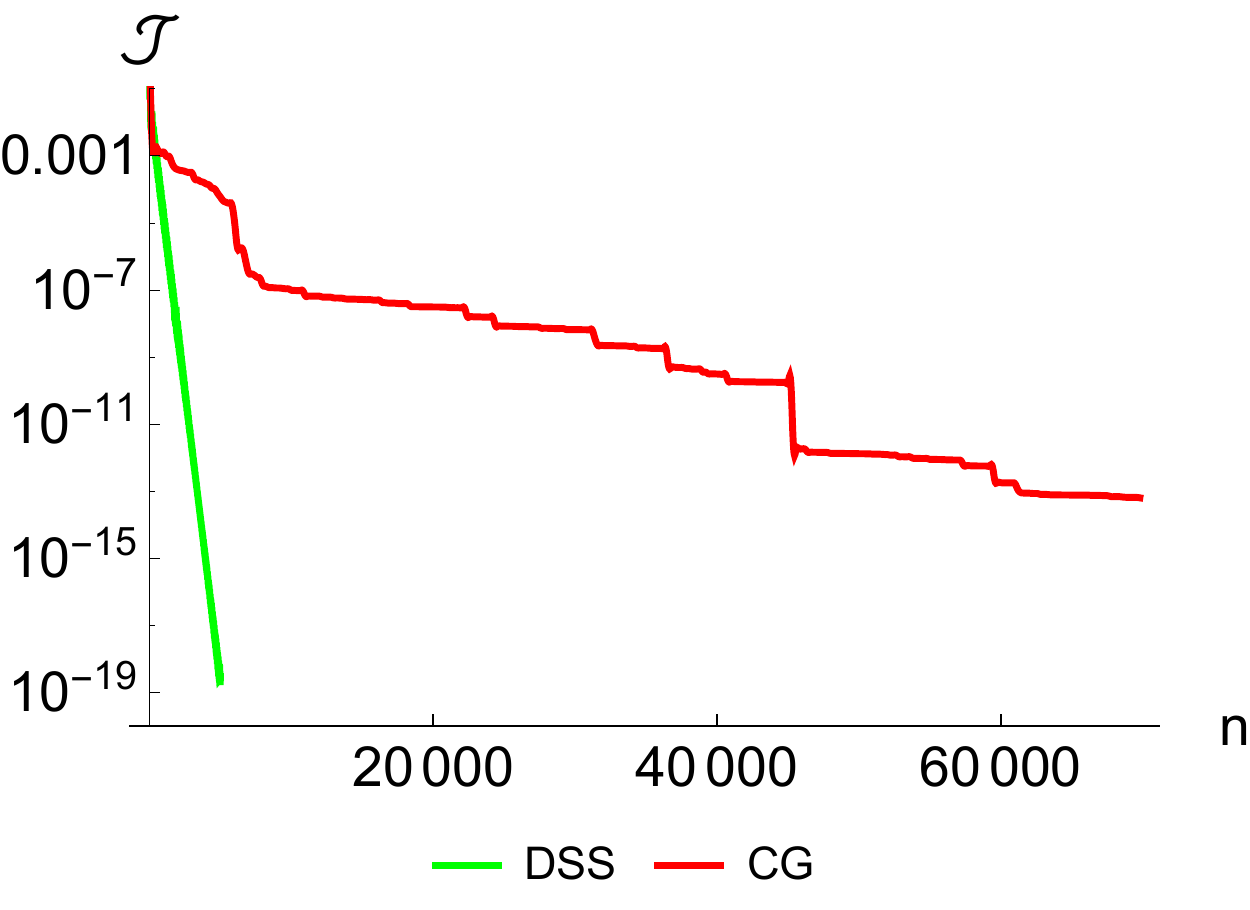}
}
\caption{Migration of $C_v$ and $C_{vv}$ from $\xi=100$ to $101$ for $\alpha=1.5$ via {\it CG} and time stepping method for CE2.5 approximation is shown in Figs. (a) \& (b) and the convergence of ${\cal J}$ as the number of iterations/time steps is in (c), where the green and red dots in (a) \& (b) give the initial guess/conditions and the terminal solutions, respectively.}
\label{disMig}
\end{figure}

We use the gradient based method to compute the fixed points of the CE2.5 approximation of the disc dynamo system. However, rather than finding the stable fixed points located by timestepping, we find that the gradient based optimization always converges to another (unstable) fixed point of the CE2.5 solution. Shown in Fig. (\ref{disMig})
is a typical solution of the {\it CG} method as compared with the solution via the time stepping method for the CE2.5 approximation of the disc dynamo. Here the thermal forcing is $\xi=100$ and $\alpha=1.5$. In this test case, we attempt to compute the stable fixed point of the dynamo system for $\xi=101$ and $\alpha=1.5$. The initial condition/guess are taken from the stable fixed point of the system for $\xi=100$ and $\alpha=1.5$ that are shown in green dots in Fig. (\ref{disMig}a \& b).  The solution for $\xi=100$ is then calculated two ways; the stable fixed points computed by the time stepping method is shown as a red dot whilst the solution found by the {\it CG} method is shown as a purple dot. It is clear that the {\it CG} method is locating an unstable fixed point. Moreover,  the convergence rate of {\it CG} is much slower than the time stepping method. For CE3 approximations, we always observe that after a few iterations, the {\it CG} method converges to a local minima of ${\cal J}$ and fails to optimize the CE3 equations.

\begin{figure}[htp]
\centering
\subfigure[${\cal J} \sim C_y$ for $\alpha=1.5$]
{
	\includegraphics[width=0.3\hsize]{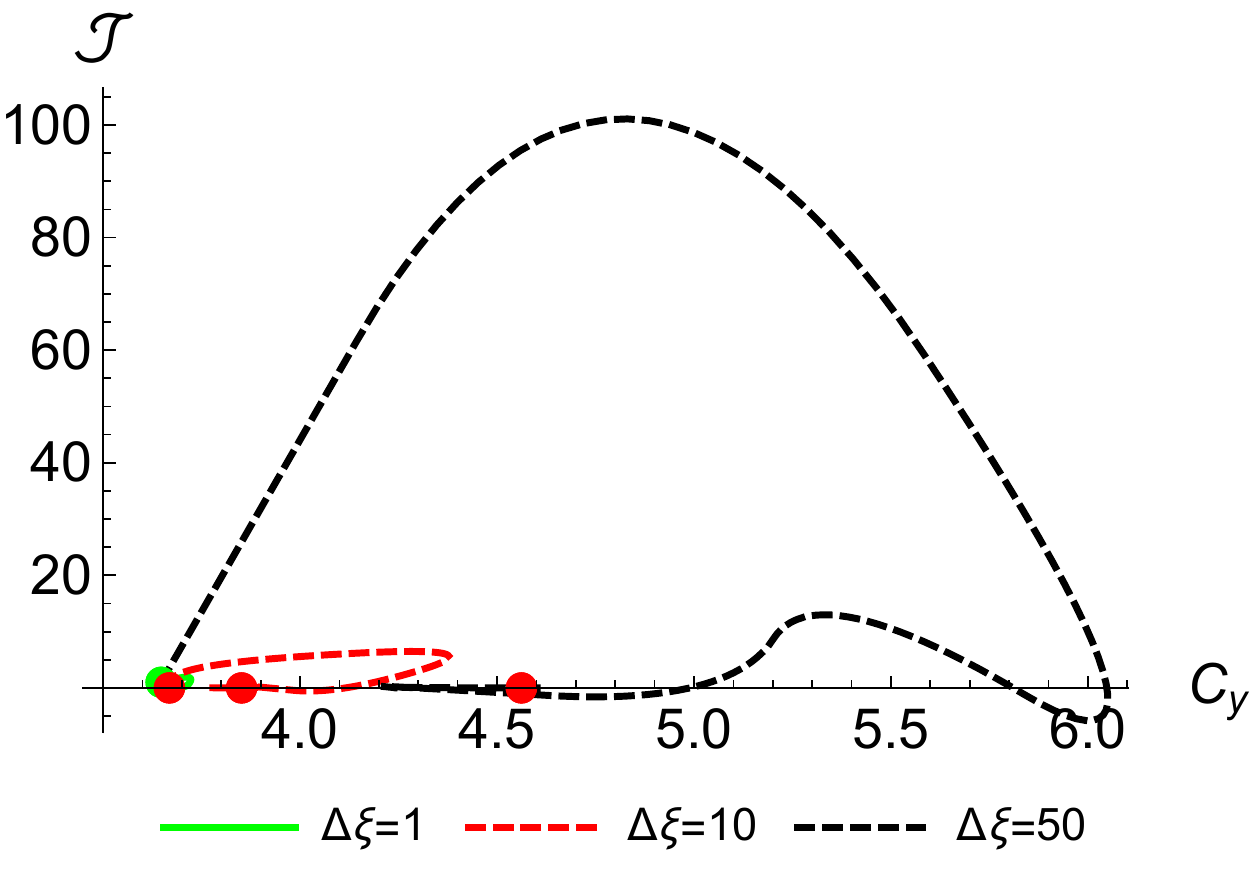}
}
\subfigure[${\cal J} \sim C_{vv}$ for $\alpha=1.5$]
{
	\includegraphics[width=0.3\hsize]{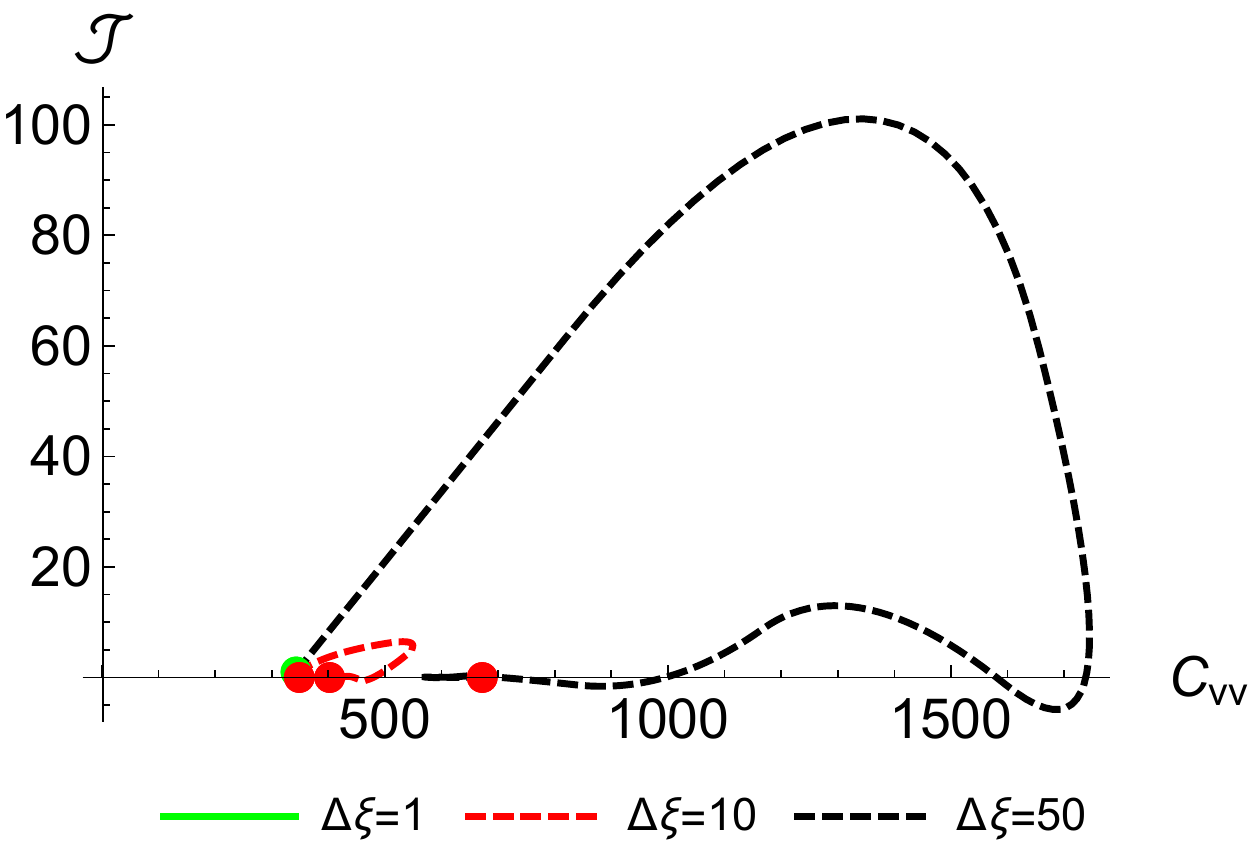}
}
\subfigure[${\cal J} \sim n$ for $\alpha=1.5$]
{
	\includegraphics[width=0.3\hsize]{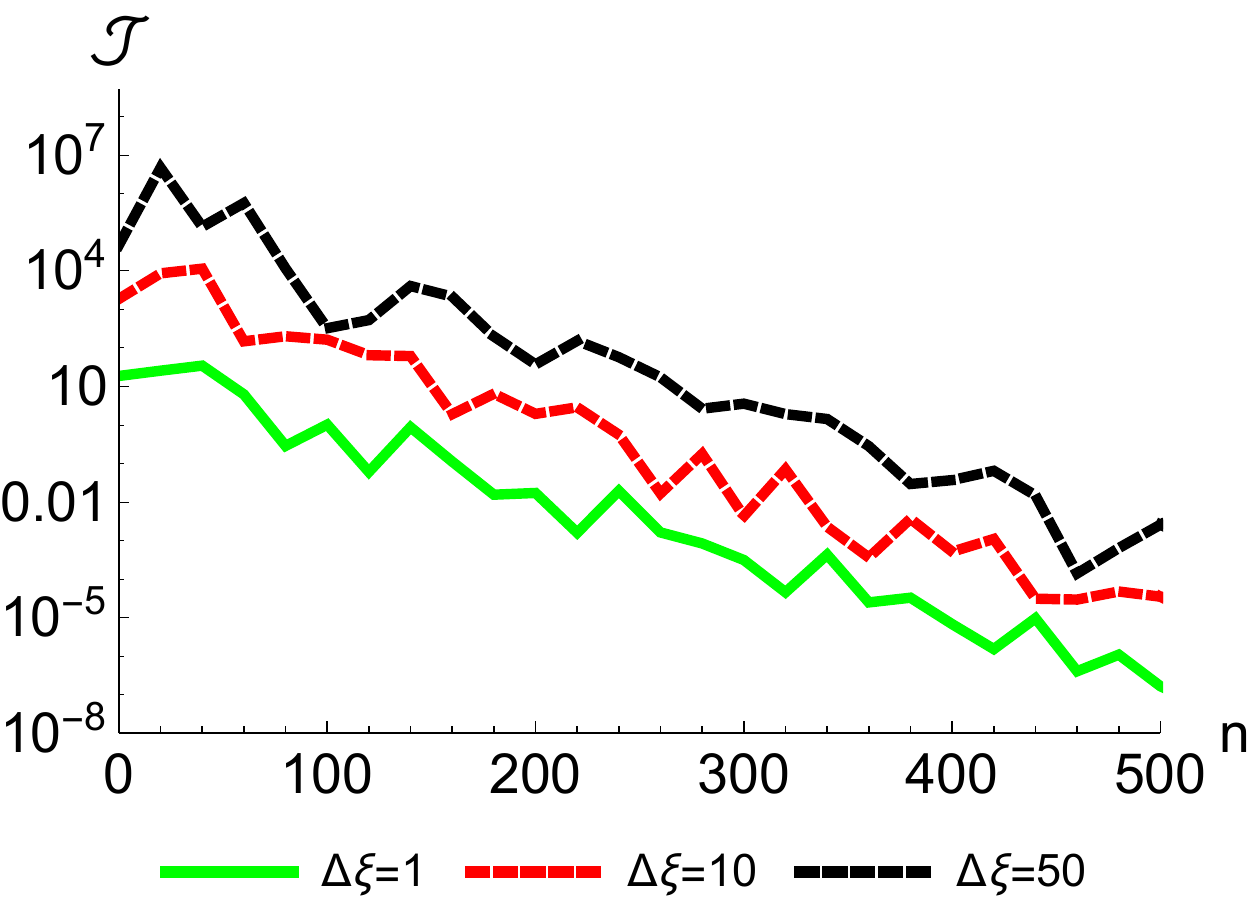}
}\\
\subfigure[${\cal J} \sim C_y$ for $\alpha=0.3$]
{
	\includegraphics[width=0.3\hsize]{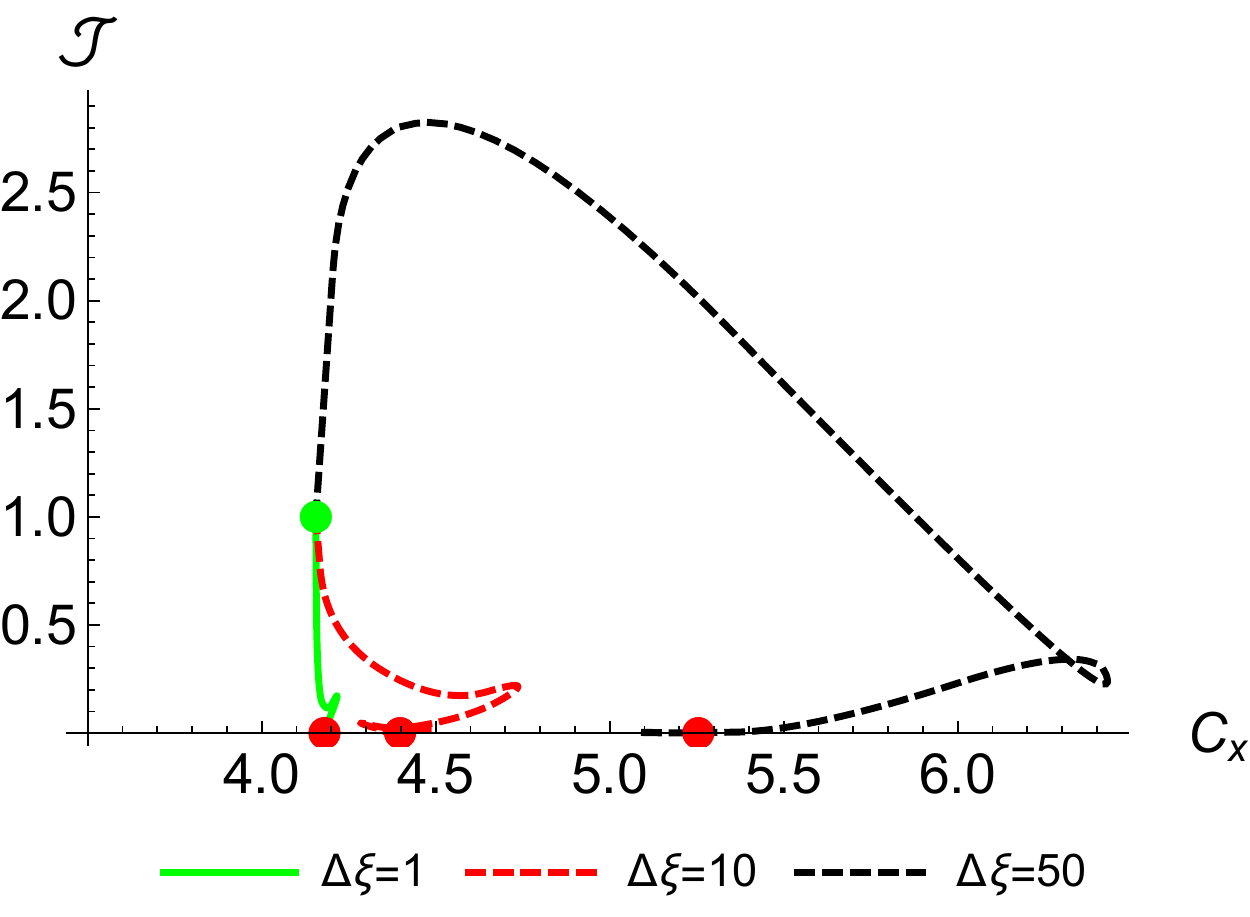}
}
\subfigure[${\cal J} \sim C_{vv}$ for $\alpha=0.3$]
{
	\includegraphics[width=0.3\hsize]{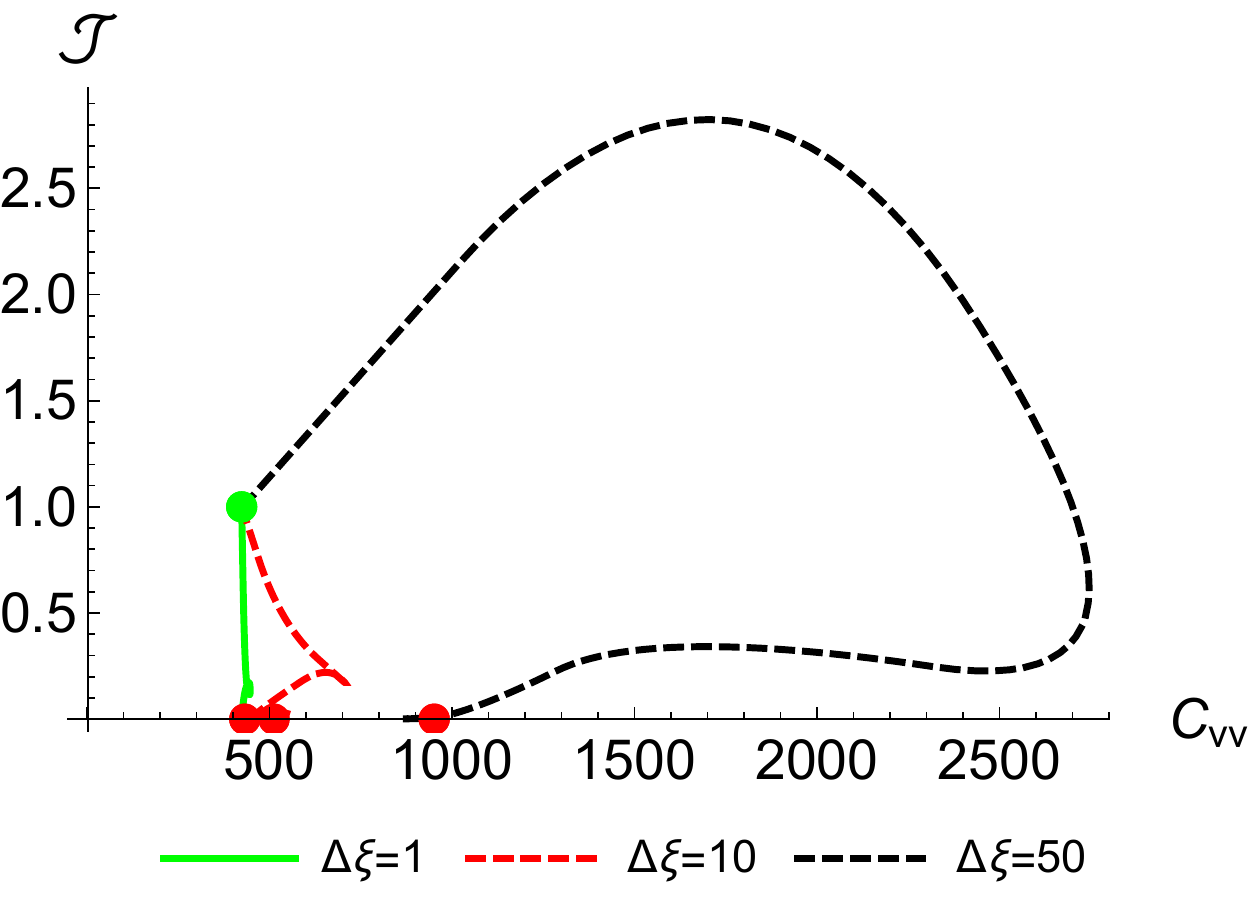}
}
\subfigure[${\cal J} \sim n$ for $\alpha=0.3$]
{
	\includegraphics[width=0.3\hsize]{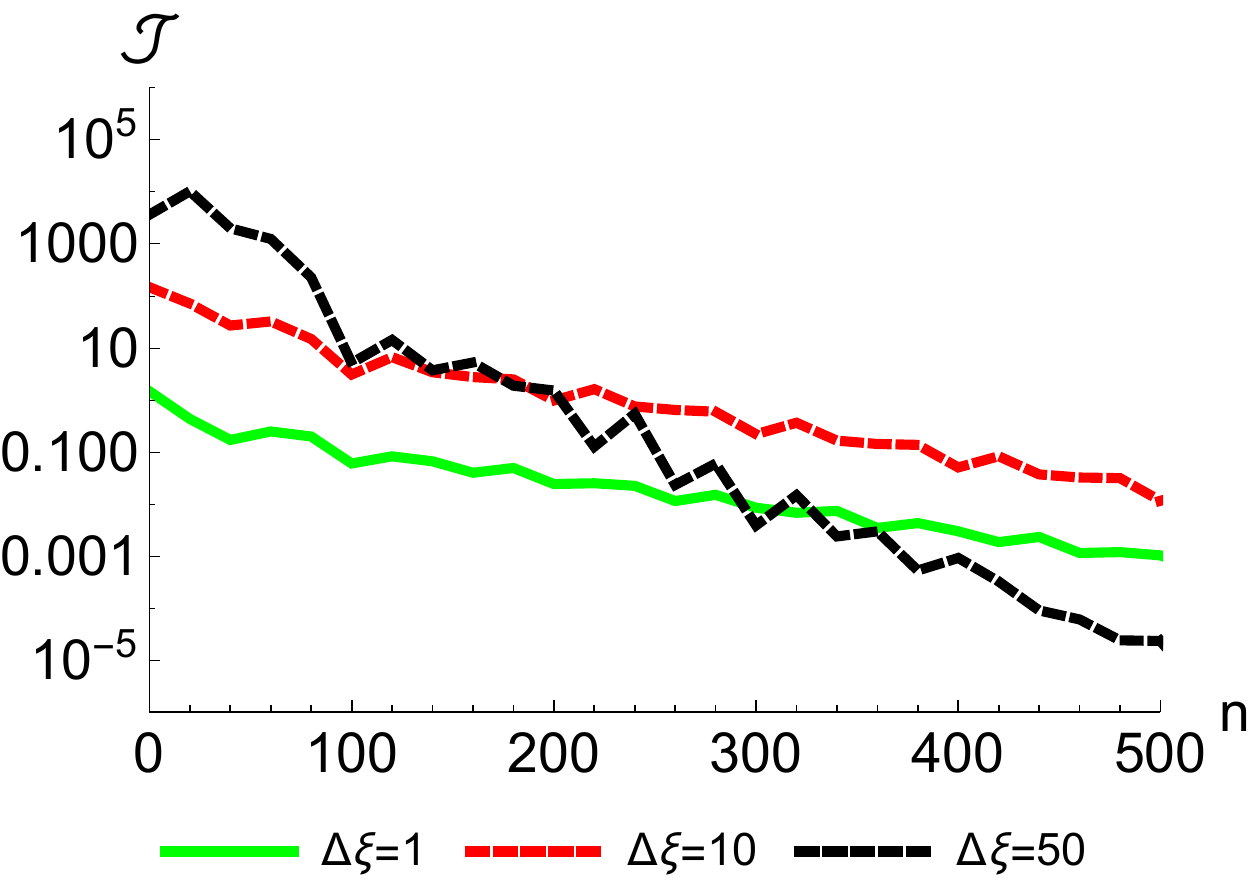}
}
\caption{Illustration of the path of the low-order statistics as the misfit, ${\cal J}$, converges to zero and the convergence of ${\cal J}$ as the number of iterations using the time stepping method for migrating of the stable time-invariant solution of the disc dynamo from $\xi=100$ to $\xi = 101, 110$ and $150$ for $\alpha=1.5$ and $\alpha=0.3$.}
\label{disMig2}
\end{figure}

Using the time stepping method, we always find the stable fixed point of the disc dynamo system, e.g., see Fig. (\ref{disMig2})
for the illustration of the path of the low-order statistics as the misfit, ${\cal J}$, converges to zero for the disc dynamo from $\xi=100$ to $\xi = 101, 110$ and $150$ for $\alpha=1.5$ and $\alpha=0.3$. The convergence rate is always exponential.

\begin{figure}
\centering
\subfigure[]
{
	\includegraphics[width=0.35\hsize]{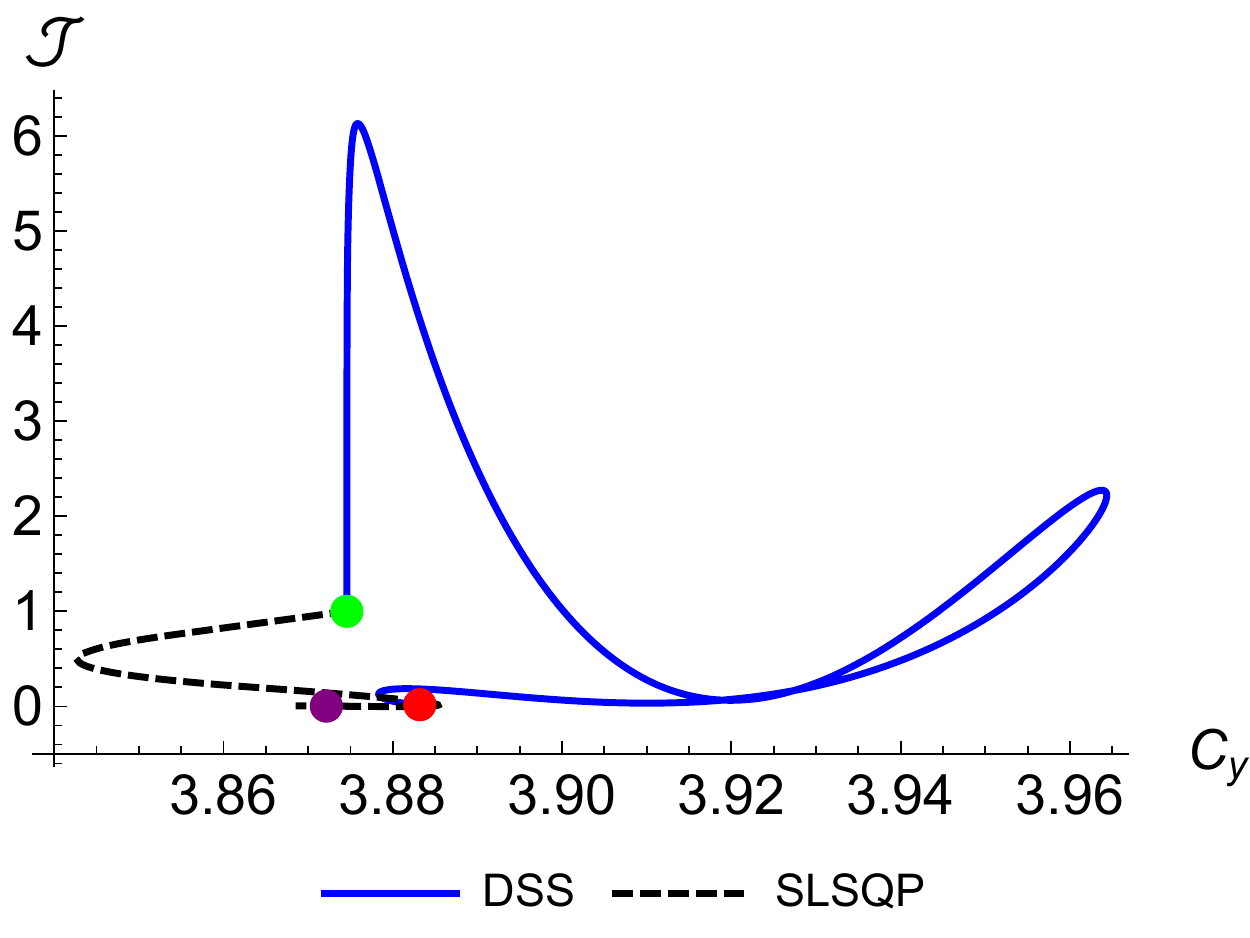}
}
\subfigure[]
{
	\includegraphics[width=0.35\hsize]{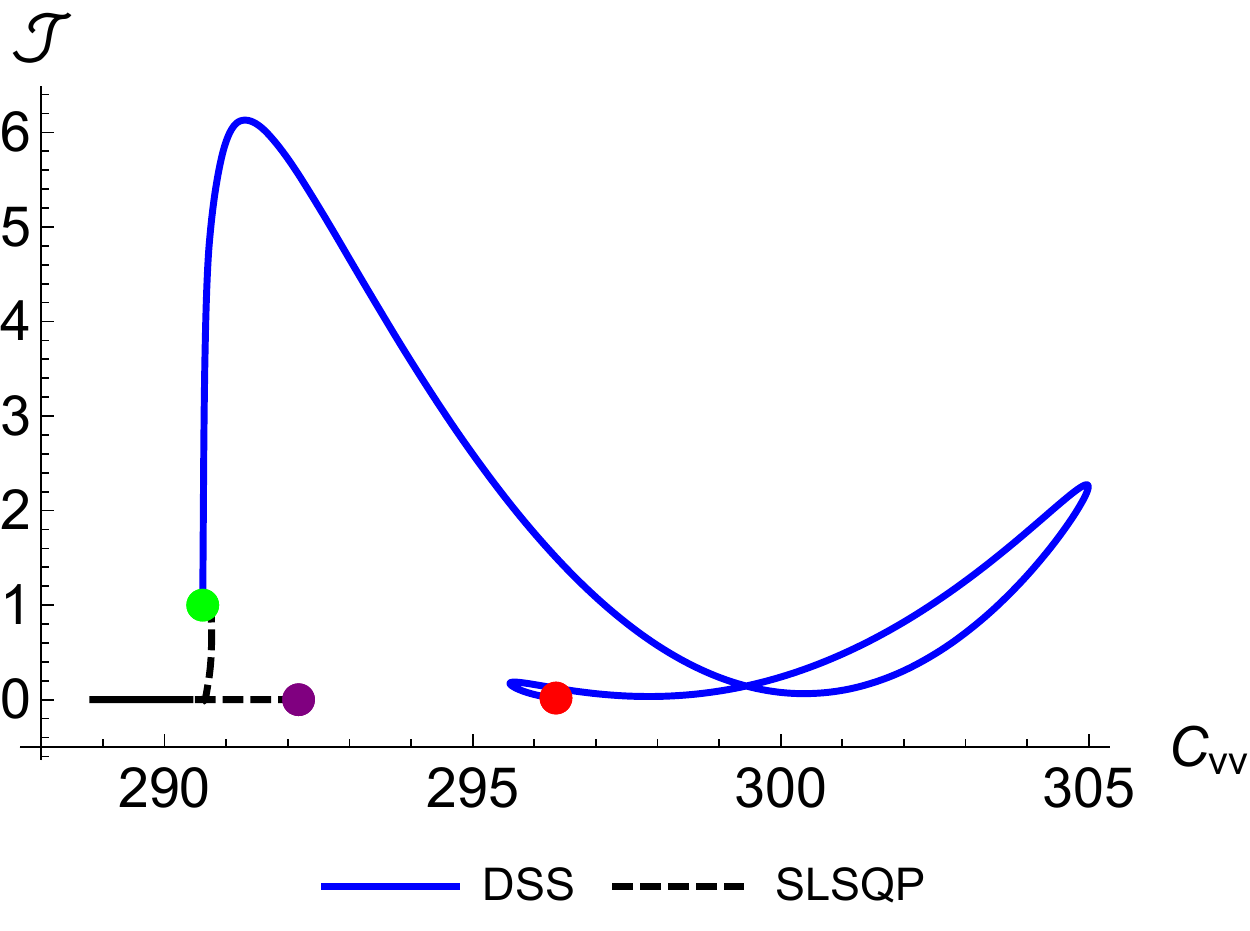}
}
\caption{Path for ${\cal J}$ as we migrate the steady solution of CE3 equation from $\xi=100$ to $101$ via the {\it SLSQP} and time stepping method for $\alpha=1.5$, where the initial guess/condition that are shown in green dots is taken from the stable fixed point of the cumulant equation of $\xi=100$, the terminal solution for {\it SLSQP} is shown in purple dots and the stable fixed point is in red dots.}
\label{disdynamoFixP2}
\end{figure}

We note here that other minimization methods, such as {\it SLSQP} or {\it trust-region}, have some limited power for computing the stable fixed point of the cumulant equations. Shown in Fig. (\ref{disdynamoFixP2})
is the optimal solution found by {\it SLSQP} method for the disc dynamo in the chaotic state for $\xi=101$ and $\alpha=1.5$, where the initial guess for the optimization is taken from the stable fixed point of the dynamical system for $\xi=100$ and $\alpha=1.5$. The first order cumulant, $C_y$, obtained by {\it SLSQP} method is only $0.3\%$ less than the true value obtained via the time stepping method. But for $C_{vv}$, the error increases to $1\%$. The misfit, ${\cal J}$, saturates at the level of ${\cal O}(10^{-4})$.
\clearpage

\section{Conclusions}

In this paper, we have implemented direct statistical simulation (DSS) of two simple models of dynamo action that yield a range of dynamical behaviour from periodic or quasiperiodic behaviour to complicated chaotic dynamics. The first low-order ODE model of the solar dynamo has cubic nonlinearities and modulation of the basic dynamo cycle occurs on a long timescale determined by interaction with the velocity. The second model involves the dynamics of a disc dynamo driven by a convective loop; here the complicated dynamics is reminiscent of that of the Lorenz63 system and involves aperiodic reversals either of the convective motion or the dynamo-generated magnetic field. We also consider the case where stochastic driving can be added to the deterministic dynamics of the first of these models.

In the absence of a stochastic force, the probability distributions of both dynamo systems in all dynamical states are strongly non-Gaussian, which is expected to cause problems for variants of direct statistical simulation that are severely truncated. We find that even for these highly non-Gaussian distributions DSS is accurate if  the cumulant equations are truncated at third order. Both CE2.5 and CE3 approximation are found numerically stable and accurately computes the long-term evolution of both systems, where the most accurate solutions are obtained for the eddy damping parameter, $\tau_d$, in the range from $10^{-2}$ to $10^{-1}$. As expected the first cumulants are more accurately computed by the truncation than the second cumulants, which may exhibit $20\%$ relative error due to the strong asymmetry of the PDFs. We find that the accuracy of the solution of the cumulant equations increases as increasing the chaoticity of the dynamo system. 

In the presence of stochastic terms the PDF for the solar dynamo system is regularised and closer to Gaussian. For $\sigma_{x,y,z}^2>1$, the solar dynamo model is closer to Gaussian and may be accurately solved by the CE2 approximation --- the most severe truncation of DSS. All of the above conclusions are obtained by timestepping forward the cumulant hierarchy equations to find stable fixed points for DSS.

We have also attempted to find directly the fixed points (both stable and unstable) of the cumulant hierarchy systems. We find that, when the 
 CE2 approximation is applicable, gradient based optimization works much more efficiently than a timestepping method to locate the stable fixed point; here the solution is as accurate as that obtained via the time stepping method. However increasing the order of the cumulant hierarchy introduces the possibility of many unstable fixed points --- this is true both for CE2.5 and CE3. We find that for the CE2.5 approximation, the gradient based method converges  to a fixed point that is either statistically unrealizable or unstable. The convergence is also slower than that for the timestepping method. Interestingly for CE3 the gradient method never succeeds in converging to a fixed point. We find that  other optimization methods, such as SLSQP, may also converge to the correct fixed point, but with larger numerical error than the time stepping method.
 
 We conclude by commenting on the consequences of our findings for DSS of more complicated models involving systems of PDEs that lead to turbulent dynamo action (e.g. \citet{sb2015}). The problems we have looked at in this paper, although described by a system of low-order ODES, are extremely nonlinear and the PDFs associated with the attractors are far from Gaussian. Moreover in some of the cases they have zero or very small first cumulants. For all these reasons one expects that DSS based on low order cumulant truncations should perform poorly. Nonetheless we have determined that sufficiently high-order DSS is able to predict the values of the low-order cumulants. We believe that 
 turbulent systems in general will be less nonlinear and more stochastic than these low-order models; it is therefore to be 
 hoped that cumulant hierarchies based on perturbations away from CE2 will give an accurate descriptions. In that case, our results show that  either timestepping or CG methods should be an  efficient method for determining the stable fixed points and continuing solutions in parameter space.
 

\section*{Acknowledgements}

This is supported in part by European Research Council (ERC) under the European Unions Horizon 2020 research and innovation program (grant agreement no. D5S-DLV-786780) and by a grant from the Simons Foundation (Grant number 662962, GF). We would also like to acknowledge discussions with Liliya Milenska.

\bibliographystyle{plainnat}
\bibliography{dss_dynamo}

\appendix

\section{The cumulant expansion of the quadratic and cubic nonlinear terms}
\label{demo}

In this section, we derive the cumulant expansions of the cubic and quadratic nonlinear terms, $x_jx_kx_l$ and $x_jx_k$. The first order cumulant expansion of $x_jx_kx_l$ is defined as the statistical average, $\left\langle x_jx_kx_l \right\rangle$, i.e.,
\begin{eqnarray}
\left\langle x_jx_kx_l \right\rangle &=& \left\langle(C_{x_j}+\delta x_j)(C_{x_k}+\delta x_k)(C_{x_l}+\delta x_l) \right\rangle \nonumber \\
&=& C_{x_j}C_{x_k}C_{x_l} + C_{x_j}C_{x_kx_l} + C_{x_jx_k}C_{x_l} + C_{x_jx_l}C_{x_k} + C_{x_jx_kx_l}.
\end{eqnarray}
Following the definition of the higher order cumulant, the second and the third order cumulant expansions are found to satisfy the following equation,
\begin{eqnarray}
\left\langle \delta x_m [x_jx_kx_l - \left\langle x_jx_kx_l \right\rangle] \right\rangle &=& C_{x_jx_kx_lx_m} + C_{x_j}C_{x_kx_lx_m} + C_{x_jx_kx_m}C_{x_l}+ C_{x_jx_lx_m}C_{x_k} \nonumber \\
&+&C_{x_j}C_{x_k}C_{x_lx_m} + C_{x_j}C_{x_kx_m}C_{x_l}  + C_{x_jx_m}C_{x_k}C_{x_l} \nonumber \\
&-& C_{x_jx_k}C_{x_lx_m}   - C_{x_jx_l}C_{x_kx_m}   - C_{x_jx_m}C_{x_kx_l}
\end{eqnarray}
and
\begin{eqnarray}
\left\langle \delta x_m \delta x_n [x_jx_kx_l - \left\langle x_jx_kx_l \right\rangle] \right\rangle &=& C_{x_j}C_{x_k}C_{x_lx_mx_n} + C_{x_j}C_{x_kx_mx_n}C_{x_l} + C_{x_jx_mx_n}C_{x_k}C_{x_l} \nonumber \\
&-& C_{x_j}C_{x_kx_l}C_{x_mx_n} - C_{x_jx_k}C_{x_l}C_{x_mx_n}  - C_{x_jx_l}C_{x_k}C_{x_mx_n} - C_{x_jx_kx_l}C_{x_mx_n}\nonumber \\
&+& C_{x_j}(-C_{x_kx_l}C_{x_mx_n} + C_{x_kx_lx_mx_n} - C_{x_kx_m}C_{x_lx_n} - C_{x_kx_n}C_{x_lx_m}) \nonumber \\
&+& C_{x_k}(-C_{x_jx_l}C_{x_mx_n} + C_{x_jx_lx_mx_n} - C_{x_jx_m}C_{x_lx_n} - C_{x_jx_n}C_{x_lx_m}) \nonumber \\
&+& C_{x_l}(-C_{x_jx_k}C_{x_mx_n} + C_{x_jx_kx_mx_n} - C_{x_jx_m}C_{x_kx_n} - C_{x_jx_n}C_{x_kx_m}) \nonumber \\ &+& \langle \delta x_j \delta x_k \delta x_l \delta x_m \delta x_n \rangle,
\label{cubicC3}
\end{eqnarray}
where $x_m$ and $x_n$ are the arbitrary components of the low-order solar dynamo system and the last term in Eq. (\ref{cubicC3}) is the 5-point correlation and is set to zero in our study.

Similarly, the expressions of the first, second and third order cumulant expansion of the quadratic nonlinear term, $x_jx_k$, are determined as follows,
\begin{eqnarray}
\left\langle x_jx_k \right\rangle &=& \left\langle(C_{x_j}+\delta x_j)(C_{x_k}+\delta x_k) \right\rangle = C_{x_j}C_{x_k} + C_{x_jx_k},
\end{eqnarray}
\begin{eqnarray}
\left\langle \delta x_m [x_jx_k - \left\langle x_jx_k \right\rangle] \right\rangle &=& C_{x_j}C_{x_kx_m}+ C_{x_jx_m}C_{x_k} + C_{x_jx_kx_m}
\end{eqnarray}
and
\begin{eqnarray}
\left\langle \delta x_m \delta x_n [x_jx_k - \left\langle x_jx_k \right\rangle] \right\rangle &=& C_{x_jx_kx_mx_n} + C_{x_j}C_{x_kx_mx_n}  + C_{x_jx_mx_n}C_{x_k} \nonumber \\
&-& 2C_{x_jx_k}C_{x_mx_n} - C_{x_jx_m}C_{x_kx_n} - C_{x_jx_n}C_{x_kx_m},
\end{eqnarray}
where $x_m$ and $x_n$ are the arbitrary components of the low-order dynamical system.
\section{The cumulant expansion of the solar dynamo up to the third order}
\label{appsolar}

Following the definition of the second order cumulant, we obtain six second order equations for the disc dynamo system, i.e.,
{\tiny
\begin{eqnarray}
d_t C_{xx} &=& 6d (C_x^2 C_{xx} + C_xC_{xxx} - 2C_xC_{xy}C_y - C_xC_{xyy} + C_{xx}^2 - C_{xx}C_y^2 - 2C_{xxy}C_y - C_{xx} C_{yy} - 2 C_{xy}^2), \nonumber \\
&+& 2a (C_xC_{xz} + C_{xx}C_z + C_{xxz}) + 2\lambda C_{xx} - 2\omega C_{xy}, \nonumber \\
d_t C_{xy} &=& 3d (2C_x^2C_{xy} + 2C_xC_{xx}C_y + 3C_xC_{xxy} - 2C_xC_yC_{yy} - C_xC_{yyy} + 4C_{xx}C_{xy} + C_{xxx}C_y - 2C_{xy}C_y^2 - 4C_{xy}C_{yy} - 3C_{xyy}C_y), \nonumber \\
&+& a (C_xC_{yz} + 2C_{xy}C_z + 2C_{xyz} + C_{xz}C_y) + \omega C_{xx}   + 2\lambda C_{xy} - \omega C_{yy} \nonumber \\
d_t C_{yy} &=& 6d( C_x^2C_{yy} + 2C_xC_{xy}C_y + 2C_xC_{xyy} + C_{xxy}*C_y - C_y^2C_{yy} - C_yC_{yyy} - C_{yy}^2 + C_{xx}C_{yy} + 2C_{xy}^2) \nonumber \\
&+& 2a (C_yC_{yz} + C_{yyz} + C_{yy}C_z) + 2\lambda C_{yy} + 2\omega C_{xy}, \nonumber \\
d_t C_{xz} &=& 3d(C_x^2C_{xz} + C_xC_{xxz} - 2C_xC_yC_{yz} - C_xC_{yyz}  + C_{xx}C_{xz} - 2C_{xy}C_{yz} + C_{xz}C_{yy} - 2C_{xyz}C_y - C_{xz}C_y^2) \nonumber \\
&+& a (C_xC_{zz} + C_{xz}C_z + C_{xzz}) + 3c(C_{xz}C_z^2 + C_{xz}C_{zz}  + C_{xzz}C_z)  - C_{xzz}  - C_{xxx} - 2C_{xy}C_y - C_{xyy} - 2C_xC_{xx}- 2C_{xz}C_z+ \lambda C_{xz} - \omega C_{yz}, \nonumber\\
d_t C_{yz} &=& 3d(C_x^2C_{yz} + 2C_xC_{xyz} + 2C_xC_{xz}C_y + C_{xxz}C_y - C_y^2C_{yz} - C_yC_{yyz} - C_{yy}C_{yz} + C_{xx}C_{yz} + 2C_{xy}C_{xz}) \nonumber \\
&+& a (C_yC_{zz} + C_{yz}C_z + C_{yzz}) + 3c(C_{yz}C_z^2 + C_{yz}C_{zz} + C_{yzz}C_z) - 2C_xC_{xy}  - C_{xxy}  - 2C_yC_{yy}  - C_{yyy}  - 2C_{yz}C_z - C_{yzz}  +\lambda C_{yz} + \omega C_{xz}, \nonumber\\
d_t C_{zz} &=& 6c(C_z^2C_{zz} + C_zC_{zzz} + C_{zz}^2) -4C_xC_{xz} - 2C_{xxz} - 4C_yC_{yz} - 2C_{yyz} - 4C_zC_{zz} - 2C_{zzz},
\label{solarc2}
\end{eqnarray}
}
and ten third order equations,
{\tiny
\begin{eqnarray}
d_t C_{xxx} &=&3d (C_x^2C_{xxx} + 6C_xC_{xx}^2 + 3C_xC_{xx}C_{yy} - 6C_xC_{xxy}C_y - 3C_xC_{xx}C_{yy} - 6C_x C_{xy}^2 - C_{xx}C_{xxx} + 3C_{xx}C_{xyy} - 3C_{xxx}C_y^2 - 12C_{xx}C_{xy}C_y) \nonumber \\
&+& 3a(C_xC_{xxz}   + 2C_{xx}C_{xz}  + C_{xxx}C_z) + 3\lambda C_{xxx} - 3\omega C_{xxy}, \nonumber \\
d_t C_{yyy} &=&
3d(3C_x^2C_{yyy} + 12C_xC_{xy}C_{yy} + 6C_xC_{xyy}C_y - 3C_{xx}C_yC_{yy} - 3C_{xxy}C_{yy} - 3C_y^2C_{yyy} - 6C_yC_{yy}^2 + 9C_yC_{xx}*C_{yy} + 6C_yC_{xy}^2 + C_{yy}C_{yyy}) \nonumber\\
&+& 3a(C_yC_{yyz}  + 2C_{yy}C_{yz} + C_{yyy}*C_z) + 3\omega C_{xyy} + 3 \lambda C_{yyy},\nonumber\\
d_t C_{zzz} &=&3c(3C_z^2C_{zzz} + 6C_zC_{zz}^2 - C_{zz}C_{zzz}) -6C_xC_{xzz} - 6C_{xz}^2 - 6C_yC_{yzz} - 6C_{yz}^2- 6C_zC_{zzz} - 6C_{zz}^2, \nonumber \\
d_t C_{xxy} &=& d (9C_x^2C_{xxy} + 24C_xC_{xx}C_{xy} + 6C_xC_{xxx}C_y - 12C_xC_{xy}C_{yy} - 12C_xC_{xyy}C_y + 6C_{xx}^2C_y - 3C_{xx}C_{xxy} + 3C_{xx}C_yC_{yy} + C_{xx}C_{yyy} - 9C_{xxy}C_y^2 \nonumber \\
&+& 12C_{xy}^2C_y + 6C_{xy}C_{xyy} - 15C_yC_{xx}C_{yy} + 30 C_yC_{xy}^2- 2C_{xxx}C_{xy}) \nonumber\\
&+& a(2C_xC_{xyz}  - C_{xx}C_{yz} + 3C_{xxy}C_z + C_{xxz}C_y - 2C_{xy}C_{xz} + 3aC_{xx}C_{yz} + 6aC_{xy}C_{xz}) + 3\lambda C_{xxy} + \omega(C_{xxx}- 2C_{xyy}), \nonumber \\
d_t C_{xyy} &=& d(9C_x^2C_{xyy} - 3C_xC_{xx}C_{yy} + 12C_xC_{xxy}C_y - 12C_xC_{xy}^2 - 6C_xC_yC_{yyy} - 6C_xC_{yy}^2 + 12C_{xx}C_{xy}C_y - C_{xxx}C_{yy} - 6C_{xxy}C_{xy} \nonumber\\
&-& 24C_{xy}C_yC_{yy} + 2C_{xy}C_{yyy} - 9C_{xyy}C_y^2 + 3C_{xyy}C_{yy} + 15C_xC_{xx}C_{yy} + 30C_xC_{xy}^2) \nonumber\\
&+& a (C_xC_{yyz}   - 2C_{xy}C_{yz} + 3C_{xyy}C_z + 2C_{xyz}C_y - C_{xz}C_{yy} + 6C_{xy}C_{yz} + 3C_{xz}C_{yy})+ \omega (2C_{xxy} -C_{yyy}) + 3\lambda C_{xyy},  \nonumber \\
d_t C_{xxz} &=& d(6C_x^2C_{xxz} + 12C_xC_{xx}C_{xz} - 2C_xC_{xxx} - 12C_xC_{xyz}C_y + 6C_xC_{xz}C_{yy} - 2C_{xxx}C_{xz} - 6C_{xxz}C_y^2 + 12C_{xy}C_{xz}C_y + 6C_{xyy}C_{xz} \nonumber \\
&-& 12C_xC_{xy}C_{yz} - 6C_xC_{xz}C_{yy} - 12C_yC_{xx}C_{yz} - 24C_yC_{xy}C_{xz}) \nonumber \\
&+& a(2C_xC_{xzz} - 2C_{xz}^2 + 2C_{xxz}C_z + 2C_{xx}C_{zz} + 4C_{xz}^2) + c(3C_zC_{xx}C_{zz} + 6C_zC_{xz}^2 - 3C_{xx}C_zC_{zz} - C_{xx}C_{zzz} + 3C_{xxz}C_z^2) \nonumber\\
&-& 2C_{xx}^2  - 2C_{xxy}C_y  - 2C_{xxz}C_z - 2C_{xy}^2 - 2C_{xz}^2 + 2\lambda C_{xxz} - 2\omega C_{xyz}, \nonumber\\
d_t C_{xyz} &=& d(6C_x^2C_{xyz} - 3C_xC_{xx}C_{yz} + 6C_xC_{xxz}C_y - 6C_xC_{xy}C_{xz} - 6C_xC_yC_{yyz} 
- 6C_xC_{yy}C_{yz}+ 6C_{xx}C_{xz}C_y - C_{xxx}C_{yz} - 3C_{xxy}C_{xz}\nonumber \\
&+& 6C_{xy}C_yC_{yz} + 3C_{xyy}C_{yz} - 6C_{xyz}C_y^2+ 3C_{xz}C_yC_{yy} + C_{xz}C_{yyy}+ 9C_x(C_{xx}C_{yz} + 2C_{xy}C_{xz}) - 9C_y(2C_{xy}C_{yz} + C_{xz}C_{yy}) \nonumber\\
&+& c(3C_{xyz}C_z^2 - 3C_{xy}C_zC_{zz} - C_{xy}C_{zzz}) + 3cC_z(C_{xy}C_{zz} + 2C_{xz}C_{yz}) \nonumber\\
&+& a(C_xC_{yzz}+ 2C_{xyz}C_z- 2C_{xz}C_{yz}+ C_{xzz}C_y + 2C_{xy}C_{zz} + 4C_{xz}C_{yz}) \nonumber\\
&-& 2C_xC_{xxy} - 2C_{xx}C_{xy} - 2C_{xy}C_{yy}  - 2C_{xyy}C_y - 2C_{xyz}C_z - 2C_{xz}C_{yz} + \omega(C_{xxz} - C_{yyz})   + 2\lambda C_{xyz} \nonumber\\
d_t C_{yyz} &=& d(6C_x^2C_{yyz} - 12C_xC_{xy}C_{yz} + 12C_xC_{xyz}C_y + 12C_x(2C_{xy}C_{yz} + C_{xz}C_{yy}) - 6C_{xx}C_yC_{yz} - 6C_{xxy}C_{yz} \nonumber\\
&-& 12C_yC_{yy}C_{yz}+ 6C_y(C_{xx}C_{yz} + 2C_{xy}C_{xz})+ 2C_{yyy}C_{yz} - 6C_y^2C_{yyz}) \nonumber\\
&+& c(- 3C_{yy}C_zC_{zz} - C_{yy}C_{zzz}  + 3C_{yyz}C_z^2 + 3C_z(C_{yy}C_{zz} + 2C_{yz}^2)) + a(2C_yC_{yzz}+ 2C_{yyz}C_z- 2C_{yz}^2 + 2(C_{yy}C_{zz} + 2C_{yz}^2)) \nonumber\\
&-& 2C_xC_{xyy}  - 2C_{xy}^2 - 2C_yC_{yyy}   - 2C_{yy}^2   - 2C_{yyz}C_z - 2C_{yz}^2 + 2\omega C_{xyz} + 2 \lambda C_{yyz}  \nonumber \\
d_t C_{xzz} &=& d(3C_x^2C_{xzz} - 3C_xC_{xx}C_{zz} - 6C_xC_yC_{yzz} + 3C_xC_{yy}C_{zz} - C_{xxx}C_{zz} + 6C_{xy}C_yC_{zz} + 3C_{xyy}C_{zz} - 3C_{xzz}C_y^2 + 3C_xC_{xx}C_{zz} \nonumber \\
&+& 6C_xC_{xz}^2 - 3C_xC_{yy}C_{zz} - 6C_xC_{yz}^2- 6C_yC_{xy}C_{zz} - 12C_yC_{xz}C_{yz}) \nonumber\\
&+& a(C_xC_{zzz} + 2C_{xz}C_{zz} + C_{xzz}C_z) + 2c(6C_{xz}C_zC_{zz} - C_{xz}C_{zzz} + 3C_{xzz}C_z^2) - 4C_xC_{xxz} - 4C_{xx}C_{xz}  \nonumber \\
&-& 4C_{xy}C_{yz}  - 4C_{xyz}C_y   - 4C_{xz}C_{zz}  - 4C_{xzz}C_z + \lambda C_{xzz}  - \omega C_{yzz}, \nonumber\\
d_t C_{yzz} &=& (3C_x^2C_{yzz} - 6C_xC_{xy}C_{zz} + 6C_xC_{xzz}C_y - 3C_{xx}C_yC_{zz} - 3C_{xxy}C_{zz} - 3C_y^2C_{yzz} + 3C_yC_{yy}C_{zz} + C_{yyy}C_{zz}+ 6C_xC_{xy}C_{zz} \nonumber \\
&+& 12C_xC_{xz}C_{yz} + 3C_yC_{xx}C_{zz} + 6C_yC_{xz}^2 - 3C_yC_{yy}C_{zz} + 6C_yC_{yz}^2) \nonumber\\
&+& 2c(6C_{yz}C_zC_{zz}c - C_{yz}C_{zzz} + 3C_{yzz}C_z^2) + a(C_yC_{zzz} + 2C_{yz}C_{zz} + C_{yzz}C_z) \nonumber \\
&-& 4C_xC_{xyz} - 4C_{xy}C_{xz} - 4C_yC_{yyz}  - 4C_{yz}C_{zz}  - 4C_{yzz}C_z - 4C_{yy}C_{yz} + \lambda C_{yzz} + \omega C_{xzz}.
\label{solarc3}
\end{eqnarray}
}
\section{The second order cumulant expansion of the self-excited disc dynamo}
\label{appdisc}

The second order cumulant expansion of the disc dynamo consists of fifteen equations as
\begin{eqnarray}
(d_t + 2\alpha \eta) C_{xx} &=& 2\alpha\omega (C_{xy}C_z + C_{xyz} + C_{xz}C_y) \\\nonumber
(d_t + 2\eta) C_{yy} &=& 2\omega(C_xC_{yz} + C_{xy}C_z + C_{xyz}) \\\nonumber
(d_t + 2\kappa) C_{zz} &=& 2\kappa(-C_xC_{yz} - C_{xyz} - C_{xz}C_y + C_{zu}) \\\nonumber
(d_t + 2) C_{uu} &=& 2(-C_{uv}*C_z - C_vC_{zu} - C_{zuv}) + 2\xi C_{zu} \\ \nonumber
(d_t + 2) C_{vv} &=& 2(C_uC_{zv} + C_{uv}C_z + C_{zuv}) \\\nonumber
(d_t + \alpha \eta + \eta) C_{xy} &=& \omega(C_xC_{xz} + C_{xx}C_z + C_{xxz}) + \alpha\omega(C_yC_{yz} + C_{yy}C_z + C_{yyz}) \\\nonumber
(d_t + \alpha \eta+\kappa) C_{xz} &=& \kappa(-C_xC_{xy} + C_{xu} - C_{xx}C_y - C_{xxy}) + \alpha\omega(C_yC_{zz} + C_{yz}C_z + C_{yzz}) \\ \nonumber
(d_t + \eta+\kappa) C_{yz} &=& \kappa(-C_xC_{yy} - C_{xy}C_y - C_{xyy} + C_{yu}) + \omega(C_xC_{zz}   + C_{xz}C_z + C_{xzz}) \\\nonumber
(d_t + \alpha \eta+1) C_{xu} &=& -C_vC_{xz} - C_{xv}C_z + \xi C_{xz} + \alpha\omega (- C_{xzv} + C_yC_{zu} + C_{yu}C_z + C_{yzu})\\\nonumber
(d_t + \eta+1) C_{yu} &=& \omega(C_xC_{zu} + C_{xu}C_z + C_{xzu}) -C_v*C_{yz} - C_{yv}C_z + \xi C_{yz}i - C_{yzv} \\\nonumber
(d_t + \kappa+1) C_{zu} &=& \kappa(C_{uu} - C_xC_{yu} - C_{xu}C_y - C_{xyu}) - C_vC_{zz} - C_zC_{zv} - C_{zzv}+ \xi C_{zz} \\\nonumber
(d_t + \alpha \eta+1) C_{xv} &=& C_u C_{xz} + C_{xu}C_z + C_{xzu} + \alpha\omega(C_yC_{zv} + C_{yv}C_z + C_{yzv}) \\\nonumber
(d_t + \eta+1) C_{yv} &=& \omega(C_xC_{zv} + C_{xv}C_z + C_{xzv}) + C_uC_{yz} +  C_{yu}C_z + C_{yzu}  \\\nonumber
(d_t + \kappa+1) C_{zv} &=& \kappa(C_{uv} - C_xC_{yv} - C_{xv}C_y - C_{xyv}) + C_uC_{zz} + C_zC_{zu} + C_{zzu} \\ \nonumber
(d_t + 2) C_{uv} &=& C_uC_{zu} + C_{uu}C_z - C_vC_{zv} - C_{vv}C_z + C_{zuu}  - C_{zvv} + \xi C_{zv},
\end{eqnarray}
where the third order terms are kept in the equations.

\end{document}